\documentclass[11pt]{article}
\usepackage{amsmath,amsfonts,amssymb,graphicx,epsfig,pdflscape,bbm,amsthm,multirow,subfig}
\usepackage{arydshln}
\usepackage{cite}
\usepackage[usenames]{color}
\usepackage{pstricks}
\numberwithin{equation}{section}
\allowdisplaybreaks
\graphicspath{{./eps/}}

\usepackage[backref=false]{hyperref}
\hypersetup{
colorlinks=true,
citecolor=red,
linkcolor=darkblue
}
\definecolor{darkblue}{rgb}{0,0,.8}
\definecolor{red}{rgb}{1,0,0}

\definecolor{purple}{rgb}{1,0,1}
\definecolor{darkpurple}{rgb}{1,.2,1}
\definecolor{pink}{rgb}{1,.7,.7}

\def\gbullet{{\color{green}\bullet}}

\newcommand{\nc}{\newcommand}
\long\def\ignore#1{}
\def\floor#1{\lfloor #1\rfloor}
\def\bfloor#1{\big\lfloor #1\big\rfloor}

\nc{\bgauss}[2]{\Big[\!\!\begin{array}{c} {#1}\\ {#2} \end{array}\!\!\Big]}

\def\trinomial#1#2#3{\left[{#1\atop #2,#3}\right]}
\def\qTrinomial#1#2#3{\left[{#1\atop #2,#3}\right]_{\!q}}

\nc\disp{\displaystyle}
\nc{\fh}{\hat{f}}
\nc{\muh}{\hat{\mu}}
\nc{\nuh}{\hat{\nu}}
\nc{\spos}[2]{\makebox(0,0)[#1]{$\sm{#2}$}}
\nc{\sm}[1]{{\scriptstyle #1}}
\nc{\qbar}{\overline{q}}
\nc{\bib}{\bibitem}
\nc{\al}{\alpha}
\nc{\g}{\gamma}
\nc{\G}{\Gamma}
\nc{\D}{\Delta}
\nc{\eps}{\epsilon}
\nc{\la}{\lambda}
\nc{\La}{\Lambda}
\nc{\var}{\varphi}
\nc{\pa}{\partial}
\nc{\nn}{\nonumber \\ }
\nc{\hf}{\frac{1}{2}}
\nc{\dz}{\frac{dz}{2\pi i}}
\nc{\bin}[2]{\left(\!\!\!\begin{array}{c} {#1}\\ {#2} \end{array}\!\!\!\right)}
\nc{\be}{\begin{equation}}
\nc{\ee}{\end{equation}}
\nc{\bea}{\begin{eqnarray}}
\nc{\eea}{\end{eqnarray}}
\nc{\bra}[1]{\langle {#1}|}
\nc{\ket}[1]{|{#1}\rangle}
\nc{\ketw}[1]{({#1})^{\phantom{a}}_{{\cal W}}}
\nc{\chit}{\raisebox{0.25ex}{$\chi$}}
\nc{\chih}{\raisebox{0.25ex}{$\hat\chi$}}
\nc{\Db}{\mbox{\boldmath $D$}}
\nc{\Hb}{\mbox{\boldmath $H$}}
\nc{\calH}{{\cal H}}
\nc{\calR}{{\cal R}}
\nc{\calL}{{\cal L}}
\nc{\calV}{{\cal V}}
\nc{\Hc}{{\cal H}}
\nc{\Rc}{{\cal R}}
\nc{\Lc}{{\cal L}}
\nc{\Vc}{{\cal V}}
\nc{\Ib}{\mbox{\boldmath $I$}}
\nc{\qb}{\bar{q}}
\nc{\Ac}{\mathcal{A}}
\nc{\Bc}{\mathcal{B}}
\nc{\Cc}{\mathcal{C}}
\nc{\Dc}{\mathcal{D}}
\nc{\Ec}{\mathcal{E}}
\nc{\Gc}{\mathcal{G}}
\nc{\Ic}{\mathcal{I}}
\nc{\Jc}{\mathcal{J}}
\nc{\Oc}{\mathcal{O}}
\nc{\Pc}{\mathcal{P}}
\nc{\Sc}{\mathcal{S}}
\nc{\Tc}{\mathcal{T}}
\nc{\Wc}{\mathcal{W}}
\nc{\Xc}{\mathcal{X}}
\nc{\Yc}{\mathcal{Y}}
\nc{\Zc}{\mathcal{Z}}
\nc{\fus}{\mbox{}\,\hat\otimes\,\mbox{}}
\nc{\Pch}{\hat{\Pc}}
\nc{\Rch}{\hat{\Rc}}
\nc{\Dh}{\hat{\Delta}}
\nc{\rh}{\hat{r}}
\nc{\sh}{\hat{s}}
\nc{\taub}{\bar{\tau}}
\nc{\Jcb}{\Jc_{\mathrm{b}}}
\nc{\rtt}{\mathtt{r}}
\nc{\stt}{\mathtt{s}}
\nc{\cosR}{\cos\frac{\pi p'rr'}{p}}
\nc{\cosS}{\cos\frac{\pi pss'}{p'}}
\nc{\sinR}{\sin\frac{\pi p'rr'}{p}}
\nc{\sinS}{\sin\frac{\pi pss'}{p'}}
\nc{\flexpr}[1]{
h_{#1}
}
\definecolor{apricot}{rgb}{1,0.9,0.7}

\nc{\Wtwo}[5]{W^{2,2}\Big.\Big(
\begin{matrix}
#4 & #3\\
#1 & #2
\end{matrix}
\Big|#5\Big)}

\nc{\Wt}[4]{W^{2,2}\!\left(
\begin{matrix}
#4 & #3\\
#1 & #2
\end{matrix}
\right)}

\nc{\Wtt}[6]{#1\Big.\Big(
\begin{matrix}
#5 & #4\\
#2 & #3
\end{matrix}
\Big|#6\Big)}

\def\vvdots{\mathinner{\mkern1mu\raise1pt\vbox{\kern7pt\hbox{.}}\mkern2mu
  \raise4pt\hbox{.}\mkern2mu\raise7pt\hbox{.}\mkern1mu}}
\nc{\gauss}[2]{\left[\!\!\begin{array}{c} {#1}\\ {#2} \end{array}\!\!\right]}
\nc{\sbin}[2]{\left\{\!\!\!\begin{array}{c} {#1}\\ {#2} 
\end{array}\!\!\!\right\}}
\nc{\sbinlr}[2]{\Big\langle\!\!\begin{array}{c} {#1}\\ {#2} 
\end{array}\!\!\Big\rangle}
\nc{\bino}[2]{\left(\!\!\begin{array}{c} {#1}\\ {#2} \end{array}\!\!\right)}
\def\half {\mbox{$\textstyle \frac{1}{2}$}}
\def\vec#1{\mbox {\boldmath $#1$}}

\definecolor{lightblue}{rgb}{.61,.61,1}
\definecolor{midblue}{rgb}{.7,.7,1}
\definecolor{lightlightblue}{rgb}{.85,.85,1}
\definecolor{lightestblue}{rgb}{.96,.96,1}
\definecolor{lightpurple}{rgb}{1,.65,1}

\def\loopa{
\psframe[linewidth=.25pt](0,0)(1,1)
\psarc[linewidth=1.5pt,linecolor=blue](1,0){.5}{90}{180}
\psarc[linewidth=1.5pt,linecolor=blue](0,1){.5}{-90}{0}
}
\def\loopb{
\psframe[linewidth=.25pt](0,0)(1,1)
\psarc[linewidth=1.5pt,linecolor=blue](0,0){.5}{0}{90}
\psarc[linewidth=1.5pt,linecolor=blue](1,1){.5}{180}{270}
}

\def\facegrid#1#2{
\psframe[fillstyle=solid,fillcolor=lightlightblue,linewidth=0pt]#1#2
\psgrid[gridlabels=0pt,subgriddiv=1]#1#2}

\nc{\ch}{{\rm ch}}
\nc{\R}{{\cal R}}
\nc{\dkk}{\delta_{j,\{k,k'\}}^{(2)}}
\nc{\drr}{\delta_{j,\{r,r'\}}^{(2)}}
\nc{\ddkk}{\delta_{j,\{k,k'\}}^{(4)}}
\nc{\dddkk}{\delta_{j,\{k,k'\}}^{(8)}}
\nc{\dnn}{\delta_{j,\{n,n'\}}^{(2)}}
\nc{\ddnn}{\delta_{j,\{n,n'\}}^{(4)}}
\nc{\dddnn}{\delta_{j,\{n,n'\}}^{(8)}}

\def\faceu#1#2#3#4#5{\ \
\begin{pspicture}[shift=-.6](0,-.25)(1,1.25)
\pspolygon[linewidth=.5pt,fillstyle=solid,fillcolor=lightlightblue](0,0)(1,0)(1,1)(0,1)(0,0)
\rput[tr](0,0){\scriptsize $#1$}
\rput[tl](1,0){\scriptsize $#2$}
\rput[bl](1,1){\scriptsize $#3$}
\rput[br](0,1){\scriptsize $#4$}
\rput(.5,.5){\small $#5$}
\end{pspicture}}
\nc{\thf}{s}
\nc{\therm}{t}
\nc{\mydot}[0]{
\pscircle[fillstyle=solid,fillcolor=black](0,0){0.08}
}

\definecolor{pink}{rgb}{1,.65,.65}

\begin{document}

\topmargin -5mm
\oddsidemargin 5mm

\setcounter{page}{1}

\vspace{8mm}
\begin{center}
{\huge {\bf One-Dimensional Sums and Finitized}}\\[10pt]
{\huge {\bf Characters of $2 \times 2$ Fused RSOS Models}}
\\[.3cm]
{\huge {\bf}}

\vspace{10mm}
{\Large Gy\"orgy Z. Feh\'er$^{1,2}$, Paul A. Pearce$^2$ and Alessandra Vittorini-Orgeas$^2$}\\[.3cm]
{\em $^1$Instituut voor Theoretische Fysica}\\
{\em Universiteit van Amsterdam, Science Park 904}\\
{\em 1098 XH Amsterdam. The Netherlands}\\[.3cm]
{\em $^1$BME ``Momentum" Statistical Field Theory Research Group}\\
{\em 1111 Budapest, Budafoki \'ut 8, Hungary}\\[.3cm]
{\em $^2$School of Mathematics and Statistics, University of Melbourne}\\
{\em Parkville, Victoria 3010, Australia}\\[.4cm]
{\tt g.feher\,@\,eik.bme.hu}, \ \ \ {\tt papearce\,@\,unimelb.edu.au}, \ \ \ \mbox{\tt alessandra.vittorini\,@\,unimelb.edu.au}

\end{center}

\vspace{8mm}
\centerline{{\bf{Abstract}}}
\vskip.4cm
\noindent
Tartaglia and Pearce have argued that the nonunitary $n\times n$ fused Forrester-Baxter $\mbox{RSOS}(m,m')$ models are described, in the continuum scaling limit, by the 
minimal models ${\cal M}(M,M',n)$ constructed as the higher-level conformal cosets 
$(A^{(1)}_1)_k\otimes (A^{(1)}_1)_n/(A^{(1)}_1)_{k+n}$  at integer fusion level $n\ge 1$ and fractional level $k=nM/(M'\!-\!M)-2$ with $(M,M')=\big(nm-(n\!-\!1)m',m'\big)$. 
These results rely on Yang-Baxter integrability and are valid in Regime~III for models determined by the crossing parameter $\lambda=(m'\!-\!m)\pi/m'$ in the interval $0<\lambda<\pi/n$. Combinatorially, Baxter's one-dimensional sums generate the finitized branching functions as weighted walks on the $A_{m'-1}$ Dynkin diagram.
The ground state walks terminate within shaded $n$-bands, consisting of $n$ contiguous shaded 1-bands. The shaded 1-bands occur between heights $(\rho,\rho+1)$ where $\rho=\rho(r)=\floor{\frac{r m'}{m}}$, $r-1,2,\ldots,m-1$. These results do not extend to the interval $\pi/n<\lambda<\pi$ since, for these models, there are no shaded $n$ bands to support the ground states. 
Here we consider the $2\times 2$ $\mbox{RSOS}(m,m')$ models in the interval $\tfrac{\pi}{2}<\lambda<\pi$ and investigate the associated one-dimensional sums. 
In this interval, we
verify that the one-dimensional sums produce new finitized Virasoro characters $\ch_{r,s}^{(N)}(q)$ of the minimal models ${\cal M}(m,m',1)$ with $m'>2m$. We further conjecture finitized bosonic forms and check that these agree with the ground state one-dimensional sums out to system sizes $N=12$. 
The $2\times 2$ $\mbox{RSOS}(m,m')$ models thus realize new Yang-Baxter integrable models in the universality classes of the minimal models ${\cal M}(m,m',1)$. 
For the series ${\cal M}(m,2m+1,1)$ with $m\ge 2$, the spin-1 one-dimensional sums were previously analysed by Jacob and Mathieu without the underlying Yang-Baxter structure.
Finitized Kac characters $\chi_{r,s}^{m,m';(N)}(q)$ for the logarithmic minimal models ${\cal LM}(p,p',1)$ are also obtained for $p'\ge 2p$ by taking the {\em logarithmic limit\/} $m,m'\to\infty$ with $m'/m\to p'/p+$. 


\renewcommand{\thefootnote}{\arabic{footnote}}
\setcounter{footnote}{0}

\newpage
\tableofcontents

\newpage
\section{Introduction}

In 1984 there appeared a remarkable concurrence of four interrelated papers. First, within the context of Conformal Field Theory (CFT), Belavin-Polyakov-Zamolodchikov~\cite{BPZ84} introduced the minimal model CFTs ${\cal M}(m,m')$ with central charges
\bea
c=c^{m,m'}=1-\frac{6(m'-m)^2}{mm'},\qquad 2\le m<m',\qquad\mbox{$m,m'$ coprime}
\label{central}
\eea
Friedan-Qiu-Shenker~\cite{FQS84} showed that unitarity restricts the values of $m,m'$ such that $m=m'\!-\!1$. 
Second, within the context of lattice statistical mechanices, Andrews-Baxter-Forrester~\cite{ABF84} solved exactly an infinite sequence of two-dimensional Restricted Solid-on-Solid (RSOS) models denoted by \mbox{$\mbox{RSOS}(m'\!-\!1,m')$}. Next, based on critical exponents, Huse~\cite{Huse84} argued (i) that the infinite sequence of lattice models $\mbox{RSOS}(m'\!-\!1,m')$ fall into the universality classes of multicritical Ising models and (ii) that, in the continuum scaling limit, they realize the unitary minimal models ${\cal M}(m'\!-\!1,m')$. Subsequently, for $m<m'\!-\!1$, it was shown that the $\mbox{RSOS}(m,m')$ lattice models~\cite{FB85} realize~\cite{Riggs89,Nakanishi90} the nonunitary minimal models ${\cal M}(m,m')$. These models include the Ising model ${\cal M}(3,4)$~\cite{Ising,Onsager,McCoyWu}, the tricritical Ising model ${\cal M}(4,5)$~\cite{Griffiths,BEG,LMC,KP91} and the Lee-Yang model ${\cal M}(2,5)$~\cite{LeeYang,Fisher78,Cardy85,FNO,BajnokDP}.

The $\mbox{RSOS}(m,m')$ lattice models~\cite{ABF84,FB85} are Yang-Baxter integrable~\cite{BaxBook} both at criticality and off-criticality. In the continuum scaling limit, the off-critical $\mbox{RSOS}(m,m')$ lattice models realize the integrable $\varphi_{1,3}$ thermally perturbed minimal models of Zamolodchikov~\cite{ZamPert}. Further Yang-Baxter integrable models, denoted by $\mbox{RSOS}(m,m')_{n\times n}$, are constructed by using fusion~\cite{KRS81} to build face weights from $n\times n$ blocks of elementary face weights of the $\mbox{RSOS}(m,m')$ lattice models. On the CFT side, the related minimal models ${\cal M}(M,M',n)$ are constructed~\cite{GKO} as the higher-level Goddard-Kent-Olive (GKO) cosets
\bea
 \mbox{COSET}(k,n):\quad \frac{(A_1^{(1)})_k\oplus(A_1^{(1)})_n}{(A_1^{(1)})_{k+n}},
 \qquad k=\frac{nM}{M'-M}-2, \qquad \mbox{gcd}\Big(\frac{M'-M}{n},M'\Big)
\label{cosetAAA}
\eea
where $n=1,2,3,\ldots$ is an integer fusion level and $k$ is a fractional fusion level. The central charges of these coset CFTs are given by
\be
 c=c_k+c_n-c_{k+n}
  =\frac{3kn(k+n+4)}{(k+2)(n+2)(k+n+2)},\qquad c_k=\frac{3k}{k+2}
\ee
where $c_k$ is the central charge of the affine current algebra $(A_1^{(1)})_k$ and ${\cal M}(m,m',1)\equiv{\cal M}(m,m')$. Recently, it was argued~\cite{TP} that the  minimal cosets ${\cal M}(M,M',n)$ are given by the continuum scaling limit of the $\mbox{RSOS}(m,m')_{n\times n}$ lattice models with
\bea
(M,M')=(nm-(n\!-\!1)m',m'),\qquad nm>(n\!-\!1)\,m'
\eea
The extended family of RSOS lattice models and their related minimal CFTs, particularly the unitary theories $(M,M')=(m'\!-\!n,m')$, have been extensively studied and form a cornerstone of our understanding of statistical mechanics and its interrelation with conformal and quantum field theory. 

In this paper, we continue the investigation initiated in \cite{TP}. Specifically, we address the question of what can happen if $nm<(n\!-\!1)\,m'$. 
In these cases, it was argued in \cite{TP} that the structure of the RSOS lattice models lacks the required ``shaded $n$-bands'' needed to support the ground states of the level-$n$ coset CFTs. 
The expectation is that the continuum scaling limit of these RSOS theories therefore defaults to a level-$n'$ coset CFT with $n'<n$. Here we use the one-dimensional sums~\cite{BaxBook,BaxterCTM} associated with Baxter's off-critical Corner Transfer Matrices (CTMs) to argue that, at least for $n=2$ and $m'>2m$, the continuum scaling limit of the $\mbox{RSOS}(m,m')_{2\times 2}$ lattice models is given by the level $n'=1$ coset CFTs ${\cal M}(m,m')$. Essentially, our arguments are based on {\em Physical Combinatorics}~\cite{PhysComb,Warnaar,FodaW,HatayamaEtAl,JacobM,Mathieu,FevPW,BFMW}.

The layout of the paper is as follows. In Section~2, we introduce the RSOS$(m,m')$ lattice models of Forrester and Baxter. From the known elliptic face weights for the $n\times n$ fused models for $n=1,2,3$ we extract the local energies with a suitable choice of gauge. We also set up Baxter's one-dimensional sums and discuss the ground states for $n=2$ and $m'>2m$. In Section~3, we consider the cases with $n=2$ and $m'=2m+1$. For these cases, we show that the one-dimensional sums agree with those of Jacob and Mathieu~\cite{JacobM} based on half-integer RSOS paths. Some emphasis is placed on the simplest models, namely, RSOS$(2,5)$ and RSOS$(3,7)$. The conformal data of the nonunitary minimal models ${\cal M}(m,m')$ is presented in Section~4. In particular, for $m'>2m$, we conjecture bosonic forms for the RSOS$(m,m')_{2\times 2}$ finitized characters. For modest system sizes $N$, these agree with the one-dimensional sums and give the standard Virasoro characters of nonunitary minimal model ${\cal M}(m,m')$ in the limit $N\to\infty$. Taking the logarithmic limit leads to conjectured bosonic forms for the the finitized characters for the logarithmic minimal models ${\cal LM}(p,p')_{2\times 2}$ with $p'\ge 2p$. We finish with some concluding remarks.

\section{Forrester-Baxter $\mbox{RSOS}(m,m')$ Models}

\subsection{RSOS$(m,m')$ lattice models}

The $\mbox{RSOS}(m,m')$ lattice model is a Restricted Solid-On-Solid (RSOS) model~\cite{ABF84,FB85} 
defined on a square lattice with heights $a=1,2,\ldots,m'-1$ restricted so that nearest neighbour heights
differ by $\pm1$. The heights thus live on the $A_{m'-1}$ Dynkin diagram.
The nonzero Boltzmann face weights are\vspace{-6pt}
\begin{subequations}
\label{Boltzmann}
\begin{align}
&\Wtt{W}{a}{a\mp1}{a}{a\pm1}u=\ \  \faceu {a}{a\!\mp\!1}{a}{a\!\pm\!1}u\ \ \ =\;\thf(\lambda-u)
\label{Boltzmann1}\\
&\Wtt{W}{a\mp1}{a}{a\pm1}{a}u=
\ \ \faceu {a\!\mp\!1}{a}{a\!\pm\!1}{a}u\ \ \ =\;-{g_{a\pm 1}\over g_{a\mp 1}}\,
\frac{\thf((a\pm 1)\lambda)}{\thf(a\lambda)}\,
\thf(u)
\label{Boltzmann2}\\
&\Wtt{W}{a\pm1}{a}{a\pm1}{a}u=
\ \ \faceu {a\!\pm\!1}{a}{a\!\pm\!1}{a}u\ \ \ =\;\frac{\thf(a\lambda\pm u)}{s(a\lambda)}
\label{Boltzmann3}
\end{align}
\label{Boltzmann}
\end{subequations}
where $\thf(u)=\vartheta_1(u,\therm)/\vartheta_1(\lambda,\therm)$ is a quotient of the standard elliptic theta functions~\cite{GR}
\bea
\vartheta_{1}(u,t)=2t^{1/4}\sin u\prod_{n=1}^{\infty}(1-2t^{2n}\cos2u+t^{4n})(1-t^{2n}),\qquad 0<u<\lambda,\quad 0<t<1
\eea
$u$ is the spectral parameter and $g_a$ are arbitrary gauge factors. Unless stated otherwise, we work in the gauge $g_a=1$. The elliptic nome $\therm=e^{-\varepsilon}$
is a temperature-like variable, with $\therm^2$ measuring the
departure from criticality corresponding to the $\varphi_{1,3}$
integrable perturbation~\cite{ZamPert}. The crossing parameter is
\begin{equation}
\lambda={(m'-m)\pi\over m'},\qquad 2\le m<m',\qquad\mbox{$m,m'$ coprime}
\label{crossing}
\end{equation}
The relevant properties of the elliptic functions are given in Appendix~\ref{AppA}. The restrictions on $u$ and $t$ mean that we are working in Regime III of \cite{ABF84,FB85}.

It was shown in \cite{ABF84,FB85} that the off-critical face weights (\ref{Boltzmann}) satisfy the Yang-Baxter equations. The RSOS$(m,m')$ lattice models are therefore exactly solvable.
At the critical point $t=0$, the Boltzmann face weights reduce to trigonometric functions. 
The algebraic structure of the solution to the Yang-Baxter equation for the critical RSOS$(m,m')_{2\times 2}$ models is discussed in Appendix~\ref{AppB}.

\subsection{RSOS$(m,m')_{2\times 2}$ face weights}

The normalized $2\times2$ fused RSOS are
\bea
\Wtwo{a}{b}{c}{d}{u}=\frac{1}{\eta^{2,2}(u)}\;
\psset{unit=1cm}
\begin{pspicture}[shift=-1](-0.2,-0.2)(2.2,2.2)
\rput(0,0){\facegrid{(0,0)}{(2,2)}}
\rput(0.5,0.5){\small $u\!-\!\lambda$}
\rput(1.5,0.5){\small $u$}
\rput(1.5,1.5){\small $u\!+\!\lambda$}
\rput(0.5,1.5){\small $u$}
\rput(0,1){\mydot}
\rput(1,1){\mydot}
\rput(1,0){\mydot}
\rput(-0.15,-0.15){$a$}
\rput(2.15,-0.15){$b$}
\rput(2.15,2.15){$c$}
\rput(-0.15,2.15){$d$}
\rput(2,1){$\times$}
\rput(1,2){$\times$}
\end{pspicture}\qquad \eta^{2,2}(u) = s(2\lambda) s(u) s(u-\lambda)
\label{2x2fusion}
\eea
The black dots indicate sums over all allowed heights at the site. The crosses indicate that the weight is independent of the allowed heights on these sites. The fused weights all have a common factor $\eta^{2,2}(u)$ which is removed so that the normalized weights are entire functions of $u$. 

The explicit formulas for the 19 normalised weights are 
\begin{subequations}
\label{eq:BoltzWeights}
\begin{align}
\Wtwo{a}{a\mp2}{a}{a\pm2}{u} &= \frac{s(\lambda-u)s(2 \lambda -u)}{s(2 \lambda )} \label{eq:bw01}\\[6pt]
\Wtwo{a}{a\pm2}{a}{a}{u} &=
\Wtwo{a}{a}{a}{a\pm2}{u} = \frac{s(\lambda-u) s((a\pm1) \lambda \mp u)}{s((a\pm1) \lambda )} \label{eq:bw03} \\[6pt]
\Wtwo{a\pm2}{a}{a}{a}{u} &= -\frac{s((a\mp1) \lambda )s(u)  s(a \lambda \pm u)}{s(2 \lambda ) s(a \lambda ) s((a\pm1)\lambda )} \label{eq:bw04} \\[6pt]
\Wtwo{a}{a}{a\pm2}{a}{u} &= -\frac{s(2 \lambda ) s((a\pm2) \lambda )s(u) s(a \lambda \pm u)}{s((a-1) \lambda ) s((a+1) \lambda )} \label{eq:bw05} \\[6pt]
\Wtwo{a\pm2}{a}{a\mp2}{a}{u} &= \frac{s((a\mp2) \lambda ) s((a\mp1) \lambda ) s(u)s(\lambda +u)}{s(2 \lambda )   s(a \lambda ) s((a\pm1) \lambda )}\label{eq:bw06}\\[6pt]
\Wtwo{a\pm2}{a}{a\pm2}{a}{u} &= \frac{s(a \lambda \pm u) s((a\pm1) \lambda \pm u)}{s(a \lambda ) s((a \pm 1) \lambda )}\label{eq:bw07}\\[6pt]
\Wtwo{a}{a\pm2}{a\pm2}{a}{u} &=\Wtwo{a}{a}{a\pm2}{a\pm2}{u}= \frac{s((a\pm3) \lambda ) s(u) s(u-\lambda )}{s(2 \lambda ) s((a\pm1) \lambda )} \label{eq:bw09}\\[6pt]
\Wtwo{a}{a}{a}{a}{u} &=\frac{s(a\lambda\pm u)s((a\pm 1)\lambda\mp u)}{s(a\lambda)s((a\pm1)\lambda)}
+\frac{s((a\pm 1)\lambda)s((a\mp2)\lambda)s(u)s(u-\lambda)}{s(2\lambda)s(a\lambda)s((a\mp1)\lambda)}
\label{eq:bw08}
\end{align}
\end{subequations}
In contrast to the other equations, in the last equation, the choice of upper or lower signs gives two equivalent expressions for the same weight.

%

\subsection{RSOS$(m,m')_{2\times 2}$ paths and shaded bands}
A $2\times2$ fused RSOS lattice path $\sigma = \{\sigma_0, \sigma_1, \ldots, \sigma_N, \sigma_{N+1} \}$ is defined as a sequence of integer heights $\sigma_j\in A_{m'-1}$, $j=0,1,\ldots,N+1$ which satisfy the same adjacency rules as neighbouring heights on the corners of the $2\times 2$ fused face weights. Explicitly, these adjacency conditions are
\bea
\sigma_{j+1}-\sigma_j = 0,\pm 2,\qquad \sigma_{j+1}+\sigma_j = 4,6,\ldots,2 m'-4,\qquad &j=0,1,2,\ldots, N
\label{adjcond}
\eea
The latter constraint implies that the heights $1$ and $m'-1$ are not allowed to be adjacent to themselves. The boundary conditions are fixed by
\be
(\sigma_0,\sigma_N,\sigma_{N+1}) = (s,\rho,\rho'),\qquad \rho'-\rho=0,\pm 2
\ee
The $2\times2$ fused RSOS lattice paths are used to define Baxter's one-dimensional configurational sum in Section~\ref{1dsums}.

\begin{figure}[htb]
\centering
\psset{unit=0.75cm}
\begin{pspicture}(-0.4,-0.4)(8,11)
\pspolygon[linewidth=0pt,fillstyle=solid,fillcolor=lightestblue](0,0)(8,0)(8,11)(0,11)
\rput(0,1){\pspolygon[linewidth=0pt,fillstyle=solid,fillcolor=lightlightblue](0,0)(8,0)(8,1)(0,1)}
\rput(0,4){\pspolygon[linewidth=0pt,fillstyle=solid,fillcolor=lightlightblue](0,0)(8,0)(8,1)(0,1)}
\rput(0,6){\pspolygon[linewidth=0pt,fillstyle=solid,fillcolor=lightlightblue](0,0)(8,0)(8,1)(0,1)}
\rput(0,9){\pspolygon[linewidth=0pt,fillstyle=solid,fillcolor=lightlightblue](0,0)(8,0)(8,1)(0,1)}
\psgrid[gridlabels=0pt,subgriddiv=1]
\rput(-.25,0){1}\rput(-.25,1){2}\rput(-.25,2){3}\rput(-.25,3){4}\rput(-.25,4){5}
\rput(-.25,5){6}\rput(-.25,6){7}\rput(-.25,7){8}\rput(-.25,8){9}\rput(-.25,9){10}
\rput(-.25,10){11}\rput(-.25,11){12}
\rput(0.1,-0.35){0}\rput(1,-0.35){1}\rput(2,-0.35){2}\rput(3,-0.35){3}\rput(4,-0.35){4}\rput(5,-0.35){5}\rput(6,-0.35){6}\rput(7,-0.35){7}\rput(8,-0.35){8}
\psline[linewidth=2pt,linecolor=blue](0,2)(2,6)
\psline[linewidth=2pt,linecolor=blue](2,6)(3,4)
\psline[linewidth=2pt,linecolor=blue](3,4)(4,4)
\psline[linewidth=2pt,linecolor=blue](4,4)(5,2)
\psline[linewidth=2pt,linecolor=blue](5,2)(6,4)
\psline[linewidth=2pt,linecolor=blue](6,4)(7,2)(8,2)
\rput(8.75,1.5){$r=1$}
\rput(8.75,4.5){$r=2$}
\rput(8.75,6.5){$r=3$}
\rput(8.75,9.5){$r=4$}
\end{pspicture}
\caption{An example path $\sigma = \{3,5,7,5,5, 3,5,3,3\}$ in model $(m,m')=(5,13)$ with boundary conditions $(s,\rho,\rho')=(3,3,3)$ and $N=7$ steps from $s=3$ to $\rho=3$. The shaded 1-bands are labelled by $r=1,2,3,\ldots$ from the bottom.}
\label{fig:patheg}
\end{figure}
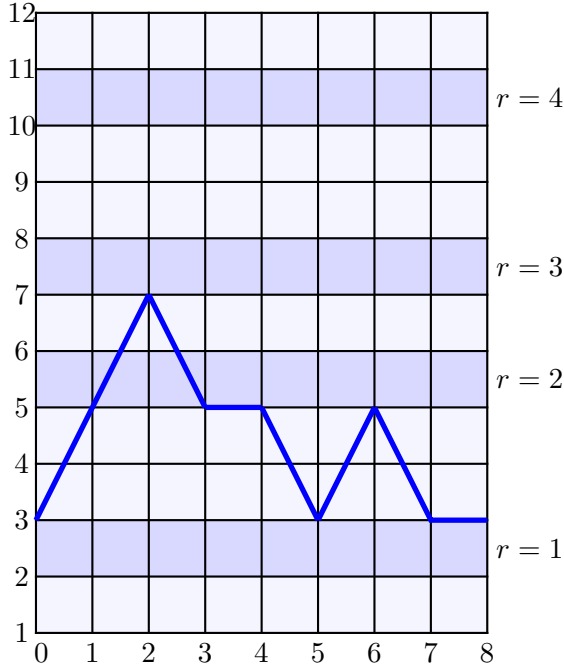

The $2\times2$ fused RSOS lattice paths on a square lattice are interpreted as ($N$+1)-step walks that start at height $s$ and then take $N$ steps, respecting the $2\times 2$ adjacency conditions (\ref{adjcond}), until height $\rho$ is reached. The last step is from height $\rho$ to height $\rho'$. Each step consists of staying at the same height or moving up or down by 2 in height. Since the heights always change by an even number, paths have a definite parity --- the heights are either all odd or all even. An example of such a path, represented as a walk on the $A_{m'-1}$ diagram, is given in Figure \ref{fig:patheg}. In this figure some bands are shaded as we now explain.

Let us define the sequence
\bea
\rho=\rho(r)=\floor {\tfrac{r m'}{m}},\qquad r=1,2,3,\ldots, m-1
\label{eq:rho}
\eea
Following \cite{FodaW}, we shade the bands in walk diagrams (such as Figure~\ref{fig:patheg}) between the heights $\rho(r)$ and $\rho(r)+1$ for $r=1,2,\ldots,m-1$.
The shaded and unshaded bands interchange under the duality $m \leftrightarrow m'$. An {\it $n$-band} consists of $n$ contiguous bands, which can be all shaded, all unshaded or mixed. An $n$-band is called a shaded $n$-band if all its 1-bands are shaded. An $n$-band is called an unshaded $n$-band if all its 1-bands are unshaded. In \cite{TP} it is shown that, for $0 < \lambda < \frac{\pi}{n}$, the number of shaded $n$-bands is
\be
\# \text{shaded }n\text{-bands}:=M-1=\begin{cases} nm-(n-1)m'-1,&0 < \lambda < \frac{\pi}{n}\\
0,&\frac{\pi}{n}<\lambda<\pi
\end{cases}
\label{eq:numSB}
\ee
For $\frac{\pi}{n}<\lambda<\pi$ there are no shaded $n$-bands.  In the unitary cases, with $\lambda= \frac{\pi}{m'}$ and $m=m'-1$, all of the 1-bands are shaded. 
For nonunitary cases, with $2\le m<m'-1$, there are both shaded and unshaded bands.

For the case of $\tfrac{\pi}{2}<\lambda<\pi$ of primary interest here, we note that there are only shaded 1-bands and no shaded $n$-bands for $n>1$. The shaded 1-bands are separated by contiguous unshaded 1-bands as in Figure~\ref{fig:patheg}. We also note that $s$ and $\rho$ have the same parity. 
For later use, it is convenient to consider the union of the two sequences $\rho(r)$ and $\rho(r)+1$ and to separate these into the union of two new sequences $\rho_0(r)$ and $\rho_1(r)$ consisting of the even and the odd members respectively.  For the example in Figure~\ref{fig:patheg}, the two new sequences are
\bea
\rho_0(r)=2,6,8,10,\qquad \rho_1(r)=3,5,7,11,\qquad r=1,2,3,4
\eea
It follows that, if we wish to end a $2\times 2$ fused RSOS lattice path on the edge of a shaded 1-band, we must have 
\bea
\rho=\rho'=\rho_\mu(r),\qquad \mu=\mbox{$s$ mod 2}
\eea
for some $r=1,2,\ldots,m-1$.



\subsection{Local energy functions}
The local energy functions $H(d,a,b)$ are extracted from the low temperature limit $t=e^{-\varepsilon}\to 1$ of the Boltzmann weights \eqref{eq:BoltzWeights} in a suitable normalization. Explicitly, with gauge factors $g_a$,
\bea
\Wtt{W^{n,n}}{a}{b}{c}{d}{u} \sim \frac{g_a g_c}{g_b g_d}\,w^{H(d,a,b)} \delta_{a,c},\qquad w=e^{-2\pi u/\varepsilon},\quad
\varepsilon \rightarrow 0,\quad u \rightarrow 0,\quad  \mbox{$\frac{u}{\varepsilon}$ fixed}
\eea
In the following subsections, we consider the cases $n=1,2,3$ separately. 
In each case, the local energies possess the reflection and height reversal symmetries
\begin{equation}
H(d,a,b)= H(b,a,d)=H(m'\!-\!d,m'\!-\!a,m'\!-\!b)
\end{equation}
These are inherited from the properties of the Boltzmann face weights.

\subsubsection{$1\times 1$ local energies}

\begin{figure}[htb]
\psset{unit=.6cm}
\begin{align}
\qquad\qquad
&\begin{pspicture}(0,.3)(2,1.5)
\psframe[linewidth=0pt,fillstyle=solid,fillcolor=lightlightblue](0,0)(2,1)
\multirput(1,0)(1,0){2}{\psline[linewidth=.5pt](0,0)(0,1)}
\psline[linewidth=2pt](0,1)(1,0)(2,1)
\end{pspicture}\ =\ 
\begin{pspicture}(0,.3)(2,1.5)
\psframe[linewidth=0pt,fillstyle=solid,fillcolor=lightlightblue](0,0)(2,1)
\multirput(1,0)(1,0){2}{\psline[linewidth=.5pt](0,0)(0,1)}
\psline[linewidth=2pt](0,0)(1,1)(2,0)
\end{pspicture}\ =\ 0
\qquad\qquad
\begin{pspicture}(0,.3)(2,1.5)
\psframe[linewidth=0pt,fillstyle=solid,fillcolor=lightestblue](0,0)(2,1)
\multirput(1,0)(1,0){2}{\psline[linewidth=.5pt](0,0)(0,1)}
\psline[linewidth=2pt](0,1)(1,0)(2,1)
\end{pspicture}\ =\ 
\begin{pspicture}(0,.3)(2,1.5)
\psframe[linewidth=0pt,fillstyle=solid,fillcolor=lightestblue](0,0)(2,1)
\multirput(1,0)(1,0){2}{\psline[linewidth=.5pt](0,0)(0,1)}
\psline[linewidth=2pt](0,0)(1,1)(2,0)
\end{pspicture}\ =\ \tfrac{1}{2}\nonumber\\[0pt]
&\begin{pspicture}[shift=-.6](0,.3)(2,2.5)
\psframe[linewidth=0pt,fillstyle=solid,fillcolor=lightlightblue](0,1)(2,2)
\psframe[linewidth=0pt,fillstyle=solid,fillcolor=lightlightblue](0,0)(2,1)
\multirput(1,0)(1,0){2}{\psline[linewidth=.5pt](0,0)(0,2)}
\psline[linewidth=2pt](0,0)(1,1)(2,2)
\end{pspicture}\ =\ 
\begin{pspicture}[shift=-.6](0,.3)(2,2.5)
\psframe[linewidth=0pt,fillstyle=solid,fillcolor=lightlightblue](0,1)(2,2)
\psframe[linewidth=0pt,fillstyle=solid,fillcolor=lightlightblue](0,0)(2,1)
\multirput(1,0)(1,0){2}{\psline[linewidth=.5pt](0,0)(0,2)}
\psline[linewidth=2pt](0,2)(1,1)(2,0)
\end{pspicture}\ =\ \tfrac{1}{2}\qquad\qquad
\begin{pspicture}[shift=-.6](0,.3)(2,2.5)
\psframe[linewidth=0pt,fillstyle=solid,fillcolor=lightestblue](0,1)(2,2)
\psframe[linewidth=0pt,fillstyle=solid,fillcolor=lightestblue](0,0)(2,1)
\multirput(1,0)(1,0){2}{\psline[linewidth=.5pt](0,0)(0,2)}
\psline[linewidth=2pt](0,0)(1,1)(2,2)
\end{pspicture}\ =\ 
\begin{pspicture}[shift=-.6](0,.3)(2,2.5)
\psframe[linewidth=0pt,fillstyle=solid,fillcolor=lightestblue](0,1)(2,2)
\psframe[linewidth=0pt,fillstyle=solid,fillcolor=lightestblue](0,0)(2,1)
\multirput(1,0)(1,0){2}{\psline[linewidth=.5pt](0,0)(0,2)}
\psline[linewidth=2pt](0,2)(1,1)(2,0)
\end{pspicture}\ =\ 0\nonumber\\[4pt]
&\begin{pspicture}[shift=-.6](0,.3)(2,2.5)
\psframe[linewidth=0pt,fillstyle=solid,fillcolor=lightestblue](0,1)(2,2)
\psframe[linewidth=0pt,fillstyle=solid,fillcolor=lightlightblue](0,0)(2,1)
\multirput(1,0)(1,0){2}{\psline[linewidth=.5pt](0,0)(0,2)}
\psline[linewidth=2pt](0,0)(1,1)(2,2)
\end{pspicture}\ =\ 
\begin{pspicture}[shift=-.6](0,.3)(2,2.5)
\psframe[linewidth=0pt,fillstyle=solid,fillcolor=lightestblue](0,1)(2,2)
\psframe[linewidth=0pt,fillstyle=solid,fillcolor=lightlightblue](0,0)(2,1)
\multirput(1,0)(1,0){2}{\psline[linewidth=.5pt](0,0)(0,2)}
\psline[linewidth=2pt](0,2)(1,1)(2,0)
\end{pspicture}\ =\ \tfrac{1}{4}\qquad\qquad
\begin{pspicture}[shift=-.6](0,.3)(2,2.5)
\psframe[linewidth=0pt,fillstyle=solid,fillcolor=lightlightblue](0,1)(2,2)
\psframe[linewidth=0pt,fillstyle=solid,fillcolor=lightestblue](0,0)(2,1)
\multirput(1,0)(1,0){2}{\psline[linewidth=.5pt](0,0)(0,2)}
\psline[linewidth=2pt](0,0)(1,1)(2,2)
\end{pspicture}\ =\ 
\begin{pspicture}[shift=-.6](0,.3)(2,2.5)
\psframe[linewidth=0pt,fillstyle=solid,fillcolor=lightlightblue](0,1)(2,2)
\psframe[linewidth=0pt,fillstyle=solid,fillcolor=lightestblue](0,0)(2,1)
\multirput(1,0)(1,0){2}{\psline[linewidth=.5pt](0,0)(0,2)}
\psline[linewidth=2pt](0,2)(1,1)(2,0)
\end{pspicture}\ =\ \tfrac{1}{4}\qquad\qquad\quad\nonumber\\[-16pt]\nonumber
\end{align}
\caption{\label{n=1H}The gauged local energies of the $1\times 1$ RSOS models in the interval $0<\lambda<\pi$. In this gauge, the local energies take the values $0,\tfrac{1}{4},\tfrac{1}{2}$.}
\end{figure}
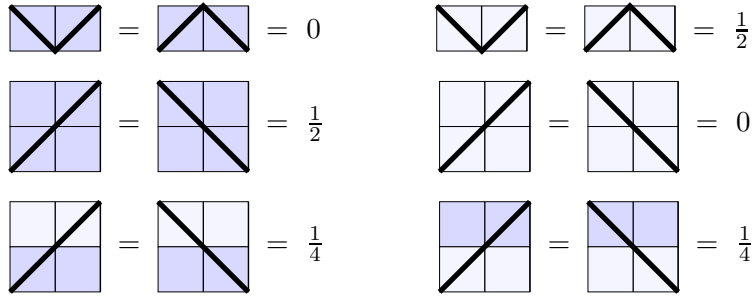
Working in the gauge $g_a=1$ for $n=1$, the local energy functions obtained by Forrester-Baxter~\cite{FB85} are
\begin{subequations}
\begin{align}
H^\text{FB}(a,a\mp 1,a)&=\pm\bfloor{\tfrac{a\lambda}{\pi}}\\
H^\text{FB}(a\!\pm\!1,a,a\!\mp\!1)&=\half
\end{align}
\end{subequations}
Changing to a more suitable gauge~\cite{TP}, the local energies for the $1\times 1$ models are given by
\begin{subequations}
\begin{align}
H(a+1,a,a+1)&=\half(h_{a+1}-h_a)\\ 
H(a-1,a,a-1)&=\half(h_a-h_{a-1})\\
H(a\pm1,a,a\mp1)&=\half-\tfrac{1}{4}(h_{a+1}-h_{a-1})
\end{align}
\label{}
\end{subequations}
These are all positive with values $0,\tfrac{1}{4},\tfrac{1}{2}$ as shown in Figure~\ref{n=1H}. The sequences
\bea
h_a=\bfloor{\frac{a(m'-m)}{m'}}=\bfloor{\frac{a\lambda}{\pi}}
\eea
count the number of unshaded $1$-bands below the height $a$. The value of $h_a$ remains unchanged within any shaded $n$-band.
Observing the duality 
\bea
m\leftrightarrow m'\!-\!m, \qquad \lambda\leftrightarrow \pi\!-\!\lambda,\qquad \mbox{shaded 1-bands}\leftrightarrow\mbox{unshaded 1-bands},\qquad h_a\leftrightarrow a\!-\!1\!-\!h_a
\label{duality}
\eea
it follows that the $n=1$ local energies satisfy 
\bea
H^{m,m',1}(a,b,c)= \half-H^{m'-m,m',1}(a,b,c)\label{n=1Hduality}
\eea

\subsubsection{$2\times 2$ local energies}
To obtain the low temperature limit of the $2\times 2$ fuesd RSOS face weights, it is convenient to first perform a \textit{conjugate modulus transformation} 
\begin{align}
\vartheta_1(u,e^{-\varepsilon})&=ie^{-\varepsilon/4}e^{-iu}E(e^{2iu},e^{-2\varepsilon})=\sqrt\frac{\pi}{\varepsilon}\, e^{-(u-\pi/2)^2/\varepsilon}E(e^{-2 \pi u/\varepsilon}, e^{-2\pi^2/\varepsilon} )
\end{align}
from nome $t=e^{-\varepsilon}$ to the conjugate nome $p=e^{-\pi^2/\varepsilon}$
where
\bea
E(w)=E(w,p)=\sum_{k=-\infty}^{\infty} (-1)^n p^{n(n-1)/2}w^n=\prod_{n=1}^\infty(1-p^{n-1}w)(1-p^nw^{-1})(1-p^n)
\eea
We introduce the variable $x = e^{-2 \pi \lambda/\varepsilon}=p^{\lambda/\pi}$ so that $p=e^{-2\pi^2/\varepsilon}=x^{\pi/\lambda}$.
The diagonal fused weights then become
\begin{subequations}
\begin{align}
&\Wt a {a\mp2} a  {a\pm2} = \frac{g_{a}^2}{g_{a-2}g_{a+2}}\, \frac{w E(x^{2}w^{-1})E(w^{-1}x)}{E(x^2)E(x)}
\\
&\Wt a {a\pm2} a a  =\Wt a a a {a\pm2}=\frac{g_{a}}{g_{a\pm2}} \frac{E(xw^{-1})E(x^{a\pm1}w^{\mp1})}{E(x)E(x^{a\pm1})}
\\
&\Wt {a\pm2} a {a\pm2} a  = \frac{g_{a\pm2}^2}{g_{a}^2} \frac{E(x^a w^{\pm1})E(x^{a\pm1}w^{\pm1})}{E(x^a)E(x^{a\pm1})}
\\
&\Wt a a a a  = \frac{xw E(w^{-1})E(xw^{-1})E(x^{a-1})E(x^{a+2})}{E(x)E(x^2)E(x^{a})E(x^{a+1})}+\frac{E(x^{a-1})E(x^aw^{-1})}{E(x^{a-1})E(x^a)}
\end{align}
\end{subequations}
where the gauge factors $g_a$ are arbitrary.

The low-temperature limit is now given by $x \rightarrow 0$ or $p\to 0$ with $w$ fixed. 
The $E$-functions satisfy the following properties
\begin{subequations}
\begin{align}
E(w,p) &= E(pw^{-1},p) = -wE(w^{-1},p)\\
E(p^nw,p) &= p^{-n(n-1)/2} (-w)^{-n} E(w,p)\\
\lim_{p\rightarrow0} E(p^a w,p^b) &= 
\begin{cases}
1, & 0<a<b\\
1-w, & a=0
\end{cases}
\end{align}
\end{subequations}
where $n$ is an integer. Another useful property is
\begin{equation}
\label{eq:hfunc}
 \lim_{x \rightarrow 0} \frac{E(x^a w^{-1})}{x^a} = w^{\floor{{a \lambda/\pi}} }
\end{equation}
These relations are used to derive the low temperature limit separately for the two intervals $0<\lambda<\tfrac{\pi}{2}$ and $\tfrac{\pi}{2}<\lambda<\pi$. 
Fixing the choice of gauge 
$
g_a = w^{a(a \lambda - \pi)/4\pi}
$
and removing the overall scale factor $\exp(-2u(\lambda-u)/\varepsilon)$, the local energy functions are
\begin{subequations}
\begin{align}
&0<\lambda<\tfrac{\pi}{2}:&H(a\pm2,a,a\mp2) &= 1 \\
&&H(a\pm2,a,a) &= H(a,a,a\pm2) = \half \pm  \flexpr{a\pm1}\\
&&H(a,a\pm2,a) &= \mp \big( \flexpr{a} + \flexpr{a\pm1}\big)\\
&&H(a,a,a) &= 
\begin{cases}
0, &\flexpr{a-1}= \flexpr{a} =  \flexpr{a+1}\qquad\quad\\[-2pt]
1, &\text{otherwise}
\end{cases}
\end{align}
\vspace{-20pt}
\end{subequations}
\begin{subequations}
\begin{align}
&\quad\tfrac{\pi}{2}<\lambda<\pi:
&H(a\pm2,a,a\mp2) &= 2 \\
&&H( a\pm2, a,a) &=  H(a,a,a\pm2 )=\half\pm h_{a\pm1} \\
&&H(a,a\pm2,a ) &=\mp(h_a+h_{a\pm1})\\
&&H(a,a,a) &= 
\begin{cases}
0, &\flexpr{a-1}= \flexpr{a}\ \text{or}\ \flexpr{a}=\flexpr{a+1}\quad \\[-2pt]
1, &\text{otherwise}
\end{cases}
\end{align}
\end{subequations}

The above expressions for the local energy functions take both positive and negative values. This is not desirable because we 
would like nonnegative local energy functions. 
To achieve this we use another gauge $g'_a=w^{G_a}$ such that the new local energies $H'(a,b,c)$ are
\be
H'(a,b,c) = H(a,b,c) + 2 G_{b} - G_a - G_c \geq0
\ee
A gauge transformation that satisfies this is
\be
G_a = 
\begin{cases}
\flexpr{1} + \flexpr{3} + \ldots \flexpr{a-1}, &a \text{ even}\\
\flexpr{2} + \flexpr{4} + \ldots \flexpr{a-1}, &a \text{ odd}
\end{cases}
\qquad\qquad
G_{a+1} - G_{a-1} = \flexpr{a}
\ee
This transformation is performed more neatly by defining $G(a,b) = G_b - G_a$ so that 
\be
H'(a,b,c) = H(a,b,c) + G(a,b) - G(b,c) \geq0, \qquad\quad G(a,b) =\half(b-a) \flexpr{\frac{a+b}{2}}
\ee 
The new expressions for local energies in this gauge, omitting the prime, are 
\begin{subequations}
\begin{align}
&0<\lambda<\tfrac{\pi}{2}:
&H(a\pm2,a,a\mp2)&= 1 - (\flexpr{a+1}-\flexpr{a-1}) \label{eq:localSuperE1}\\
&&H(a\pm2,a,a)&= H(a,a,a\pm2) = \half \label{eq:localSuperE2}\\
&&H(a,a\pm2,a)&= \pm \big(\flexpr{a\pm1} -  \flexpr{a}\big) \label{eq:localSuperE3}\\
&&H(a,a,a)&= 
\begin{cases}
0, &\flexpr{a-1}= \flexpr{a} =  \flexpr{a+1}\qquad\qquad\\[-2pt]
1, &\text{otherwise}
\end{cases} \label{eq:localSuperE4}
\end{align}
\label{eq:localSuperE}
\end{subequations}
\vspace{-20pt}
\begin{subequations}
\begin{align}
&\quad\tfrac{\pi}{2}<\lambda<\pi:
&H(a\pm2,a,a\mp2) &= 2 - (\flexpr{a+1}-\flexpr{a-1})  \label{eq:localE1}\\
&&\quad H( a\pm2, a,a) &=  H(a,a,a\pm2 )=\half  \label{eq:localE2}\\
&&\quad H(a,a\pm2,a ) &=\pm(h_{a\pm1}-h_a) \label{eq:localE3}\\
&&\quad H(a,a,a) &= 
\begin{cases}
0, &\flexpr{a-1}= \flexpr{a}\ \text{or}\ \flexpr{a}=\flexpr{a+1}\qquad\\[-2pt]
1, &\text{otherwise} 
\end{cases} \label{eq:localE4}
\end{align}
\label{eq:localE}
\end{subequations}
These are shown in Figures~\ref{superLocalEs} and \ref{LocalEs}.

\begin{figure}[p]
\begin{align}
&\psset{unit=0.5cm}
\begin{pspicture}[shift=-1.85](0,0)(2,4)
\pspolygon[fillstyle=solid,fillcolor=lightlightblue](0,1)(2,1)(2,4)(0,4)
\pspolygon[fillstyle=solid,fillcolor=lightlightblue](0,2)(2,2)(2,3)(0,3)
\pspolygon[fillstyle=solid,fillcolor=lightestblue](0,1)(2,1)(2,2)(0,2)
\pspolygon[fillstyle=solid,fillcolor=lightlightblue](0,0)(2,0)(2,1)(0,1)
\psgrid[gridlabels=0pt,subgriddiv=1]
\psline[linewidth=2pt,linecolor=blue](0,0)(2,4)
\end{pspicture} = 
\psset{unit=0.5cm}
\begin{pspicture}[shift=-1.85](0,0)(2,4)
\pspolygon[fillstyle=solid,fillcolor=lightlightblue](0,2)(2,2)(2,3)(0,3)
\pspolygon[fillstyle=solid,fillcolor=lightestblue](0,1)(2,1)(2,2)(0,2)
\pspolygon[fillstyle=solid,fillcolor=lightlightblue](0,0)(2,0)(2,1)(0,1)
\pspolygon[fillstyle=solid,fillcolor=lightestblue](0,3)(2,3)(2,4)(0,4)
\psgrid[gridlabels=0pt,subgriddiv=1]
\psline[linewidth=2pt,linecolor=blue](0,0)(2,4)
\end{pspicture} = 
\psset{unit=0.5cm}
\begin{pspicture}[shift=-1.85](0,0)(2,4)
\pspolygon[fillstyle=solid,fillcolor=lightlightblue](0,0)(2,0)(2,4)(0,4)
\pspolygon[fillstyle=solid,fillcolor=lightlightblue](0,3)(0,4)(2,4)(2,3)
\pspolygon[fillstyle=solid,fillcolor=lightestblue](0,2)(2,2)(2,3)(0,3)
\pspolygon[fillstyle=solid,fillcolor=lightlightblue](0,1)(2,1)(2,2)(0,2)
\psgrid[gridlabels=0pt,subgriddiv=1]
\psline[linewidth=2pt,linecolor=blue](0,0)(2,4)
\end{pspicture} = 
\psset{unit=0.5cm}
\begin{pspicture}[shift=-1.85](0,0)(2,4)
\pspolygon[fillstyle=solid,fillcolor=lightestblue](0,0)(2,0)(2,4)(0,4)
\pspolygon[fillstyle=solid,fillcolor=lightlightblue](0,3)(0,4)(2,4)(2,3)
\pspolygon[fillstyle=solid,fillcolor=lightestblue](0,2)(2,2)(2,3)(0,3)
\pspolygon[fillstyle=solid,fillcolor=lightlightblue](0,1)(2,1)(2,2)(0,2)
\psgrid[gridlabels=0pt,subgriddiv=1]
\psline[linewidth=2pt,linecolor=blue](0,0)(2,4)
\end{pspicture} = 
\psset{unit=0.5cm}
\begin{pspicture}[shift=-1.85](0,0)(2,4)
\pspolygon[fillstyle=solid,fillcolor=lightestblue](0,0)(2,0)(2,4)(0,4)
\pspolygon[fillstyle=solid,fillcolor=lightlightblue](0,2)(2,2)(2,3)(0,3)
\pspolygon[fillstyle=solid,fillcolor=lightestblue](0,1)(2,1)(2,2)(0,2)
\pspolygon[fillstyle=solid,fillcolor=lightlightblue](0,0)(2,0)(2,1)(0,1)
\psgrid[gridlabels=0pt,subgriddiv=1]
\psline[linewidth=2pt,linecolor=blue](0,4)(2,0)
\end{pspicture} = 
\psset{unit=0.5cm}
\begin{pspicture}[shift=-1.85](0,0)(2,4)
\pspolygon[fillstyle=solid,fillcolor=lightlightblue](0,0)(2,0)(2,4)(0,4)
\pspolygon[fillstyle=solid,fillcolor=lightlightblue](0,2)(2,2)(2,3)(0,3)
\pspolygon[fillstyle=solid,fillcolor=lightestblue](0,1)(2,1)(2,2)(0,2)
\pspolygon[fillstyle=solid,fillcolor=lightlightblue](0,0)(2,0)(2,1)(0,1)
\psgrid[gridlabels=0pt,subgriddiv=1]
\psline[linewidth=2pt,linecolor=blue](0,4)(2,0)
\end{pspicture} = 
\psset{unit=0.5cm}
\begin{pspicture}[shift=-1.85](0,0)(2,4)
\pspolygon[fillstyle=solid,fillcolor=lightestblue](0,0)(2,0)(2,4)(0,4)
\pspolygon[fillstyle=solid,fillcolor=lightlightblue](0,3)(0,4)(2,4)(2,3)
\pspolygon[fillstyle=solid,fillcolor=lightestblue](0,2)(2,2)(2,3)(0,3)
\pspolygon[fillstyle=solid,fillcolor=lightlightblue](0,1)(2,1)(2,2)(0,2)
\psgrid[gridlabels=0pt,subgriddiv=1]
\psline[linewidth=2pt,linecolor=blue](0,4)(2,0)
\end{pspicture} =
\psset{unit=0.5cm}
\begin{pspicture}[shift=-1.85](0,0)(2,4)
\pspolygon[fillstyle=solid,fillcolor=lightestblue](0,0)(2,0)(2,4)(0,4)
\pspolygon[fillstyle=solid,fillcolor=lightlightblue](0,0)(2,0)(2,4)(0,4)
\pspolygon[fillstyle=solid,fillcolor=lightlightblue](0,3)(0,4)(2,4)(2,3)
\pspolygon[fillstyle=solid,fillcolor=lightestblue](0,2)(2,2)(2,3)(0,3)
\pspolygon[fillstyle=solid,fillcolor=lightlightblue](0,1)(2,1)(2,2)(0,2)
\psgrid[gridlabels=0pt,subgriddiv=1]
\psline[linewidth=2pt,linecolor=blue](0,4)(2,0)
\end{pspicture} = 0\nonumber
\end{align}
\nobreak\vspace{-12pt}
\begin{align}
&\psset{unit=0.5cm}
\begin{pspicture}[shift=-1.85](0,0)(2,4)
\pspolygon[fillstyle=solid,fillcolor=lightlightblue](0,0)(2,0)(2,4)(0,4)
\pspolygon[fillstyle=solid,fillcolor=lightlightblue](0,1)(2,1)(2,3)(0,3)
\psgrid[gridlabels=0pt,subgriddiv=1]
\psline[linewidth=2pt,linecolor=blue](0,0)(2,4)
\end{pspicture} =
\psset{unit=0.5cm}
\begin{pspicture}[shift=-1.85](0,0)(2,4)
\pspolygon[fillstyle=solid,fillcolor=lightestblue](0,0)(2,0)(2,4)(0,4)
\pspolygon[fillstyle=solid,fillcolor=lightlightblue](0,1)(2,1)(2,3)(0,3)
\psgrid[gridlabels=0pt,subgriddiv=1]
\psline[linewidth=2pt,linecolor=blue](0,0)(2,4)
\end{pspicture} =
\psset{unit=0.5cm}
\begin{pspicture}[shift=-1.85](0,0)(2,4)
\pspolygon[fillstyle=solid,fillcolor=lightestblue](0,0)(2,0)(2,4)(0,4)
\pspolygon[fillstyle=solid,fillcolor=lightlightblue](0,1)(2,1)(2,4)(0,4)
\psgrid[gridlabels=0pt,subgriddiv=1]
\psline[linewidth=2pt,linecolor=blue](0,0)(2,4)
\end{pspicture} =
\psset{unit=0.5cm}
\begin{pspicture}[shift=-1.85](0,0)(2,4)
\pspolygon[fillstyle=solid,fillcolor=lightestblue](0,0)(2,0)(2,4)(0,4)
\pspolygon[fillstyle=solid,fillcolor=lightlightblue](0,0)(2,0)(2,3)(0,3)
\psgrid[gridlabels=0pt,subgriddiv=1]
\psline[linewidth=2pt,linecolor=blue](0,0)(2,4)
\end{pspicture} =
\psset{unit=0.5cm}
\begin{pspicture}[shift=-1.85](0,0)(2,4)
\pspolygon[fillstyle=solid,fillcolor=lightlightblue](0,0)(2,0)(2,4)(0,4)
\pspolygon[fillstyle=solid,fillcolor=lightlightblue](0,1)(2,1)(2,3)(0,3)
\psgrid[gridlabels=0pt,subgriddiv=1]
\psline[linewidth=2pt,linecolor=blue](0,4)(2,0)
\end{pspicture} = 
\psset{unit=0.5cm}
\begin{pspicture}[shift=-1.85](0,0)(2,4)
\pspolygon[fillstyle=solid,fillcolor=lightestblue](0,0)(2,0)(2,4)(0,4)
\pspolygon[fillstyle=solid,fillcolor=lightlightblue](0,1)(2,1)(2,3)(0,3)
\psgrid[gridlabels=0pt,subgriddiv=1]
\psline[linewidth=2pt,linecolor=blue](0,4)(2,0)
\end{pspicture} =
\psset{unit=0.5cm}
\begin{pspicture}[shift=-1.85](0,0)(2,4)
\pspolygon[fillstyle=solid,fillcolor=lightestblue](0,0)(2,0)(2,4)(0,4)
\pspolygon[fillstyle=solid,fillcolor=lightlightblue](0,1)(2,1)(2,4)(0,4)
\psgrid[gridlabels=0pt,subgriddiv=1]
\psline[linewidth=2pt,linecolor=blue](0,4)(2,0)
\end{pspicture} = 
\psset{unit=0.5cm}
\begin{pspicture}[shift=-1.85](0,0)(2,4)
\pspolygon[fillstyle=solid,fillcolor=lightestblue](0,0)(2,0)(2,4)(0,4)
\pspolygon[fillstyle=solid,fillcolor=lightlightblue](0,0)(2,0)(2,3)(0,3)
\psgrid[gridlabels=0pt,subgriddiv=1]
\psline[linewidth=2pt,linecolor=blue](0,4)(2,0)
\end{pspicture} =1\nonumber
\end{align}
\nobreak\vspace{-12pt}
\begin{align}
&\psset{unit=0.5cm}
\begin{pspicture}[shift=-0.85](0,0)(2,2)
\pspolygon[fillstyle=solid,fillcolor=lightlightblue](0,0)(2,0)(2,2)(0,2)
\psgrid[gridlabels=0pt,subgriddiv=1]
\psline[linewidth=2pt,linecolor=blue](0,0)(1,0)
\psline[linewidth=2pt,linecolor=blue](1,0)(2,2)
\end{pspicture}  =
\psset{unit=0.5cm}
\begin{pspicture}[shift=-0.85](0,0)(2,2)
\pspolygon[fillstyle=solid,fillcolor=lightestblue](0,0)(2,0)(2,2)(0,2)
\pspolygon[fillstyle=solid,fillcolor=lightlightblue](0,0)(2,0)(2,1)(0,1)
\psgrid[gridlabels=0pt,subgriddiv=1]
\psline[linewidth=2pt,linecolor=blue](0,0)(1,0)
\psline[linewidth=2pt,linecolor=blue](1,0)(2,2)
\end{pspicture}  =
\psset{unit=0.5cm}
\begin{pspicture}[shift=-0.85](0,0)(2,2)
\pspolygon[fillstyle=solid,fillcolor=lightestblue](0,0)(2,0)(2,2)(0,2)
\pspolygon[fillstyle=solid,fillcolor=lightlightblue](0,1)(2,1)(2,2)(0,2)
\psgrid[gridlabels=0pt,subgriddiv=1]
\psline[linewidth=2pt,linecolor=blue](0,0)(1,0)
\psline[linewidth=2pt,linecolor=blue](1,0)(2,2)
\end{pspicture}  =
\psset{unit=0.5cm}
\begin{pspicture}[shift=-0.85](0,0)(2,2)
\pspolygon[fillstyle=solid,fillcolor=lightlightblue](0,0)(2,0)(2,2)(0,2)
\psgrid[gridlabels=0pt,subgriddiv=1]
\psline[linewidth=2pt,linecolor=blue](0,2)(1,2)
\psline[linewidth=2pt,linecolor=blue](1,2)(2,0)
\end{pspicture} = 
\psset{unit=0.5cm}
\begin{pspicture}[shift=-0.85](0,0)(2,2)
\pspolygon[fillstyle=solid,fillcolor=lightestblue](0,0)(2,0)(2,2)(0,2)
\pspolygon[fillstyle=solid,fillcolor=lightlightblue](0,0)(2,0)(2,1)(0,1)
\psgrid[gridlabels=0pt,subgriddiv=1]
\psline[linewidth=2pt,linecolor=blue](0,2)(1,2)
\psline[linewidth=2pt,linecolor=blue](1,2)(2,0)
\end{pspicture} =
\psset{unit=0.5cm}
\begin{pspicture}[shift=-0.85](0,0)(2,2)
\pspolygon[fillstyle=solid,fillcolor=lightestblue](0,0)(2,0)(2,2)(0,2)
\pspolygon[fillstyle=solid,fillcolor=lightlightblue](0,1)(2,1)(2,2)(0,2)
\psgrid[gridlabels=0pt,subgriddiv=1]
\psline[linewidth=2pt,linecolor=blue](0,2)(1,2)
\psline[linewidth=2pt,linecolor=blue](1,2)(2,0)
\end{pspicture} =\phantom{\half}\nonumber\\[-30pt]\nonumber
\end{align}
\begin{align}
&\psset{unit=0.5cm}
\begin{pspicture}[shift=-0.85](0,0)(2,2)
\pspolygon[fillstyle=solid,fillcolor=lightlightblue](0,0)(2,0)(2,2)(0,2)
\psgrid[gridlabels=0pt,subgriddiv=1]
\psline[linewidth=2pt,linecolor=blue](1,2)(2,2)
\psline[linewidth=2pt,linecolor=blue](0,0)(1,2)
\end{pspicture} =
\psset{unit=0.5cm}
\begin{pspicture}[shift=-0.85](0,0)(2,2)
\pspolygon[fillstyle=solid,fillcolor=lightestblue](0,0)(2,0)(2,1)(0,1)
\pspolygon[fillstyle=solid,fillcolor=lightlightblue](0,0)(2,0)(2,1)(0,1)
\psgrid[gridlabels=0pt,subgriddiv=1]
\psline[linewidth=2pt,linecolor=blue](1,2)(2,2)
\psline[linewidth=2pt,linecolor=blue](0,0)(1,2)
\end{pspicture} =
\psset{unit=0.5cm}
\begin{pspicture}[shift=-0.85](0,0)(2,2)
\pspolygon[fillstyle=solid,fillcolor=lightestblue](0,0)(2,0)(2,1)(0,1)
\pspolygon[fillstyle=solid,fillcolor=lightlightblue](0,1)(2,1)(2,2)(0,2)
\psgrid[gridlabels=0pt,subgriddiv=1]
\psline[linewidth=2pt,linecolor=blue](1,2)(2,2)
\psline[linewidth=2pt,linecolor=blue](0,0)(1,2)
\end{pspicture} =
\psset{unit=0.5cm}
\begin{pspicture}[shift=-0.85](0,0)(2,2)
\pspolygon[fillstyle=solid,fillcolor=lightlightblue](0,0)(2,0)(2,2)(0,2)
\psgrid[gridlabels=0pt,subgriddiv=1]
\psline[linewidth=2pt,linecolor=blue](1,0)(2,0)
\psline[linewidth=2pt,linecolor=blue](0,2)(1,0)
\end{pspicture} = 
\psset{unit=0.5cm}
\begin{pspicture}[shift=-0.85](0,0)(2,2)
\pspolygon[fillstyle=solid,fillcolor=lightestblue](0,0)(2,0)(2,1)(0,1)
\pspolygon[fillstyle=solid,fillcolor=lightlightblue](0,0)(2,0)(2,1)(0,1)
\psgrid[gridlabels=0pt,subgriddiv=1]
\psline[linewidth=2pt,linecolor=blue](1,0)(2,0)
\psline[linewidth=2pt,linecolor=blue](0,2)(1,0)
\end{pspicture} =
\psset{unit=0.5cm}
\begin{pspicture}[shift=-0.85](0,0)(2,2)
\pspolygon[fillstyle=solid,fillcolor=lightestblue](0,0)(2,0)(2,1)(0,1)
\pspolygon[fillstyle=solid,fillcolor=lightlightblue](0,1)(2,1)(2,2)(0,2)
\psgrid[gridlabels=0pt,subgriddiv=1]
\psline[linewidth=2pt,linecolor=blue](1,0)(2,0)
\psline[linewidth=2pt,linecolor=blue](0,2)(1,0)
\end{pspicture} = \half \nonumber
\end{align}
\nobreak\vspace{-12pt}
\begin{align}
\psset{unit=0.5cm}
\qquad\quad
\begin{pspicture}[shift=-0.85](0,0)(2,2)
\pspolygon[fillstyle=solid,fillcolor=lightlightblue](0,0)(2,0)(2,2)(0,2)
\psgrid[gridlabels=0pt,subgriddiv=1]
\psline[linewidth=2pt,linecolor=blue](0,0)(1,2)
\psline[linewidth=2pt,linecolor=blue](1,2)(2,0)
\end{pspicture} &= 
\psset{unit=0.5cm}
\begin{pspicture}[shift=-0.85](0,0)(2,2)
\pspolygon[fillstyle=solid,fillcolor=lightestblue](0,0)(2,0)(2,2)(0,2)
\pspolygon[fillstyle=solid,fillcolor=lightlightblue](0,0)(2,0)(2,2)(0,2)
\psgrid[gridlabels=0pt,subgriddiv=1]
\psline[linewidth=2pt,linecolor=blue](0,2)(1,0)
\psline[linewidth=2pt,linecolor=blue](1,0)(2,2)
\end{pspicture} = 
\psset{unit=0.5cm}
\begin{pspicture}[shift=-0.85](0,0)(2,2)
\pspolygon[fillstyle=solid,fillcolor=lightestblue](0,0)(2,0)(2,2)(0,2)
\pspolygon[fillstyle=solid,fillcolor=lightlightblue](0,0)(2,0)(2,1)(0,1)
\psgrid[gridlabels=0pt,subgriddiv=1]
\psline[linewidth=2pt,linecolor=blue](0,0)(1,2)
\psline[linewidth=2pt,linecolor=blue](1,2)(2,0)
\end{pspicture} =
\psset{unit=0.5cm}
\begin{pspicture}[shift=-0.85](0,0)(2,2)
\pspolygon[fillstyle=solid,fillcolor=lightestblue](0,0)(2,0)(2,2)(0,2)
\pspolygon[fillstyle=solid,fillcolor=lightlightblue](0,1)(2,1)(2,2)(0,2)
\psgrid[gridlabels=0pt,subgriddiv=1]
\psline[linewidth=2pt,linecolor=blue](0,2)(1,0)
\psline[linewidth=2pt,linecolor=blue](1,0)(2,2)
\end{pspicture} =0
& 
\psset{unit=0.5cm}
\begin{pspicture}[shift=-0.85](0,0)(2,2)
\pspolygon[fillstyle=solid,fillcolor=lightlightblue](0,0)(2,0)(2,2)(0,2)
\pspolygon[fillstyle=solid,fillcolor=lightestblue](0,0)(2,0)(2,1)(0,1)
\psgrid[gridlabels=0pt,subgriddiv=1]
\psline[linewidth=2pt,linecolor=blue](0,0)(1,2)
\psline[linewidth=2pt,linecolor=blue](1,2)(2,0)
\end{pspicture} = 
\psset{unit=0.5cm}
\begin{pspicture}[shift=-0.85](0,0)(2,2)
\pspolygon[fillstyle=solid,fillcolor=lightestblue](0,0)(2,0)(2,2)(0,2)
\pspolygon[fillstyle=solid,fillcolor=lightlightblue](0,0)(2,0)(2,1)(0,1)
\psgrid[gridlabels=0pt,subgriddiv=1]
\psline[linewidth=2pt,linecolor=blue](0,2)(1,0)
\psline[linewidth=2pt,linecolor=blue](1,0)(2,2)
\end{pspicture} = 1\nonumber
\end{align}
\nobreak\vspace{-12pt}
\begin{align}
\psset{unit=0.5cm}
\begin{pspicture}[shift=-0.85](0,0)(2,2)
\pspolygon[fillstyle=solid,fillcolor=lightlightblue](0,0)(2,0)(2,2)(0,2)
\psgrid[gridlabels=0pt,subgriddiv=1]
\psline[linewidth=2pt,linecolor=blue](0,1)(2,1)
\end{pspicture} = 0  \qquad\ \ \ \ 
\psset{unit=0.5cm}
\begin{pspicture}[shift=-0.85](0,0)(2,2)
\pspolygon[fillstyle=solid,fillcolor=lightlightblue](0,1)(2,1)(2,2)(0,2)
\pspolygon[fillstyle=solid,fillcolor=lightestblue](0,0)(2,0)(2,1)(0,1)
\psgrid[gridlabels=0pt,subgriddiv=1]
\psline[linewidth=2pt,linecolor=blue](0,1)(2,1)
\end{pspicture} = 
\psset{unit=0.5cm}
\begin{pspicture}[shift=-0.85](0,0)(2,2)
\pspolygon[fillstyle=solid,fillcolor=lightestblue](0,1)(2,1)(2,2)(0,2)
\pspolygon[fillstyle=solid,fillcolor=lightlightblue](0,0)(2,0)(2,1)(0,1)
\psgrid[gridlabels=0pt,subgriddiv=1]
\psline[linewidth=2pt,linecolor=blue](0,1)(2,1)
\end{pspicture} =1 \nonumber
\end{align}
\caption{Local energies for $2\times 2$ fused RSOS models in the interval $0<\lambda<\tfrac{\pi}{2}$.\label{superLocalEs}}
\end{figure}
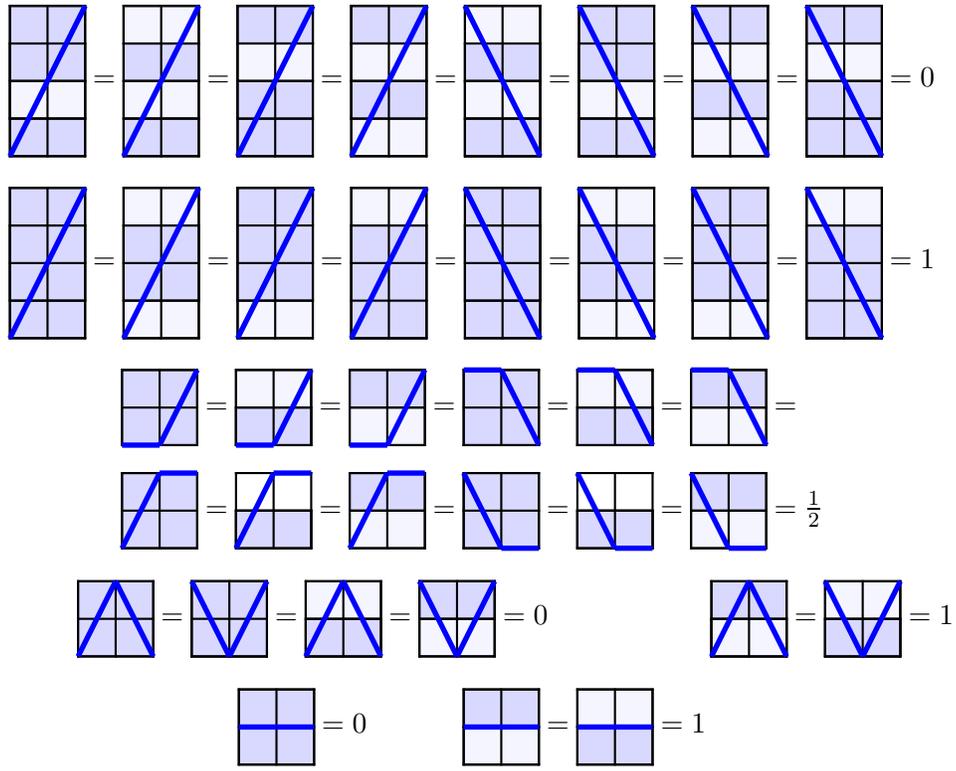%
\begin{figure}[p]
\begin{align}
\psset{unit=0.5cm}
\begin{pspicture}[shift=-1.85](0,0)(2,4)
\pspolygon[fillstyle=solid,fillcolor=lightestblue](0,0)(2,0)(2,4)(0,4)
\psgrid[gridlabels=0pt,subgriddiv=1]
\psline[linewidth=2pt,linecolor=blue](0,0)(2,4)
\end{pspicture} =
\psset{unit=0.5cm}
\begin{pspicture}[shift=-1.85](0,0)(2,4)
\pspolygon[fillstyle=solid,fillcolor=lightestblue](0,0)(2,0)(2,4)(0,4)
\pspolygon[fillstyle=solid,fillcolor=lightlightblue](0,3)(2,3)(2,4)(0,4)
\pspolygon[fillstyle=solid,fillcolor=lightlightblue](0,0)(2,0)(2,1)(0,1)
\psgrid[gridlabels=0pt,subgriddiv=1]
\psline[linewidth=2pt,linecolor=blue](0,0)(2,4)
\end{pspicture} =
\psset{unit=0.5cm}
\begin{pspicture}[shift=-1.85](0,0)(2,4)
\pspolygon[fillstyle=solid,fillcolor=lightestblue](0,0)(2,0)(2,4)(0,4)
\pspolygon[fillstyle=solid,fillcolor=lightlightblue](0,0)(2,0)(2,1)(0,1)
\psgrid[gridlabels=0pt,subgriddiv=1]
\psline[linewidth=2pt,linecolor=blue](0,0)(2,4)
\end{pspicture} =
\psset{unit=0.5cm}
\begin{pspicture}[shift=-1.85](0,0)(2,4)
\pspolygon[fillstyle=solid,fillcolor=lightestblue](0,0)(2,0)(2,4)(0,4)
\pspolygon[fillstyle=solid,fillcolor=lightlightblue](0,3)(2,3)(2,4)(0,4)
\psgrid[gridlabels=0pt,subgriddiv=1]
\psline[linewidth=2pt,linecolor=blue](0,0)(2,4)
\end{pspicture} =
\psset{unit=0.5cm}
\begin{pspicture}[shift=-1.85](0,0)(2,4)
\pspolygon[fillstyle=solid,fillcolor=lightestblue](0,0)(2,0)(2,4)(0,4)
\psgrid[gridlabels=0pt,subgriddiv=1]
\psline[linewidth=2pt,linecolor=blue](0,4)(2,0)
\end{pspicture} =
\psset{unit=0.5cm}
\begin{pspicture}[shift=-1.85](0,0)(2,4)
\pspolygon[fillstyle=solid,fillcolor=lightestblue](0,0)(2,0)(2,4)(0,4)
\pspolygon[fillstyle=solid,fillcolor=lightlightblue](0,3)(2,3)(2,4)(0,4)
\pspolygon[fillstyle=solid,fillcolor=lightlightblue](0,0)(2,0)(2,1)(0,1)
\psgrid[gridlabels=0pt,subgriddiv=1]
\psline[linewidth=2pt,linecolor=blue](0,4)(2,0)
\end{pspicture} =
\psset{unit=0.5cm}
\begin{pspicture}[shift=-1.85](0,0)(2,4)
\pspolygon[fillstyle=solid,fillcolor=lightestblue](0,0)(2,0)(2,4)(0,4)
\pspolygon[fillstyle=solid,fillcolor=lightlightblue](0,0)(2,0)(2,1)(0,1)
\psgrid[gridlabels=0pt,subgriddiv=1]
\psline[linewidth=2pt,linecolor=blue](0,4)(2,0)
\end{pspicture} =
\psset{unit=0.5cm}
\begin{pspicture}[shift=-1.85](0,0)(2,4)
\pspolygon[fillstyle=solid,fillcolor=lightestblue](0,0)(2,0)(2,4)(0,4)
\pspolygon[fillstyle=solid,fillcolor=lightlightblue](0,3)(2,3)(2,4)(0,4)
\psgrid[gridlabels=0pt,subgriddiv=1]
\psline[linewidth=2pt,linecolor=blue](0,4)(2,0)
\end{pspicture} = 0\nonumber
\end{align}
\nobreak\vspace{-12pt}
\begin{align}
\psset{unit=0.5cm}
\begin{pspicture}[shift=-1.85](0,0)(2,4)
\pspolygon[fillstyle=solid,fillcolor=lightestblue](0,0)(2,0)(2,4)(0,4)
\pspolygon[fillstyle=solid,fillcolor=lightlightblue](0,1)(2,1)(2,2)(0,2)
\psgrid[gridlabels=0pt,subgriddiv=1]
\psline[linewidth=2pt,linecolor=blue](0,0)(2,4)
\end{pspicture} =
\psset{unit=0.5cm}
\begin{pspicture}[shift=-1.85](0,0)(2,4)
\pspolygon[fillstyle=solid,fillcolor=lightestblue](0,0)(2,0)(2,4)(0,4)
\pspolygon[fillstyle=solid,fillcolor=lightlightblue](0,3)(2,3)(2,4)(0,4)
\pspolygon[fillstyle=solid,fillcolor=lightlightblue](0,1)(2,1)(2,2)(0,2)
\psgrid[gridlabels=0pt,subgriddiv=1]
\psline[linewidth=2pt,linecolor=blue](0,0)(2,4)
\end{pspicture} &=
\psset{unit=0.5cm}
\begin{pspicture}[shift=-1.85](0,0)(2,4)
\pspolygon[fillstyle=solid,fillcolor=lightestblue](0,0)(2,0)(2,4)(0,4)
\pspolygon[fillstyle=solid,fillcolor=lightlightblue](0,2)(2,2)(2,3)(0,3)
\psgrid[gridlabels=0pt,subgriddiv=1]
\psline[linewidth=2pt,linecolor=blue](0,0)(2,4)
\end{pspicture} =
\psset{unit=0.5cm}
\begin{pspicture}[shift=-1.85](0,0)(2,4)
\pspolygon[fillstyle=solid,fillcolor=lightestblue](0,0)(2,0)(2,4)(0,4)
\pspolygon[fillstyle=solid,fillcolor=lightlightblue](0,0)(2,0)(2,1)(0,1)
\pspolygon[fillstyle=solid,fillcolor=lightlightblue](0,2)(2,2)(2,3)(0,3)
\psgrid[gridlabels=0pt,subgriddiv=1]
\psline[linewidth=2pt,linecolor=blue](0,0)(2,4)
\end{pspicture} = 
\psset{unit=0.5cm}
\begin{pspicture}[shift=-1.85](0,0)(2,4)
\pspolygon[fillstyle=solid,fillcolor=lightestblue](0,0)(2,0)(2,4)(0,4)
\pspolygon[fillstyle=solid,fillcolor=lightlightblue](0,3)(2,3)(2,4)(0,4)
\pspolygon[fillstyle=solid,fillcolor=lightlightblue](0,1)(2,1)(2,2)(0,2)
\psgrid[gridlabels=0pt,subgriddiv=1]
\psline[linewidth=2pt,linecolor=blue](0,4)(2,0)
\end{pspicture} = 
\psset{unit=0.5cm}
\begin{pspicture}[shift=-1.85](0,0)(2,4)
\pspolygon[fillstyle=solid,fillcolor=lightestblue](0,0)(2,0)(2,4)(0,4)
\pspolygon[fillstyle=solid,fillcolor=lightlightblue](0,1)(2,1)(2,2)(0,2)
\psgrid[gridlabels=0pt,subgriddiv=1]
\psline[linewidth=2pt,linecolor=blue](0,4)(2,0)
\end{pspicture} = 
\psset{unit=0.5cm}
\begin{pspicture}[shift=-1.85](0,0)(2,4)
\pspolygon[fillstyle=solid,fillcolor=lightestblue](0,0)(2,0)(2,4)(0,4)
\pspolygon[fillstyle=solid,fillcolor=lightlightblue](0,0)(2,0)(2,1)(0,1)
\pspolygon[fillstyle=solid,fillcolor=lightlightblue](0,2)(2,2)(2,3)(0,3)
\psgrid[gridlabels=0pt,subgriddiv=1]
\psline[linewidth=2pt,linecolor=blue](0,4)(2,0)
\end{pspicture} =
\psset{unit=0.5cm}
\begin{pspicture}[shift=-1.85](0,0)(2,4)
\pspolygon[fillstyle=solid,fillcolor=lightestblue](0,0)(2,0)(2,4)(0,4)
\pspolygon[fillstyle=solid,fillcolor=lightlightblue](0,2)(2,2)(2,3)(0,3)
\psgrid[gridlabels=0pt,subgriddiv=1]
\psline[linewidth=2pt,linecolor=blue](0,4)(2,0)
\end{pspicture} =1\nonumber
\end{align}
\nobreak\vspace{-12pt}
\begin{align}
\psset{unit=0.5cm}
\begin{pspicture}[shift=-0.85](0,0)(2,2)
\pspolygon[fillstyle=solid,fillcolor=lightestblue](0,0)(2,0)(2,2)(0,2)
\psgrid[gridlabels=0pt,subgriddiv=1]
\psline[linewidth=2pt,linecolor=blue](0,0)(1,0)
\psline[linewidth=2pt,linecolor=blue](1,0)(2,2)
\end{pspicture}  =
\psset{unit=0.5cm}
\begin{pspicture}[shift=-0.85](0,0)(2,2)
\pspolygon[fillstyle=solid,fillcolor=lightestblue](0,0)(2,0)(2,2)(0,2)
\pspolygon[fillstyle=solid,fillcolor=lightlightblue](0,0)(2,0)(2,1)(0,1)
\psgrid[gridlabels=0pt,subgriddiv=1]
\psline[linewidth=2pt,linecolor=blue](0,0)(1,0)
\psline[linewidth=2pt,linecolor=blue](1,0)(2,2)
\end{pspicture}  =
\begin{pspicture}[shift=-0.85](0,0)(2,2)
\pspolygon[fillstyle=solid,fillcolor=lightestblue](0,0)(2,0)(2,2)(0,2)
\pspolygon[fillstyle=solid,fillcolor=lightlightblue](0,1)(2,1)(2,2)(0,2)
\psgrid[gridlabels=0pt,subgriddiv=1]
\psline[linewidth=2pt,linecolor=blue](0,0)(1,0)
\psline[linewidth=2pt,linecolor=blue](1,0)(2,2)
\end{pspicture} =
\psset{unit=0.5cm}
\begin{pspicture}[shift=-0.85](0,0)(2,2)
\pspolygon[fillstyle=solid,fillcolor=lightestblue](0,0)(2,0)(2,2)(0,2)
\psgrid[gridlabels=0pt,subgriddiv=1]
\psline[linewidth=2pt,linecolor=blue](0,2)(1,2)
\psline[linewidth=2pt,linecolor=blue](1,2)(2,0)
\end{pspicture} = 
\psset{unit=0.5cm}
\begin{pspicture}[shift=-0.85](0,0)(2,2)
\pspolygon[fillstyle=solid,fillcolor=lightestblue](0,0)(2,0)(2,2)(0,2)
\pspolygon[fillstyle=solid,fillcolor=lightlightblue](0,0)(2,0)(2,1)(0,1)
\psgrid[gridlabels=0pt,subgriddiv=1]
\psline[linewidth=2pt,linecolor=blue](0,2)(1,2)
\psline[linewidth=2pt,linecolor=blue](1,2)(2,0)
\end{pspicture} = 
\psset{unit=0.5cm}
\begin{pspicture}[shift=-0.85](0,0)(2,2)
\pspolygon[fillstyle=solid,fillcolor=lightestblue](0,0)(2,0)(2,2)(0,2)
\pspolygon[fillstyle=solid,fillcolor=lightlightblue](0,1)(2,1)(2,2)(0,2)
\psgrid[gridlabels=0pt,subgriddiv=1]
\psline[linewidth=2pt,linecolor=blue](0,2)(1,2)
\psline[linewidth=2pt,linecolor=blue](1,2)(2,0)
\end{pspicture} = \phantom{\half}\nonumber\\[-30pt]\nonumber
\end{align}
\begin{align}
\psset{unit=0.5cm}
\begin{pspicture}[shift=-0.85](0,0)(2,2)
\pspolygon[fillstyle=solid,fillcolor=lightestblue](0,0)(2,0)(2,2)(0,2)
\psgrid[gridlabels=0pt,subgriddiv=1]
\psline[linewidth=2pt,linecolor=blue](1,2)(2,2)
\psline[linewidth=2pt,linecolor=blue](0,0)(1,2)
\end{pspicture} =
\psset{unit=0.5cm}
\begin{pspicture}[shift=-0.85](0,0)(2,2)
\pspolygon[fillstyle=solid,fillcolor=lightestblue](0,0)(2,0)(2,2)(0,2)
\pspolygon[fillstyle=solid,fillcolor=lightlightblue](0,0)(2,0)(2,1)(0,1)
\psgrid[gridlabels=0pt,subgriddiv=1]
\psline[linewidth=2pt,linecolor=blue](1,2)(2,2)
\psline[linewidth=2pt,linecolor=blue](0,0)(1,2)
\end{pspicture} =
\psset{unit=0.5cm}
\begin{pspicture}[shift=-0.85](0,0)(2,2)
\pspolygon[fillstyle=solid,fillcolor=lightestblue](0,0)(2,0)(2,2)(0,2)
\pspolygon[fillstyle=solid,fillcolor=lightlightblue](0,1)(2,1)(2,2)(0,2)
\psgrid[gridlabels=0pt,subgriddiv=1]
\psline[linewidth=2pt,linecolor=blue](1,2)(2,2)
\psline[linewidth=2pt,linecolor=blue](0,0)(1,2)
\end{pspicture} =
\psset{unit=0.5cm}
\begin{pspicture}[shift=-0.85](0,0)(2,2)
\pspolygon[fillstyle=solid,fillcolor=lightestblue](0,0)(2,0)(2,2)(0,2)
\psgrid[gridlabels=0pt,subgriddiv=1]
\psline[linewidth=2pt,linecolor=blue](1,0)(2,0)
\psline[linewidth=2pt,linecolor=blue](0,2)(1,0)
\end{pspicture} = 
\psset{unit=0.5cm}
\begin{pspicture}[shift=-0.85](0,0)(2,2)
\pspolygon[fillstyle=solid,fillcolor=lightestblue](0,0)(2,0)(2,2)(0,2)
\pspolygon[fillstyle=solid,fillcolor=lightlightblue](0,0)(2,0)(2,1)(0,1)
\psgrid[gridlabels=0pt,subgriddiv=1]
\psline[linewidth=2pt,linecolor=blue](1,0)(2,0)
\psline[linewidth=2pt,linecolor=blue](0,2)(1,0)
\end{pspicture} = 
\psset{unit=0.5cm}
\begin{pspicture}[shift=-0.85](0,0)(2,2)
\pspolygon[fillstyle=solid,fillcolor=lightestblue](0,0)(2,0)(2,2)(0,2)
\pspolygon[fillstyle=solid,fillcolor=lightlightblue](0,1)(2,1)(2,2)(0,2)
\psgrid[gridlabels=0pt,subgriddiv=1]
\psline[linewidth=2pt,linecolor=blue](1,0)(2,0)
\psline[linewidth=2pt,linecolor=blue](0,2)(1,0)
\end{pspicture} =\half\nonumber
\end{align}
\nobreak\vspace{-12pt}
\begin{align}
\psset{unit=0.5cm}
\begin{pspicture}[shift=-0.85](0,0)(2,2)
\pspolygon[fillstyle=solid,fillcolor=lightestblue](0,0)(2,0)(2,2)(0,2)
\psgrid[gridlabels=0pt,subgriddiv=1]
\psline[linewidth=2pt,linecolor=blue](0,0)(1,2)
\psline[linewidth=2pt,linecolor=blue](1,2)(2,0)
\end{pspicture} =
\psset{unit=0.5cm}
\begin{pspicture}[shift=-0.85](0,0)(2,2)
\pspolygon[fillstyle=solid,fillcolor=lightestblue](0,0)(2,0)(2,2)(0,2)
\psgrid[gridlabels=0pt,subgriddiv=1]
\psline[linewidth=2pt,linecolor=blue](0,2)(1,0)
\psline[linewidth=2pt,linecolor=blue](1,0)(2,2)
\end{pspicture} =
\begin{pspicture}[shift=-0.85](0,0)(2,2)
\pspolygon[fillstyle=solid,fillcolor=lightestblue](0,0)(2,0)(2,2)(0,2)
\pspolygon[fillstyle=solid,fillcolor=lightlightblue](0,1)(2,1)(2,2)(0,2)
\psgrid[gridlabels=0pt,subgriddiv=1]
\psline[linewidth=2pt,linecolor=blue](0,0)(1,2)
\psline[linewidth=2pt,linecolor=blue](1,2)(2,0)
\end{pspicture} = 
\psset{unit=0.5cm}
\begin{pspicture}[shift=-0.85](0,0)(2,2)
\pspolygon[fillstyle=solid,fillcolor=lightestblue](0,0)(2,0)(2,2)(0,2)
\pspolygon[fillstyle=solid,fillcolor=lightlightblue](0,0)(2,0)(2,1)(0,1)
\psgrid[gridlabels=0pt,subgriddiv=1]
\psline[linewidth=2pt,linecolor=blue](0,2)(1,0)
\psline[linewidth=2pt,linecolor=blue](1,0)(2,2)
\end{pspicture} = 1\qquad
\psset{unit=0.5cm}
\begin{pspicture}[shift=-0.85](0,0)(2,2)
\pspolygon[fillstyle=solid,fillcolor=lightestblue](0,0)(2,0)(2,2)(0,2)
\pspolygon[fillstyle=solid,fillcolor=lightlightblue](0,0)(2,0)(2,1)(0,1)
\psgrid[gridlabels=0pt,subgriddiv=1]
\psline[linewidth=2pt,linecolor=blue](0,0)(1,2)
\psline[linewidth=2pt,linecolor=blue](1,2)(2,0)
\end{pspicture} =
\psset{unit=0.5cm}
\begin{pspicture}[shift=-0.85](0,0)(2,2)
\pspolygon[fillstyle=solid,fillcolor=lightestblue](0,0)(2,0)(2,2)(0,2)
\pspolygon[fillstyle=solid,fillcolor=lightlightblue](0,1)(2,1)(2,2)(0,2)
\psgrid[gridlabels=0pt,subgriddiv=1]
\psline[linewidth=2pt,linecolor=blue](0,2)(1,0)
\psline[linewidth=2pt,linecolor=blue](1,0)(2,2)
\end{pspicture} =0\nonumber
\end{align}
\nobreak\vspace{-12pt}
\begin{align}
\psset{unit=0.5cm}
\begin{pspicture}[shift=-0.85](0,0)(2,2)
\pspolygon[fillstyle=solid,fillcolor=lightestblue](0,0)(2,0)(2,2)(0,2)
\psgrid[gridlabels=0pt,subgriddiv=1]
\psline[linewidth=2pt,linecolor=blue](0,1)(2,1)
\end{pspicture} = 1  \qquad
\psset{unit=0.5cm}
\begin{pspicture}[shift=-0.85](0,0)(2,2)
\pspolygon[fillstyle=solid,fillcolor=lightestblue](0,0)(2,0)(2,2)(0,2)
\pspolygon[fillstyle=solid,fillcolor=lightlightblue](0,0)(2,0)(2,1)(0,1)
\psgrid[gridlabels=0pt,subgriddiv=1]
\psline[linewidth=2pt,linecolor=blue](0,1)(2,1)
\end{pspicture} =
\psset{unit=0.5cm}
\begin{pspicture}[shift=-0.85](0,0)(2,2)
\pspolygon[fillstyle=solid,fillcolor=lightestblue](0,0)(2,0)(2,2)(0,2)
\pspolygon[fillstyle=solid,fillcolor=lightlightblue](0,1)(2,1)(2,2)(0,2)
\psgrid[gridlabels=0pt,subgriddiv=1]
\psline[linewidth=2pt,linecolor=blue](0,1)(2,1)
\end{pspicture}=0\nonumber
\end{align}
\caption{Local energies for $2\times 2$ fused RSOS models in the interval $\tfrac{\pi}{2}<\lambda<\pi$.\label{LocalEs}}
\end{figure}
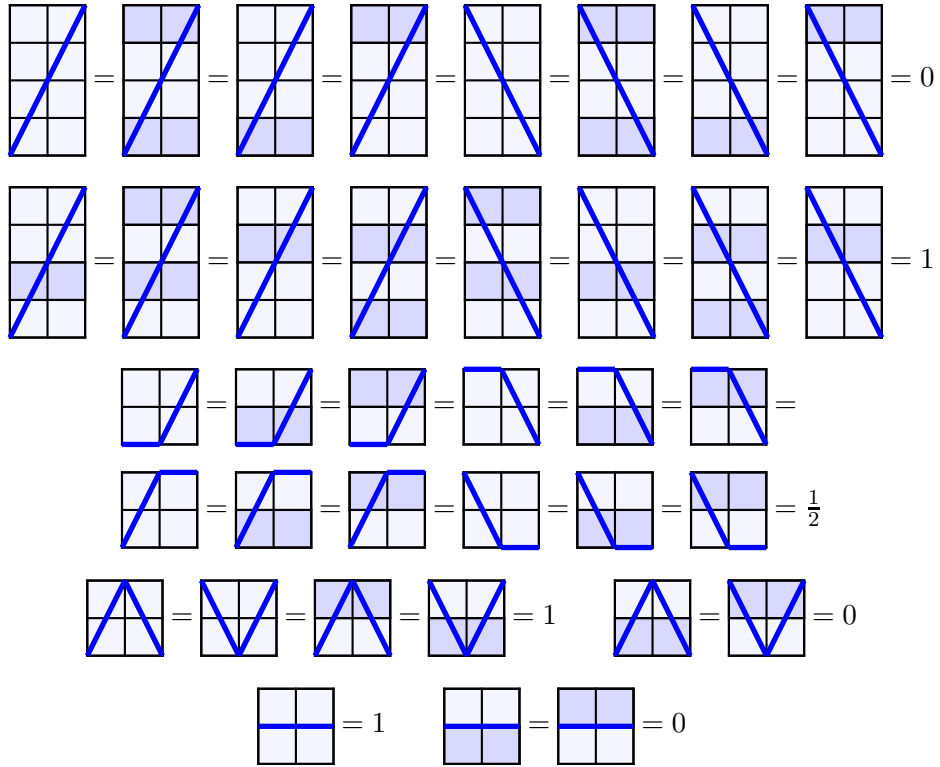

The local energy functions are now nonnegative $H(a,b,c)\ge 0$ and take only the values $0,\half,1$, which is consistent with a spin-1 interpretation. 
The local energies only depend on the heights $a$ through the shading of the 1-bands since they depend only on the differences
\bea
\delta_a=h_{a+1}-h_a=\begin{cases} 
0,&\mbox{the 1-band $(a,a+1)$ is shaded}\\
1,&\mbox{the 1-band $(a,a+1)$ is unshaded}
\end{cases}
\eea
The shaded and unshaded 1-bands for the two intervals  $0<\lambda<\tfrac{\pi}{2}$ and $\tfrac{\pi}{2}<\lambda<\pi$ are interchanged under duality (\ref{duality}) with 
$\delta_a\leftrightarrow 1\!-\!\delta_a$. 
For $0<\lambda<\tfrac{\pi}{2}$, there are no unshaded 2-bands and the local energies are shown in Figure~\ref{superLocalEs}. 
By contrast, in the dual interval $\tfrac{\pi}{2}<\lambda<\pi$, there are no shaded 2-bands and the local energies are shown in Figure~\ref{LocalEs}. 
Under duality (\ref{duality}), the local energies are related by
\bea
H^{m,m',2}(a,b,c)=1-H^{m'-m,m',2}(a,b,c)
\eea

\subsubsection{$3\times 3$ local energies}

Starting with the $3\times 3$ local energies for $0<\lambda<\frac{\pi}{3}$ in \cite{TP}, the local energies for $\frac{2\pi}{3}<\lambda<\pi$ can be obtained by applying, for $n=3$, the conjectured duality relation
\bea
H^{m,m'}(a,b,c)=\tfrac{n}{2}-H^{m'-m,m'}(a,b,c),\qquad n=1,2,3,\ldots
\eea


\subsection{Energy statistic and one-dimensional sums}
\label{1dsums}
Following Baxter~\cite{BaxBook,BaxterCTM}, the energy statistic of RSOS paths is
\bea
E(\sigma)=\sum_{j=1}^N j H(\sigma_{j-1},\sigma_j,\sigma_{j+1})
\eea
The associated one-dimensional sums are defined as
\be
X_{abc}^{(N)}(q) = \sum_{\sigma} q^{E(\sigma)}, \qquad \sigma_0 = a, \ \sigma_N = b, \ \sigma_{N+1} = c
\label{eq:oneDconfigsums}
\ee
where the sum is over all RSOS paths $\sigma=\{\sigma_0,\sigma_1,\ldots,\sigma_N,\sigma_{N+1}\}$ at the given fusion level $n$.
These sums satisfy the recursion
\bea
X_{abc}^{(N)}(q) =\sum_{d\sim b} q^{N H(d,b,c)} X_{adb}^{(N-1)}(q) 
\label{recursion}
\eea
subject to the boundary conditions
\bea
X_{a0c}^{(N)}(q) =X_{a{m'}c}^{(N)}(q)= 0,\qquad
X_{abc}^{(0)}(q) &= \delta_{a,b}
\eea
where $d\sim b$ denotes that the heights $d$ and $b$ are adjacent at fusion level $n$.

\subsection{Ground states and sectors for $n=2$ and $m'>2m$}

\begin{figure}[htb]
\centering
\subfloat[Flat and alternating ground states for $0 < \lambda < \pi/2$.]{
\psset{unit=0.75cm}
\begin{pspicture}(-0.4,-0.5)(6,9)
\pspolygon[linewidth=0pt,fillstyle=solid,fillcolor=lightestblue](0,0)(6,0)(6,9)(0,9)
\pspolygon[linewidth=0pt,fillstyle=solid,fillcolor=lightlightblue](0,0)(6,0)(6,1)(0,1)
\pspolygon[linewidth=0pt,fillstyle=solid,fillcolor=lightlightblue](0,2)(6,2)(6,3)(0,3)
\pspolygon[linewidth=0pt,fillstyle=solid,fillcolor=lightlightblue](0,3)(6,3)(6,4)(0,4)
\pspolygon[linewidth=0pt,fillstyle=solid,fillcolor=lightlightblue](0,5)(6,5)(6,6)(0,6)
\pspolygon[linewidth=0pt,fillstyle=solid,fillcolor=lightlightblue](0,6)(6,6)(6,7)(0,7)
\pspolygon[linewidth=0pt,fillstyle=solid,fillcolor=lightlightblue](0,8)(6,8)(6,9)(0,9)
\psgrid[gridlabels=0pt,subgriddiv=1]
\rput(-.25,0){1}\rput(-.25,1){2}\rput(-.25,2){3}\rput(-.25,3){4}\rput(-.25,4){5}
\rput(-.25,5){6}\rput(-.25,6){7}\rput(-.25,7){8}\rput(-.25,8){9}\rput(-.35,9){10}
\rput(0.1,-0.35){0}\rput(1,-0.35){1}\rput(2,-0.35){2}\rput(3,-0.35){3}\rput(4,-0.35){4}\rput(5,-0.35){5}\rput(6,-0.35){6}
\psline[linewidth=1.5pt,linecolor=purple](0,3)(6,3)
\psline[linewidth=1.5pt,linecolor=blue](0,2)(1,4)(2,2)(3,4)(4,2)(5,4)(6,2)
\psline[linewidth=1.5pt,linecolor=apricot](0,4)(1,2)(2,4)(3,2)(4,4)(5,2)(6,4)
\psline[linewidth=1.5pt,linecolor=purple](0,6)(6,6)
\psline[linewidth=1.5pt,linecolor=blue](0,5)(1,7)(2,5)(3,7)(4,5)(5,7)(6,5)
\psline[linewidth=1.5pt,linecolor=apricot](0,7)(1,5)(2,7)(3,5)(4,7)(5,5)(6,7)
\rput(6.7,3){\textcolor{red}{$r\!=\!1$}}
\rput(6.7,6){\textcolor{red}{$r\!=\!2$}}
\end{pspicture}
}\hspace{2cm}
\subfloat[Flat ground states for  $\pi/2 < \lambda < \pi$.]{
\psset{unit=0.75cm}
\begin{pspicture}(-0.4,-0.5)(6,9)
\pspolygon[linewidth=0pt,fillstyle=solid,fillcolor=lightestblue](0,0)(6,0)(6,9)(0,9)
\pspolygon[linewidth=0pt,fillstyle=solid,fillcolor=lightlightblue](0,1)(6,1)(6,2)(0,2)
\pspolygon[linewidth=0pt,fillstyle=solid,fillcolor=lightlightblue](0,4)(6,4)(6,5)(0,5)
\pspolygon[linewidth=0pt,fillstyle=solid,fillcolor=lightlightblue](0,7)(6,7)(6,8)(0,8)
\psgrid[gridlabels=0pt,subgriddiv=1]
\rput(-.25,0){1}\rput(-.25,1){2}\rput(-.25,2){3}\rput(-.25,3){4}\rput(-.25,4){5}
\rput(-.25,5){6}\rput(-.25,6){7}\rput(-.25,7){8}\rput(-.25,8){9}\rput(-.35,9){10}
\rput(0.1,-0.35){0}\rput(1,-0.35){1}\rput(2,-0.35){2}\rput(3,-0.35){3}\rput(4,-0.35){4}\rput(5,-0.35){5}\rput(6,-0.35){6}
\psline[linewidth=1.5pt,linecolor=purple](0,2)(6,2)
\psline[linewidth=1.5pt,linecolor=purple](0,4)(6,4)
\psline[linewidth=1.5pt,linecolor=blue](0,7)(6,7)
\psline[linewidth=1.5pt,linecolor=blue](0,1)(6,1)
\psline[linewidth=1.5pt,linecolor=blue](0,5)(6,5)
\psline[linewidth=1.5pt,linecolor=purple](0,8)(6,8)
\rput(6.6,1){\textcolor{black}{$\rho_0(1)$}}\rput(6.6,2){\textcolor{black}{$\rho_1(1)$}}\rput(6.6,4){\textcolor{black}{$\rho_1(2)$}}\rput(6.6,5){\textcolor{black}{$\rho_0(2)$}}\rput(6.6,7){\textcolor{black}{$\rho_0(3)$}}\rput(6.6,8){\textcolor{black}{$\rho_1(3)$}}
\rput(7.8,1.5){\textcolor{red}{$r\!=\!1$}}
\rput(7.8,4.5){\textcolor{red}{$r\!=\!2$}}
\rput(7.8,7.5){\textcolor{red}{$r\!=\!2$}}
\end{pspicture}\hspace{1.5cm}
}
\caption{A comparison of ground state RSOS paths for dual $\mbox{RSOS}(m,m')_{2\times 2}$ models with (a)~$(m,m')=(7,11)$ and (b) $(m,m')=(4,11)$: 
(a) For $(m,m')=(7,11)$, the shaded 1-bands occur at heights $\left \lfloor \frac{11r}{7} \right\rfloor = 1,3,4,6,7,9$ for $r=1,2,\ldots,6$. 
The 6 ground state (shaded) 2-bands occur centered at heights $a= 4,7$. The ground states are either flat of the form $\{a,a,\ldots,a\}$ or alternating of the form $\{a\pm1,a\mp1,a\pm1,a\mp1,\ldots \}$. (b) For $(m,m')=(4,11)$, there are no shaded 2-bands. The ground state (shaded) 1-bands occur at heights $\left \lfloor \frac{11r}{4} \right\rfloor = 2,5,8$ for $r=1,2,3$. 
The 6 ground states are flat of the form $\{a,a,\ldots,a\}$ with $a$ belonging to the even or odd sequences $\rho_0(r)=2,6,8$ or $\rho_1(r)=3,5,9$.}
\label{fig:GSpaths}
\end{figure}
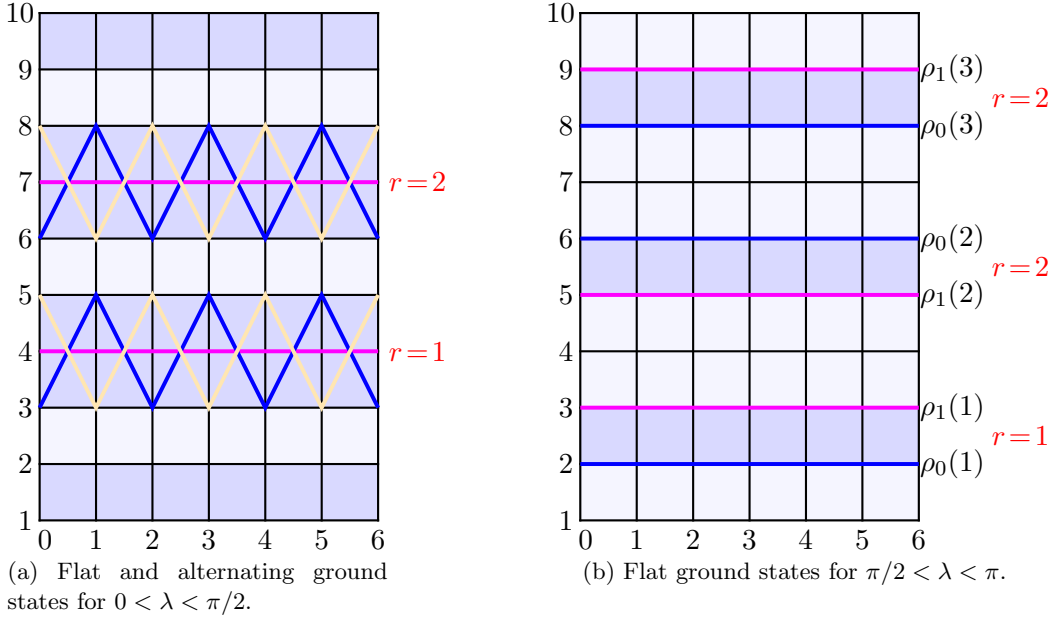

For the interval $0<\lambda<\tfrac{\pi}{2}$, there are shaded 2-bands and the associated ground states relate to the superconformal minimal models in the Neveu-Schwarz and Ramond sectors as discussed in \cite{TP}.
In contrast, for $\tfrac{\pi}{2}<\lambda<\pi$, there are no shaded 2-bands only shaded 1-bands. So, for these models, the superconformal groundstates are not supported and we find that the ground states are associated to shaded 1-bands. A comparison of ground states in the two different intervals is shown in Figure~\ref{fig:GSpaths}. Since the local energies are nonnegative $H(a,b,c)\ge 0$, any one-dimensional RSOS path $\sigma$ with $E(\sigma)=0$ is a ground state. The only RSOS paths with energy $E(\sigma)=0$ are the $2(m-1)$ flat paths $\sigma=\{\rho,\rho,\ldots,\rho\}$ with $\rho=\rho_0(r)$ or $\rho=\rho_1(r)$ for some $r$.

Generically, for suitable choices of $(a,b,c)$, the one-dimensional sums (\ref{eq:oneDconfigsums}) are interpreted as finitized conformal characters
\bea
\chi^{(N)}_\Delta(q)=q^{-c/24+\Delta}X_{abc}^{(N)}(q)=q^{-c/24+\Delta}\sum_{\sigma} q^{E(\sigma)}
\eea
where $E(\sigma)$ are conformal energies of the infinite system. These are the spectrum generating functions for a finite truncated set of conformal energy levels in a given sector labelled by $(a,b,c)$. The connection with characters is made by choosing the last step $(b,c)$ of the one-dimensional walks to agree with a ground state RSOS path labelled by $r$. Restricting to models with $\tfrac{\pi}{2}<\lambda<\pi$, the precise connection between $(a,b,c)$ and the conformal Kac quantum numbers $(r,s)$ is given by
\bea
a=s,\qquad b=c=\rho_\mu(r),\qquad \mu=\mbox{$s$ mod 2}
\eea

\section{$\mbox{RSOS}(m,2m+1)_{2\times 2}$ One-Dimensional Sums}

In \cite{JacobM,BFMW}, a study has been carried out of the one-dimensional sums associated with a particular choice of local energy functions for RSOS lattice paths with half-integer steps. We call these half-integer RSOS paths JM paths. In these papers, the local energy functions of JM paths were not related to integrable lattice models. However, in the thermodynamic limit, these one-dimensional sums quite remarkably reproduce the Virasoro characters of the sequences of nonunitary minimal models ${\cal M}(m,2m+ 1)$ with $m=2,3,4,\dots$. In this section, we show that that there exists an energy-preserving bijection between the JM paths and $\mbox{RSOS}(m,2m+1)_{2\times 2}$ paths. More precisely, we present a one-to-one mapping of RSOS spin-1 paths into equivalent spin-$\half$ JM paths with equivalent local energies. We conclude that the JM paths are described by the $\mbox{RSOS}(m,2m+1)_{2\times 2}$ Yang-Baxter integrable lattice models. This explains the remarkable observed properties of the JM one-dimensional sums.

We observe that when $m'=2m+1$, the sequence of integers~\eqref{eq:rho} defining the shaded bands simplifies to
\bea
\rho=\floor {\tfrac{r (2m+1)}{m}}=\floor{2r+{\tfrac{r}{m}}}=2r\qquad r=1,2,\dots,m-1
\eea
This means that every second band starting from the lowest height $a=1$ is shaded and
\bea
\delta_a=
\begin{cases}
0\quad\mbox{$a$ even}\\
1\quad\mbox{$a$ odd}
\end{cases}
\qquad\mbox{$a=1,\dots,m'-2$}
\label{eq:identity}
\eea
Since the adjacency graph decomposes for the $2\times 2$ fused models, it suffices to restrict $a,b,c$ to be odd in these cases.


In the next two subsections, we consider the two simplest nonunitary RSOS$(m,m')$ models with crossing parameter in the interval $\tfrac{\pi}{2}<\lambda<\pi$, namely, the RSOS(2,5) (Yang-Lee model) and RSOS(3,7). The heights $a=1,2,\dots...,m'-1$ live on the $A_{m'-1}$ Dynkin diagrams which possess a $\mathbb{Z}_2$ symmetry. Specifically, the Boltzmann face weights are invariant under the height reversal $\sigma\leftrightarrow m'-\sigma$. As a consequence of this symmetry the Dynkin diagrams, which encode the adjacency rules among heights, can be folded into tadpole diagrams. Example foldings are shown in Figures~\ref{fig:tadpoleM(2,5)1} and~\ref{fig:tadpoleM(3,7)1} for the $n=1$ adjacency. 
For RSOS$(2,5)_{1 \times 1}$, the nodes of the tadpole are interpreted as particles $\bullet$ or vacancies $\circ$ with nearest neighbour particle exclusion. Specifically, a particle $\bullet=1=4$ is allowed next to a vacancy $\circ=2=3$ but not allowed next to another particle. For RSOS$(3,7)_{1 \times 1}$, there are two types of particles $\bullet=1=6$ and $\gbullet=2=5$ in addition to the vacancy $\circ=3=4$. 
At fusion level $n=2$, the RSOS$(2,5)_{2 \times 2}$ model is equivalent to two independent folded copies of the original unfused model RSOS$(2,5)_{1 \times 1}$ as indicated in Figure~\ref{fig:tadpoleM(2,5)2}. The situation is different for the RSOS$(3,7)_{2 \times 2}$ fused model. The new tadpole diagram $T_3^{(n=2)}$, shown in Figure~\ref{fig:tadpoleM(3,7)2}, acquires an additional loop corresponding to the particle $\circ=3$ and $\circ=4$.

\begin{figure}
%
\subfloat[RSOS$(2,5)_{1\times1}$: folding of the $A_4=A_4^{(n=1)}$ and the corresponding $T_2^{(n=1)}$ tadpole.] {
\psset{unit=1cm}
\begin{pspicture}(-1.5,-1)(2,4)
\psline(0,0)(0,3)
\psarc[linewidth=1.5pt,arrowsize=8pt]{<->}(0.05,1.5){1.4}{305}{55}
\rput(1.8,1.5){\scalebox{1.2}{$\mathbb{Z}_2$}}
\rput(0.5,0){1}\rput(0.5,1){2}\rput(0.5,2){3}\rput(0.5,3){4}
\rput(0,0){$\pscircle[fillstyle=solid,fillcolor=black](0,0){.12}$}\rput(0,1){$\pscircle[fillstyle=solid,fillcolor=white](0,0){.12}$}\rput(0,2){$\pscircle[fillstyle=solid,fillcolor=white](0,0){.12}$}\rput(0,3){$\pscircle[fillstyle=solid,fillcolor=black](0,0){.12}$}
\end{pspicture}
\psset{unit=1cm}
\begin{pspicture}(-1,-1)(2.5,4)
\psline(0,0)(0,1)
\rput(0.5,0){1=4}\rput(0.5,.8){2=3}
\rput(0,0){$\pscircle[fillstyle=solid,fillcolor=black](0,0){.12}$}
\rput(0,1){$\pscircle[fillstyle=solid,fillcolor=white](0,.27){.30}$}
\rput(0,1){$\pscircle[fillstyle=solid,fillcolor=white](0,0){.12}$}
\end{pspicture}
\label{fig:tadpoleM(2,5)1}
}
\hspace{1 cm}
\subfloat[RSOS$(2,5)_{2\times2}$: decomposition of $A_4^{(n=2)}$ into the tensor product of two $T_2^{(n=2)}$ tadpoles.] {
\psset{unit=1cm}
\begin{pspicture}(-1.,-1)(1,4)
\psarc[linewidth=1pt,arrowsize=0pt]{<->}(-1.2,1){1.55}{318}{42}
\rput(0,0){$\pscircle[fillstyle=solid,fillcolor=black](0,0){.12}$}
\rput(0,2){$\pscircle[fillstyle=solid,fillcolor=white](0.35,0){.30}$}
\rput(0,2){$\pscircle[fillstyle=solid,fillcolor=white](0,0){.12}$}
\rput(0.3,0){1}
\rput(-0.22,2){3}
\psarc[linewidth=1pt,arrowsize=0pt]{<->}(1.2,2){1.55}{138}{222}
\rput(0,3){$\pscircle[fillstyle=solid,fillcolor=black](0,0){.12}$}
\rput(0,1){$\pscircle[fillstyle=solid,fillcolor=white](-0.35,0){.30}$}
\rput(0,1){$\pscircle[fillstyle=solid,fillcolor=white](0,0){.12}$}
\rput(0.3,3){4}
\rput(0.22,1){2}
\rput(1.2,1){\scalebox{2}{$=$}}
\end{pspicture}
\psset{unit=1cm}
\begin{pspicture}(-1,-1)(4,4)
\multiput(0,0)(2,0){2}{\psline(0,0)(0,2)}
\rput(0.5,0){1}\rput(0.5,1.8){3}
\rput(0,0){$\pscircle[fillstyle=solid,fillcolor=black](0,0){.12}$}\rput(0,2){$\pscircle[fillstyle=solid,fillcolor=white](0,0.35){.30}$}\rput(0,2){$\pscircle[fillstyle=solid,fillcolor=white](0,0){.12}$}
\rput(2.5,0){4}\rput(2.5,1.8){2}
\rput(2,0){$\pscircle[fillstyle=solid,fillcolor=black](0,0){.12}$}\rput(2,2){$\pscircle[fillstyle=solid,fillcolor=white](0,0.35){.30}$}\rput(2,2){$\pscircle[fillstyle=solid,fillcolor=white](0,0){.12}$}
\rput(1,1){\scalebox{2}{$\otimes$}}
\end{pspicture}
\label{fig:tadpoleM(2,5)2}
}
\\
\psset{unit=1cm}
\subfloat[RSOS$(3,7)_{1\times1}$: folding of the $A_6=A_6^{(n=1)}$ and the corresponding $T_3^{(n=1)}$ tadpole.]{
\begin{pspicture}(-1.5,-1)(2,6)
\psline(0,0)(0,5)
\psarc[linewidth=1.5pt,arrowsize=8pt]{<->}(0.05,2.5){1.4}{305}{55}
\rput(1.8,2.5){\scalebox{1.2}{$\mathbb{Z}_2$}}
\rput(0.5,0){1}\rput(0.5,1){2}\rput(0.5,2){3}\rput(0.5,3){4}\rput(0.5,4){5}\rput(0.5,5){6}
 \rput(0,0){$\pscircle[fillstyle=solid,fillcolor=black](0,0){.12}$}\rput(0,1){$\pscircle[fillstyle=solid,fillcolor=green](0,0){.12}$}\rput(0,2){$\pscircle[fillstyle=solid,fillcolor=white](0,0){.12}$}\rput(0,3){$\pscircle[fillstyle=solid,fillcolor=white](0,0){.12}$}\rput(0,4){$\pscircle[fillstyle=solid,fillcolor=green](0,0){.12}$}\rput(0,5){$\pscircle[fillstyle=solid,fillcolor=black](0,0){.12}$}
\end{pspicture}
\psset{unit=1cm}
\begin{pspicture}(-1,-1)(2.5,6)
\psline(0,0)(0,2)
\rput(0.5,0){1=6}\rput(0.5,1){2=5}\rput(0.5,1.8){3=4}
\rput(0,0){$\pscircle[fillstyle=solid,fillcolor=black](0,0){.12}$}\rput(0,1){$\pscircle[fillstyle=solid,fillcolor=green](0,0){.12}$}\rput(0,2){$\pscircle[fillstyle=solid,fillcolor=white](0,0.28){.30}$}\rput(0,2){$\pscircle[fillstyle=solid,fillcolor=white](0,0){.12}$}
\end{pspicture}
\label{fig:tadpoleM(3,7)1}
}
\hspace{1 cm}
\subfloat[RSOS$(3,7)_{2\times2}$: decomposition of $A_6^{(n=2)}$ into the tensor product of two $T_3^{(n=2)}$ tadpoles.]{
\psset{unit=1cm}
\begin{pspicture}(-1.,-1)(1,6)
\psarc[linewidth=1pt,arrowsize=0pt]{<->}(-1.2,1){1.55}{318}{42}
\psarc[linewidth=1pt,arrowsize=0pt]{<->}(-1.2,3){1.55}{318}{42}
\rput(0,0){$\pscircle[fillstyle=solid,fillcolor=black](0,0){.12}$}
\rput(0,2){$\pscircle[fillstyle=solid,fillcolor=white](0.35,0){.30}$}
\rput(0,2){$\pscircle[fillstyle=solid,fillcolor=white](0,0){.12}$}
\rput(0,4){$\pscircle[fillstyle=solid,fillcolor=white](0.35,0){.30}$}
\rput(0,4){$\pscircle[fillstyle=solid,fillcolor=green](0,0){.12}$}
\rput(0.3,0){1}
\rput(-0.22,2){3}
\rput(-0.22,4){5}
\psarc[linewidth=1pt,arrowsize=0pt]{<->}(1.2,2){1.55}{138}{222}
\psarc[linewidth=1pt,arrowsize=0pt]{<->}(1.2,4){1.55}{138}{222}
\rput(0,5){$\pscircle[fillstyle=solid,fillcolor=black](0,0){.12}$}
\rput(0,3){$\pscircle[fillstyle=solid,fillcolor=white](-0.35,0){.30}$}
\rput(0,3){$\pscircle[fillstyle=solid,fillcolor=white](0,0){.12}$}
\rput(0,1){$\pscircle[fillstyle=solid,fillcolor=white](-0.35,0){.30}$}
\rput(0,1){$\pscircle[fillstyle=solid,fillcolor=green](0,0){.12}$}
\rput(0.22,1){2}
\rput(0.22,3){4}
\rput(0.25,5){6}
\rput(1.2,2){\scalebox{2}{$=$}}
\end{pspicture}
\psset{unit=1cm}
\begin{pspicture}(-1,-1)(4,6)
\multiput(0.2,0)(2,0){2}{\psline(0,0)(0,4)}
\rput(0.5,0){1}\rput(0.5,2){3}\rput(0.5,4){5}
\rput(1.9,0){6}\rput(1.9,2){4}\rput(1.9,4){2}
\rput(0.2,0){$\pscircle[fillstyle=solid,fillcolor=black](0,0){.12}$}
\rput(0.2,2){$\pscircle[fillstyle=solid,fillcolor=white](-0.3,0){.30}$}
\rput(0.2,2){$\pscircle[fillstyle=solid,fillcolor=white](0,0){.12}$}
\rput(0.2,4){$\pscircle[fillstyle=solid,fillcolor=white](-0.3,0){.30}$}
\rput(0.2,4){$\pscircle[fillstyle=solid,fillcolor=green](0,0){.12}$}
 \rput(2.2,0){$\pscircle[fillstyle=solid,fillcolor=black](0,0){.12}$}
 \rput(2.2,2){$\pscircle[fillstyle=solid,fillcolor=white](0.3,0){.30}$}
 \rput(2.2,2){$\pscircle[fillstyle=solid,fillcolor=white](0,0){.12}$}
 \rput(2.2,4){$\pscircle[fillstyle=solid,fillcolor=white](0.3,0){.30}$}
 \rput(2.2,4){$\pscircle[fillstyle=solid,fillcolor=green](0,0){.12}$}
 \rput(1.2,2){\scalebox{2}{$\otimes$}}
\end{pspicture}
\label{fig:tadpoleM(3,7)2}
}
\caption{The RSOS$(2,5)$ and RSOS$(3,7)$ lattice models are identified with tadpole diagrams which encode the adjacency rules between heights. The RSOS$(2,5)_{1\times1}$ (a)  and RSOS$(2,5)_{2\times2}$ (b) models share the same tadpole diagram. This results from the fact that the $2 \times 2$ lattice fusion gives back the original lattice model. In contrast, the RSOS$(3,7)_{1\times1}$(c)  and RSOS$(3,7)_{2\times2}$ (d) models show different tadpole diagrams because the $2 \times 2$ lattice fusion produces a new lattice model. }
\end{figure}
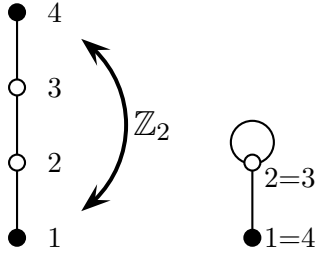
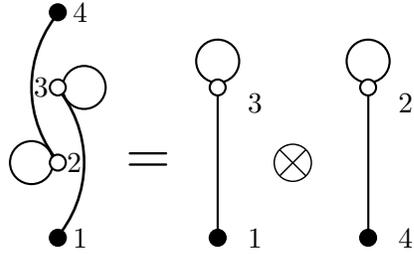
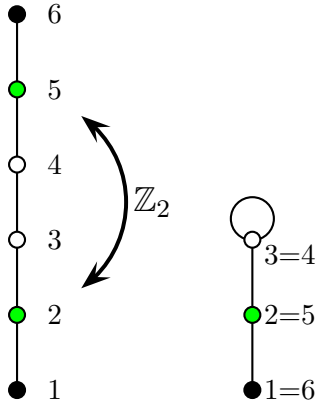
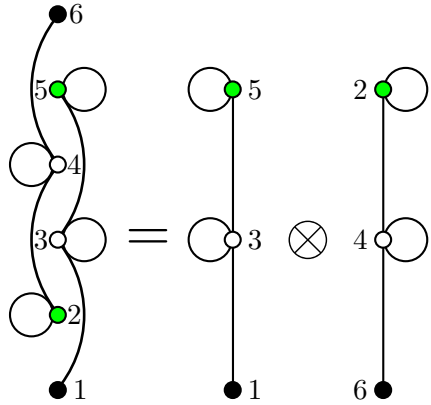

\subsection{$\mbox{RSOS}(2,5)$}

\label{LeeYang}
\begin{figure}[htb]
\centering
\psset{unit=0.75cm}
\subfloat[ $\mbox{RSOS}(2,5)_{1\times 1}$ has a single shaded 1-band and the 2 ground states corresponds to alternating paths inside it.]{
\begin{pspicture}(-0.4,-0.5)(6,3)
\pspolygon[linewidth=0pt,fillstyle=solid,fillcolor=lightestblue](0,0)(6,0)(6,3)(0,3)
\pspolygon[linewidth=0pt,fillstyle=solid,fillcolor=lightlightblue](0,1)(6,1)(6,2)(0,2)
\psgrid[gridlabels=0pt,subgriddiv=1]
\rput(0.1,-0.35){0}\rput(1,-0.35){1}\rput(2,-0.35){2}\rput(3,-0.35){3}\rput(4,-0.35){4}\rput(5,-0.35){5}\rput(6,-0.35){6}
\psline[linewidth=1.5pt,linecolor=blue](0,1)(1,2)(2,1)(3,2)(4,1)(5,2)(6,1)
\psline[linewidth=1.5pt,linecolor=purple](0,2)(1,1)(2,2)(3,1)(4,2)(5,1)(6,2)
\rput(-.35,0){1}\rput(-.35,1){2}\rput(-.35,2){3}\rput(-.35,3){4}\rput(0,0){$\bullet$}\rput(0,1){$\pscircle[fillstyle=solid,fillcolor=white](0,0){.12}$}\rput(0,2){$\pscircle[fillstyle=solid,fillcolor=white](0,0){.12}$}\rput(0,3){$\bullet$}
\label{fig:YL-GS1}
\end{pspicture}
}
\hspace{2cm}
\subfloat[ $\mbox{RSOS}(2,5)_{2\times 2}$ has a single shaded 1-band and 2 flat ground states corresponding to the lower and upper height of this band.]{
\psset{unit=0.75cm}
\begin{pspicture}(-0.4,-0.5)(6,3)
\pspolygon[linewidth=0pt,fillstyle=solid,fillcolor=lightestblue](0,0)(6,0)(6,3)(0,3)
\pspolygon[linewidth=0pt,fillstyle=solid,fillcolor=lightlightblue](0,1)(6,1)(6,2)(0,2)
\psgrid[gridlabels=0pt,subgriddiv=1]
\rput(0.1,-0.35){0}\rput(1,-0.35){1}\rput(2,-0.35){2}\rput(3,-0.35){3}\rput(4,-0.35){4}\rput(5,-0.35){5}\rput(6,-0.35){6}
\psline[linewidth=1.5pt,linecolor=purple](0,2)(6,2)
\psline[linewidth=1.5pt,linecolor=blue](0,1)(6,1)
\rput(6.6,1){\textcolor{black}{$\rho_0(1)$}}\rput(6.6,2){\textcolor{black}{$\rho_1(1)$}}
\rput(-.35,0){1}\rput(-.35,1){2}\rput(-.35,2){3}\rput(-.35,3){4}\rput(0,0){$\bullet$}\rput(0,1){$\pscircle[fillstyle=solid,fillcolor=white](0,0){.12}$}\rput(0,2){$\pscircle[fillstyle=solid,fillcolor=white](0,0){.12}$}\rput(0,3){$\bullet$}
\label{fig:YL-GS2}
\end{pspicture}
}
\caption{Ground state configurations of RSOS$(2,5)_{1\times 1}$ and RSOS$(2,5)_{2\times2}$ lattice models.}
\end{figure}
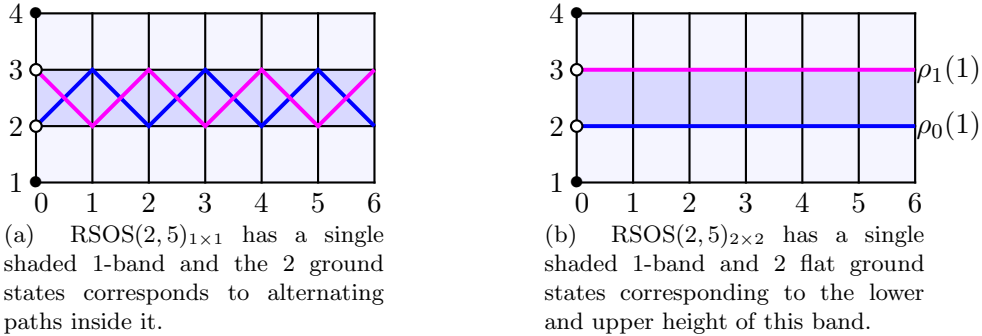

The Yang-Lee (YL) model~\cite{LeeYang} is associated with the exactly solvable $\mbox{RSOS(2,5)}$ model. It is defined on a square lattice and its heights live on  a $A_4$ Dynkin diagram. 
There is a single shaded 1-band between heights 2 and 3, and ground state configurations are alternating paths inside it as shown in Figure~\ref{fig:YL-GS1}.
Identifying the heights related by this ${\Bbb Z}_2$ symmetry and using the notation $\bullet=1, 4$, $\circ= 2, 3$, the face weights (\ref{Boltzmann}) with $g_a=1$ become
\begin{subequations}
\begin{align}
\Wtt{W}{\circ}{\bullet}{\circ}{\circ} u&=\Wtt{W}{\circ}{\circ}{\circ}{\bullet} u = s(\lambda-u)
\\
\Wtt{W}{\bullet}{\circ}{\circ}{\circ} u &= -s(u)
\\
\Wtt{W}{\circ}{\circ}{\bullet}{\circ} u &= -\frac{s(u)}{s(2\lambda)}
\\
\Wtt{W}{\bullet}{\circ}{\bullet}{\circ} u &= \frac{s(2\lambda-u)}{s(2\lambda)}
\\
\Wtt{W}{\circ}{\bullet}{\circ}{\bullet} u & = s(\lambda+u)
\\
\Wtt{W}{\circ}{\circ}{\circ}{\circ} u & = \frac{s(2\lambda+u) }{s(2\lambda)}
\end{align}
\label{1x1weights}
\end{subequations}
In particle notation for RSOS$(2,5)$, $\bullet=1$ is used for an occupied site and $\circ=0$ for an unoccupied site. In this notation, the adjacency graph is the tadpole $T_2$. The local energy functions can be taken to be
\bea
H(a,b,c) =
\begin{cases}1,\quad &(a,b,c)=(\circ,\bullet,\circ)\\
0,&\mbox{other allowed triples}
\end{cases}
\eea

Implementing the $2\times2$ fusion as in (\ref{2x2fusion}) with a gauge yields the following face weights for the $\mbox{RSOS(2,5)}_{2\times2}$ model
\begin{subequations}
\begin{align}
\Wtwo{\circ}{\bullet}{\circ}{\circ} u&=\Wtwo{\circ}{\circ}{\circ}{\bullet} u =\frac{s(\lambda-u)s(2\lambda+u)}{s(2\lambda)}
\\
\Wtwo{\bullet}{\circ}{\circ}{\circ} u & = -\frac{g_0}{g_1} \frac{s(u) s(2\lambda+u)}{s(2\lambda)^3}
\\
\Wtwo{\circ}{\circ}{\bullet}{\circ} u & = -\frac{g_1}{g_0}\,s(u)s(2\lambda+u)
\\
\Wtwo{\bullet}{\circ}{\bullet}{\circ} u &= \frac{s(2\lambda-u)s(2\lambda+u)}{s(2\lambda)^2}
\\
\Wtwo{\circ}{\bullet}{\circ}{\bullet} u & =\frac{s(\lambda+u)s(2\lambda+u)}{s(2\lambda)}
\\
\Wtwo{\circ}{\circ}{\circ}{\circ} u & = \frac{s(2\lambda+u)^2}{s(2\lambda)^2}
\end{align}
\end{subequations}
Fixing the gauge to be $g_{0}/g_{1}=s(2\lambda)^2$ and removing the overall scale factor $s(2\lambda+u)/s(2\lambda)$, these weights coincide with the $1\times 1$ weights above (\ref{1x1weights}). Consequently  RSOS$(2,5)_{1\times1}$ and RSOS$(2,5)_{2 \times 2}$ coincide as lattice models.

The new local energies for the $2\times2$ fused model are 
\bea
H(\circ,\bullet,\circ) =H(\circ,\circ,\circ)=0,\qquad
H( \circ, \circ, \bullet) =H( \bullet, \circ, \circ)=\half,\qquad
H(\bullet,\circ,\bullet)=1
\eea
The ground state configurations for the fused model are flat paths corresponding to the lower and upper height of the single shaded 1-band as seen in Figure~\ref{fig:YL-GS2}.

%

\subsection{$\mbox{RSOS}(3,7)$}
The second simplest example of nonunitary RSOS model is $\mbox{RSOS(3,7)}$, with heights living on the $A_{6}$ Dynkin diagram. Identifying the heights related by the ${\Bbb Z}_2$ symmetry and using the notation $\bullet= 1,6$, $\circ=3,4$, $\gbullet=5,2$, the face weights (\ref{Boltzmann}) with $g_a=1$ become\goodbreak
\begin{subequations}
\begin{align}
\Wtt{W}{\gbullet}{\bullet}{\gbullet}{\circ} u&=\Wtt{W}{\gbullet}{\circ}{\gbullet}{\bullet} u = \Wtt{W}{\circ}{\circ}{\circ}{\gbullet} u=\Wtt{W}{\circ}{\gbullet}{\circ}{\circ} u= s(\lambda-u)
\\
\Wtt{W}{\bullet}{\gbullet}{\circ}{\gbullet} u &= -\frac{s(3\lambda)s(u)}{s(2\lambda)}
\\
\Wtt{W}{\circ}{\gbullet}{\bullet}{\gbullet} u &= -\frac{s(u)}{s(2\lambda)}
\\
\Wtt{W}{\gbullet}{\circ}{\circ}{\circ} u&=- s(u)
\\
\Wtt{W}{\circ}{\circ}{\gbullet}{\circ} u &= -\frac{s(2\lambda)s(u)}{s(3\lambda)}
\\
\Wtt{W}{\bullet}{\gbullet}{\bullet}{\gbullet} u &= \frac{s(2\lambda-u)}{s(2\lambda)}
\\
\Wtt{W}{\gbullet}{\bullet}{\gbullet}{\bullet} u & = s(\lambda+u)
\\
\Wtt{W}{\gbullet}{\circ}{\gbullet}{\circ} u &= \frac{s(3\lambda-u)}{s(3\lambda)}
\\
\Wtt{W}{\circ}{\gbullet}{\circ}{\gbullet} u & = \frac{s(2\lambda+u) }{s(2\lambda)}
\\
\Wtt{W}{\circ}{\circ}{\circ}{\circ} u &= \frac{s(3\lambda+u)}{s(3\lambda)}
%
\end{align}
\label{1x1weights}
\end{subequations}
In particle notation $\bullet=1$, $\circ=3$ and $\gbullet=5$ are used for the three different occupation states. In this notation, the adjacency graph is the tadpole $T_3$. The RSOS$(3,7)_{1\times 1}$ local energy functions are
\begin{subequations}
\begin{align}
H(\gbullet,\circ,\gbullet) &=H(\circ,\gbullet,\circ)=0\\
H(\circ,\circ,\gbullet) &=H(\gbullet,\circ,\circ)=H(\circ,\gbullet,\bullet)=H(\bullet,\gbullet,\circ)=\tfrac{1}{4}\\
H(\circ,\circ,\circ) &=H(\gbullet,\bullet,\gbullet)=H(\bullet,\gbullet,\bullet)=\half
\end{align}
\end{subequations}
There are two shaded 1-bands, one between heights 2 and 3, the other between heights 4 and 5. The ground state configurations are alternating paths inside each shaded 1-band as shown in Figure~\ref{fig:3,7-GS1}.

Implementing the $2\times2$ fusion as in (\ref{2x2fusion}) and allowing gauge factors yields the following face weights for the $\mbox{RSOS(3,7)}_{2\times2}$ model
\begin{subequations}
\begin{align}
\Wtwo{\circ}{\bullet}{\circ}{\gbullet} u & =\Wtwo{\circ}{\gbullet}{\circ}{\bullet} u= \frac{s(\lambda-u)s(2\lambda-u)}{s(2\lambda)}
\\
\Wtwo{\circ}{\bullet}{\circ}{\circ} u&=\Wtwo{\circ}{\circ}{\circ}{\bullet} u =\frac{s(\lambda-u)s(2\lambda+u)}{s(2\lambda)}
\\
\Wtwo{\circ}{\gbullet}{\circ}{\circ} u & = \Wtwo{\circ}{\circ}{\circ}{\gbullet} u= \frac{s(\lambda-u) s(3\lambda+u)}{s(3\lambda)}
\\
\Wtwo{\gbullet}{\circ}{\gbullet}{\gbullet} u & =\Wtwo{\gbullet}{\gbullet}{\gbullet}{\circ} u= \frac{s(\lambda-u)s(3\lambda-u)}{s(3\lambda)}
\\
\Wtwo{\bullet}{\circ}{\circ}{\circ} u & = -\frac{g_2}{g_1}\,\frac{s(u)s(3\lambda-u)}{s(2\lambda)^2}
\\
\Wtwo{\gbullet}{\circ}{\circ}{\circ} u & =-\frac{g_2}{g_3}\,\frac{s(u)s(3\lambda+u)}{s(3\lambda)^2}
\\
\Wtwo{\circ}{\gbullet}{\gbullet}{\gbullet} u & =-\frac{g_3}{g_2}\,\frac{s(u)s(2\lambda+u)}{s(2\lambda)s(3\lambda)}
\\
\Wtwo{\circ}{\circ}{\bullet}{\circ} u &= -\frac{g_1}{g_2}\,\frac{s(u)s(3\lambda-u)}{s(3\lambda)}
\\
\Wtwo{\circ}{\circ}{\gbullet}{\circ} u & =-\frac{g_3}{g_2}\,\frac{s(u)s(3\lambda+u)s(2\lambda)}{s(3\lambda)}
\\
\Wtwo{\gbullet}{\gbullet}{\circ}{\gbullet}u & =-\frac{g_2}{g_3}\,s(u)s(2\lambda+u)s(2\lambda)
\\
\Wtwo{\gbullet}{\circ}{\bullet}{\circ} u & =\frac{g_1}{g_3}\,\frac{s(u)s(\lambda+u)}{s(3\lambda)^2}
\\
\Wtwo{\bullet}{\circ}{\gbullet}{\circ} u & =\frac{g_3}{g_1}\,\frac{s(u)s(\lambda+u)}{s(2\lambda)}
\\
\Wtwo{\circ}{\bullet}{\circ}{\bullet} u&=\frac{s(\lambda+u)s(2\lambda+u)}{s(2\lambda)}
\\
\Wtwo{\bullet}{\circ}{\bullet}{\circ} u & =\frac{s(2\lambda-u)s(3\lambda-u)}{s(2\lambda)s(3\lambda)}
\\
\Wtwo{\circ}{\gbullet}{\circ}{\gbullet}  u & =\frac{s(2\lambda+u)s(3\lambda+u)}{s(2\lambda)s(3\lambda)}
\\
\Wtwo{\gbullet}{\circ}{\gbullet}{\circ} u & =\frac{s(3\lambda+u)s(3\lambda-u)}{s(3\lambda)^2}
\\
\Wtwo{\circ}{\gbullet}{\gbullet}{\circ} u & =\Wtwo{\circ}{\circ}{\gbullet}{\gbullet} u= \frac{g_3}{g_2}\,\frac{s(u)s(u-\lambda)}{s(2\lambda)s(3\lambda)}
\\
\Wtwo{\gbullet}{\circ}{\circ}{\gbullet} u & =\Wtwo{\gbullet}{\gbullet}{\circ}{\circ} u =\frac{g_2}{g_3}\,\frac{s(u)s(u-\lambda)}{s(3\lambda)}
\\
\Wtwo{\circ}{\circ}{\circ}{\circ} u & = \frac{s(2\lambda+u)s(3\lambda-u)}{s(2\lambda)s(3\lambda)}
+\frac{s(2\lambda)s(u)s(u-\lambda)}{s(3\lambda)^2}
\\
\Wtwo{\gbullet}{\gbullet}{\gbullet}{\gbullet} u & =\frac{s(2\lambda+u)s(3\lambda-u)}{s(2\lambda)s(3\lambda)}
\end{align}
\end{subequations}
In contrast to the Yang-Lee case, no choice of gauge factors can map the fused weights onto the unfused ones. This implies that the RSOS$(3,7)_{1\times1}$ and RSOS$(3,7)_{2 \times 2}$ models are distinct Yang-Baxter integrable lattice models even though we will argue that they lie in the same universality class.

The local energies for the RSOS$(3,7)_{2 \times 2}$ model are 
\begin{subequations}
\begin{align}
H(\bullet,\circ,\gbullet) =H(\gbullet,\circ,\bullet)=H(\bullet,\circ,\bullet)=H(\circ,\gbullet,\circ)&=1\\
H(\bullet,\circ,\circ)=H(\circ,\circ,\bullet)=H(\circ,\circ,\gbullet)=H(\gbullet,\circ,\circ)=H(\circ,\gbullet,\gbullet)=H(\gbullet,\gbullet,\circ)&=\half\\
H(\circ,\bullet,\circ)=H(\gbullet,\circ,\gbullet)=H(\circ,\circ,\circ)=H(\gbullet,\gbullet,\gbullet)&=0
\end{align}
\end{subequations}
The ground state configurations for this model are flat paths corresponding to the lower and upper height of each shaded 1-band as seen in Figure~\ref{fig:3,7-GS2}.
\begin{figure}[htb]
\centering
\psset{unit=0.75cm}
\subfloat[ $\mbox{RSOS}(3,7)_{1\times 1}$ has two shaded 1-bands and the 4 ground states corresponds to alternating paths inside them.]{
\begin{pspicture}(-0.4,-0.5)(6,5)
\pspolygon[linewidth=0pt,fillstyle=solid,fillcolor=lightestblue](0,0)(6,0)(6,5)(0,5)
\pspolygon[linewidth=0pt,fillstyle=solid,fillcolor=lightlightblue](0,1)(6,1)(6,2)(0,2)
\pspolygon[linewidth=0pt,fillstyle=solid,fillcolor=lightlightblue](0,3)(6,3)(6,4)(0,4)
\psgrid[gridlabels=0pt,subgriddiv=1]
\rput(0.1,-0.35){0}\rput(1,-0.35){1}\rput(2,-0.35){2}\rput(3,-0.35){3}\rput(4,-0.35){4}\rput(5,-0.35){5}\rput(6,-0.35){6}
\psline[linewidth=1.5pt,linecolor=blue](0,1)(1,2)(2,1)(3,2)(4,1)(5,2)(6,1)
\psline[linewidth=1.5pt,linecolor=purple](0,2)(1,1)(2,2)(3,1)(4,2)(5,1)(6,2)
\psline[linewidth=1.5pt,linecolor=blue](0,3)(1,4)(2,3)(3,4)(4,3)(5,4)(6,3)
\psline[linewidth=1.5pt,linecolor=purple](0,4)(1,3)(2,4)(3,3)(4,4)(5,3)(6,4)
\rput(-.35,0){1}\rput(-.35,1){2}\rput(-.35,2){3}\rput( -.35,3){4}\rput( -.35,4){5}\rput( -.35,5){6}\rput(0,0){$\bullet$}\rput(0,1){$\pscircle[fillstyle=solid,fillcolor=green](0,0){.12}$}\rput(0,2){$\pscircle[fillstyle=solid,fillcolor=white](0,0){.12}$}\rput(0,3){$\pscircle[fillstyle=solid,fillcolor=white](0,0){.12}$}\rput(0,4){$\pscircle[fillstyle=solid,fillcolor=green](0,0){.12}$}\rput(0,5){$\bullet$}
\label{fig:3,7-GS1}
\end{pspicture}
}
\hspace{2cm}
\subfloat[ $\mbox{RSOS}(3,7)_{2\times 2}$ has two shaded 1-bands and 4 flat ground states corresponding to the lower and upper height of these bands.]{
\psset{unit=0.75cm}
\begin{pspicture}(-0.4,-0.5)(6,5)
\pspolygon[linewidth=0pt,fillstyle=solid,fillcolor=lightestblue](0,0)(6,0)(6,3)(0,3)
\pspolygon[linewidth=0pt,fillstyle=solid,fillcolor=lightlightblue](0,1)(6,1)(6,2)(0,2)
\pspolygon[linewidth=0pt,fillstyle=solid,fillcolor=lightlightblue](0,3)(6,3)(6,4)(0,4)
\psgrid[gridlabels=0pt,subgriddiv=1]
\rput(0.1,-0.35){0}\rput(1,-0.35){1}\rput(2,-0.35){2}\rput(3,-0.35){3}\rput(4,-0.35){4}\rput(5,-0.35){5}\rput(6,-0.35){6}
\psline[linewidth=1.5pt,linecolor=purple](0,2)(6,2)
\psline[linewidth=1.5pt,linecolor=blue](0,1)(6,1)
\psline[linewidth=1.5pt,linecolor=purple](0,4)(6,4)
\psline[linewidth=1.5pt,linecolor=blue](0,3)(6,3)
\rput(6.6,1){\textcolor{black}{$\rho_0(1)$}}\rput(6.6,2){\textcolor{black}{$\rho_1(1)$}}\rput(6.6,3){\textcolor{black}{$\rho_0(2)$}}\rput(6.6,4){\textcolor{black}{$\rho_1(2)$}}
\rput(-.35,0){1}\rput(-.35,1){2}\rput(-.35,2){3}\rput( -.35,3){4}\rput( -.35,4){5}\rput( -.35,5){6}\rput(0,0){$\bullet$}\rput(0,1){$\pscircle[fillstyle=solid,fillcolor=green](0,0){.12}$}\rput(0,2){$\pscircle[fillstyle=solid,fillcolor=white](0,0){.12}$}\rput(0,3){$\pscircle[fillstyle=solid,fillcolor=white](0,0){.12}$}\rput(0,4){$\pscircle[fillstyle=solid,fillcolor=green](0,0){.12}$}\rput(0,5){$\bullet$}
\label{fig:3,7-GS2}
\end{pspicture}
}
\caption{Ground state configurations for the $1\times1$ unfused and $2\times2$ fused $A_6$ RSOS models.}
\end{figure}
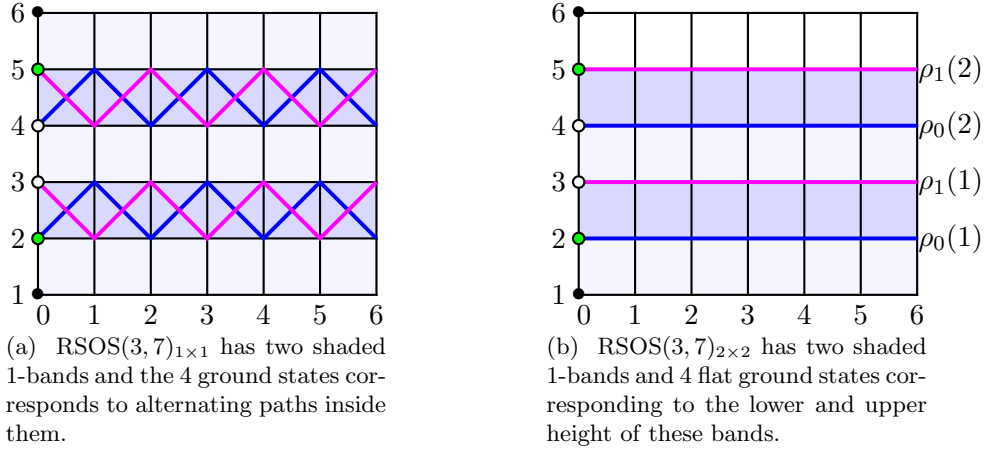

\subsection{RSOS$(m,2m+1)$ and JM$(m,2m+1)$}

\begin{figure}[t]
\vspace{-1cm}
\begin{center}
\subfloat[JM(2,5) and RSOS$(2,5)_{2\times2}$ paths with $k=1$ and $N=17$ integer steps: $\sigma=1,2$ corresponds to $1=\bullet$ and $2=\circ$ in the $T_2^{(n=2)}$ tadpole description of RSOS models.]{
\psset{unit=0.8cm}
\begin{pspicture}(0,-0.8)(18.5,2)
\centering
\pspolygon[linewidth=1pt,linecolor=lightestblue,fillstyle=solid,fillcolor=lightestblue](0,0)(18.,0)(18.,1)(0,1)
\multiput(0.5,0)(1,0){18}{\psline[linewidth=.5pt,linestyle=dashed,dash=2pt](0,0)(0,1)}
\multiput(1,0)(1,0){18}{\psline(0,0)(0,1)}
\multiput(0,0.5)(0,1){1}{\psline[linewidth=.5pt,linestyle=dashed,dash=2pt](0,0)(18.,0)}
\multiput(0,1)(0,1){1}{\psline[linestyle=solid](0,0)(18.,0)}
\psline[linewidth=1pt,arrows=->, arrowsize=6pt,arrowinset=0.2]{->}(0,0)(18.,0)
\psline[linewidth=1pt](0,0)(0,1)
\rput(-.25,0){\small 1}\rput(-.25,1){\small 2}
\rput(-.85,0.5){$\sigma$}
\rput(1,-0.35){\small 1}\rput(2,-0.35){\small 2}\rput(3,-0.35){\small 3}\rput(4,-0.35){\small 4}\rput(5,-0.35){\small 5}\rput(6,-0.35){\small 6}
\rput(7,-0.35){\small 7}\rput(8,-0.35){\small 8}\rput(9,-0.35){$\dots$}\rput(16,-0.35){\dots}
\rput(17,-0.35){\small$N$}
\rput(18.,-0.35){$j$}
\psline[linewidth=1.5pt,linecolor=blue](0,0)(1,1)(1.5,0.5)(2,1)(2.5,0.5)(3,0)(3.5,0.5)(4,1)(4.5,0.5)(5,0)(5.5,0.5)(6,1)(6.5,0.5)(7,1)(7.5,0.5)(8,0)(8.5,0.5)(9,1)(9.5,0.5)(10,1)(10.5,0.5)(11,0)(11.5,0.5)(12,1)(12.5,0.5)(13,0)(13.5,0.5)(14,1)(14.5,0.5)(15,1)(15.5,0.5)(16,0)(16.5,0.5)(17,1)(17.5,0.5)
\psline[linewidth=1.5pt,linecolor=black](1,1)(1.5,0.5)(2,1)
\psline[linewidth=1.5pt,linecolor=black](6,1)(6.5,0.5)(7,1)
\psline[linewidth=1.5pt,linecolor=black](9,1)(9.5,0.5)(10,1)
\psline[linewidth=1.5pt,linecolor=black](14,1)(14.5,0.5)(15,1)
\psline[linewidth=1.5pt,linecolor=purple](1,1)(2,1)
\psline[linewidth=1.5pt,linecolor=purple](6,1)(7,1)
\psline[linewidth=1.5pt,linecolor=purple](9,1)(10,1)
\psline[linewidth=1.5pt,linecolor=purple](14,1)(15,1)
\psline[linewidth=1.5pt,linecolor=purple](17,1)(18,1)
\psline[linewidth=1.5pt,linecolor=black](17,1)(17.5,0.5)
\psline[linestyle=dotted,dotsep=1.5pt,linewidth=1.5pt,linecolor=black](17.5,0.5)(18,1)
\label{fig:JMpath1}
\end{pspicture}
}
\\
\vspace{-1cm}
\subfloat[JM(3,7) and RSOS$(3,7)_{2\times2}$ paths with $k=2$ and $N=17$ integer steps: $\sigma_j=1,2,3$ corresponds to $1=\bullet$, $2=\gbullet$ and $3=\circ$ in the $T_3^{(n=2)}$ tadpole description of RSOS models.]{
\psset{unit=0.8cm}
\begin{pspicture}(0,-0.8)(18.5,4)
\pspolygon[linewidth=1pt,linecolor=lightestblue,fillstyle=solid,fillcolor=lightestblue](0,0)(18.,0)(18.,2)(0,2)
\multiput(0.5,0)(1,0){18}{\psline[linewidth=.5pt,linestyle=dashed,dash=2pt](0,0)(0,2)}
\multiput(1,0)(1,0){18}{\psline[linestyle=solid](0,0)(0,2)}
\multiput(0,0.5)(0,1){2}{\psline[linewidth=.5pt,linestyle=dashed,dash=2pt](0,0)(18.,0)}
\multiput(0,1)(0,1){2}{\psline[linestyle=solid](0,0)(18.,0)}
\psline[linewidth=1pt,arrows=->, arrowsize=6pt,arrowinset=0.2]{->}(0,0)(18.,0)
\psline[linewidth=1pt](0,0)(0,2)
\rput(-.25,0){1}\rput(-.25,1){2}\rput(-.25,2){3}
\rput(-.85,1){$\sigma$}
\rput(1,-0.35){\small 1}\rput(2,-0.35){\small 2}\rput(3,-0.35){\small 3}\rput(4,-0.35){\small 4}\rput(5,-0.35){\small 5}\rput(6,-0.35){\small 6}\rput(9,-0.35){\dots}
\rput(7,-0.35){\small 7}\rput(8,-0.35){\small 8}\rput(16,-0.35){\dots}
\rput(17,-0.35){\small $N$}
\rput(18.,-0.35){$j$}
\psline[linewidth=1.5pt,linecolor=blue](0,0)(1,1)(1.5,0.5)(3,2)(4,1)(5,2)(6.5,0.5)(7,1)(7.5,0.5)(9,2)(9.5,1.5)(10,2)(10.5,1.5)(11,2)(13,0)(14,1)(14.5,0.5)(16,2)(16.5,1.5)(17,1)(17.5,0.5)
\psline[linewidth=1.5pt,linecolor=black](1,1)(1.5,0.5)(2,1)
\psline[linewidth=1.5pt,linecolor=black](6,1)(6.5,0.5)(7,1)(7.5,0.5)(8,1)
\psline[linewidth=1.5pt,linecolor=black](14,1)(14.5,0.5)(15,1)
\psline[linewidth=1.5pt,linecolor=purple](1,1)(2,1)
\psline[linewidth=1.5pt,linecolor=purple](6,1)(8,1)
\psline[linewidth=1.5pt,linecolor=purple](14,1)(15,1)
\psline[linewidth=1.5pt,linecolor=purple](17,1)(18,1)
\psline[linewidth=1.5pt,linecolor=black](17,1)(17.5,0.5)
\psline[linewidth=1.5pt,linecolor=black](9,2)(9.5,1.5)(10,2)
\psline[linewidth=1.5pt,linecolor=purple](9,2)(10,2)
\psline[linewidth=1.5pt,linecolor=black](10,2)(10.5,1.5)(11,2)
\psline[linewidth=1.5pt,linecolor=purple](10,2)(11,2)
\psline[linestyle=dotted,dotsep=1.5pt,linewidth=1.5pt,linecolor=black](17.5,0.5)(18,1)
\label{fig:JMpath2}
\end{pspicture}
}
\caption{Two examples of the bijection between JM and RSOS paths. The edges of JM paths are black while their equivalent in the RSOS description are purple. The shared path is shown in blue. The bijection between JM and RSOS paths preserves the contour of the path except at local minima at half-integer heights where two consecutive half-integer steps in the JM path are replaced with a single horizontal step in the RSOS path. The last half-integer step of a JM path must be down and we add an additional half-integer up step shown dashed. The last RSOS step is therefore flat in the tadpole representation.}
\label{fig:JMpath}
\end{center}
\end{figure}
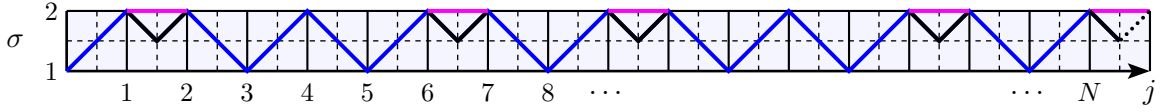
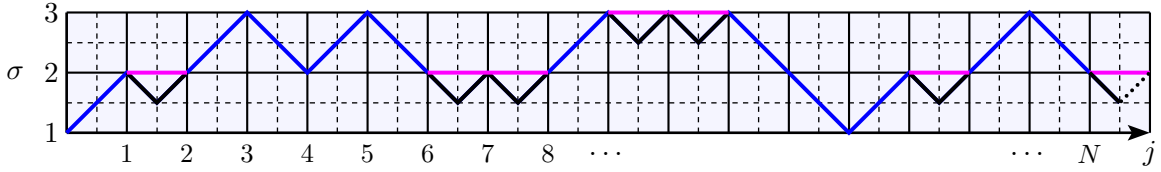

Fix $k\in{\Bbb N}_{\ge 1}$ with $k=m-1$. A JM$(k+1,2k+3)$ path~\cite{JacobM}
\bea
\sigma=\{\sigma_0,\sigma_{\frac{1}{2}},\sigma_1,\ldots,\sigma_N,\sigma_{N+\frac{1}{2}}\},\qquad
\sigma_j\in\{1, \tfrac{3}{2},2,\ldots, k+\tfrac{1}{2},k+1\}
\eea
is then a sequence of heights $\sigma_j\in\half{\Bbb N}$ subject to the constraints
\begin{subequations}
\bea
&\sigma_0, \sigma_{N}\in{\Bbb N},\quad \sigma_{N+\frac{1}{2}}=\sigma_N-\half,\quad \sigma_N=\sigma_{N+1},\quad 
\sigma_{j+\frac{1}{2}}-\sigma_j=\pm \half,\  j=0, \half, 1, \tfrac{3}{2}, \ldots,N&\\
&(j,\sigma_j)\in{\Bbb N}^2\ \ \mbox{at all local peaks}&
\eea
\end{subequations}
Examples are shown in Figure~\ref{fig:JMpath} for $k=1,2$. The JM paths are thus defined on lattices with half-integer spacing and half-integer heights, while in the RSOS description, only integer values of steps and heights are considered. Therefore, to map the energy of JM paths into the local energy functions of the RSOS paths with integer heights, we need to sum out (decimate) the half odd integer heights. 

The energy of a JM path is given by 
\bea
E(\sigma)=\sum_{j=\frac{1}{2},\,j\in\frac{1}{2}{\Bbb N}}^{N}j w(j),\qquad w(j)=\half \big| \sigma_{j+\frac{1}{2}}-\sigma_{j-\frac{1}{2}} \big|
\label{eq:JMenergy}
\eea
Using $w(N+\half)=0$, this is gauge equivalent to the energy of the corresponding RSOS path with local energies (\ref{eq:localE})
\begin{align}
E(\sigma)&=\sum_{j=\frac{1}{2},\,j\in\frac{1}{2}{\Bbb N}}^{N+\frac{1}{2}} jw(j)
=\tfrac{1}{4}w(\tfrac{1}{2})+
\tfrac{1}{2}\sum_{j=1}^{N}\big[(j\!-\!\half)w(j\!-\!\half)+{\tiny 2}jw(j)+(j+\half)w(j\!+\!\half)\big]\nonumber\\[-2pt]
&=\tfrac{1}{4}w(\tfrac{1}{2})+
\tfrac{1}{2}\sum_{j=1}^{N}j\big[w(j\!-\!\half)+{\tiny 2}w(j)+w(j\!+\!\half)\big]+\tfrac{1}{4}\sum_{j=1}^{N}\big[w(j\!+\!\half)-w(j\!-\!\half)\big]\\[-2pt]
&=
\tfrac{1}{2}\sum_{j=1}^{N}j\big[w(j\!-\!\half)+{\tiny 2}w(j)+w(j\!+\!\half)\big]
=\sum_{j=1}^{N}j\big[\tilde H(\sigma_{j-1},\sigma_{j},\sigma_{j+1})\!+\!G_{j-1} \!-\! 2 G_{j} \!+\! G_{j+1}]\!-\!G_0\!+\!G_N\nonumber
\end{align}
Here $G_{j}=g(\sigma_{j})=\tfrac{1}{4}\sigma_j$ and $G_{N+1}=G_N$ is a suitable gauge such that
\begin{align}
\tilde H(\sigma_{j-1},\sigma_{j},\sigma_{j+1})=\half \big[w(j-\half)+2w(j)+w(j+\half)\big]-G_{j-1}+ 2 G_{j}  - G_{j+1}
\label{eq:GaugeEq}
\end{align}
and the constant shift of energy by $G_0-G_N$ is irrelevant. Explicitly,
%
%
\begin{subequations}
\begin{align}
\tilde H(a\pm1,a,a\mp1) = 1-g(a-1)+2g(a)-g(a+1)&=1  \label{eq:localE1oddg}\\[-0pt]
\tilde H( a+1, a,a) = \tilde H(a,a,a+1 )=\tfrac{3}{4}+g(a)-g(a+1)&=\half  \label{eq:localE2oddg}\\[-0pt]
\tilde H( a-1, a,a) =\tilde  H(a,a,a-1 )=\tfrac{1}{4}+g(a)-g(a-1)&=\half  \label{eq:localE3oddg}\\[-0pt]
\tilde H(a-1,a,a-1 ) =\half+2g(a)-2g(a-1)&=1 \label{eq:localE4oddg}\\[-0pt]
\tilde H(a+1,a,a+1 ) =\half+2g(a)-2g(a+1)&=0 \label{eq:localE5oddg}\\[-0pt]
\tilde H(a,a,a ) &=0 \label{eq:localE6oddg}
\end{align}
\label{eq:localEoddg}
\end{subequations}
\goodbreak
The intermediate half-integer heights $a\pm \half$ needed in (\ref{eq:GaugeEq}) are uniquely determined by the adjacent integer heights. 
The local energies $\tilde H(\tilde a,\tilde b,\tilde c)$ are related to the local energies (\ref{eq:localE}) by
\bea
\tilde H(\tilde a,\tilde b,\tilde c)=\half H(a,b,c),\qquad a=2\tilde a-1=1,3,5,\ldots,2k+1
\eea
The factor of $\half$ arises because the fused weights leading to (\ref{eq:GaugeEq}) are calculated with integer fundamental steps rather than the half-integer fundamental steps of the JM paths, that is, $a-b=0,\pm2$ compared with $\tilde a-\tilde b=0,\pm 1$.
%

\section{Nonunitary Minimal Models ${\cal M} (m,m')$}

\subsection{Conformal data and characters}

In the continuum scaling limit, the RSOS lattice models are described by the rational minimal models. 
The central charges of the $n=1$ minimal models ${\cal M}(m,m')$, with $m<m'$ and $m,m'$ coprime, are given by (\ref{central}). The conformal weights and associated Virasoro characters are
\begin{align}
\Delta_{r,s}^{m,m'}&={(rm'-sm)^2-(m-m')^2\over 4mm'},\quad 1\le r\le m-1,\ \ 1\le s\le m'-1\\
\mbox{ch}^{m,m'}_{r,s}(q)&={q^{-c/24+\Delta_{r,s}^{m,m'}}\over
(q)_\infty}\!\!\! \sum_{k=-\infty}^\infty \big[q^{k(k m
m'+m'r-ms)}-q^{(km+r)(km'+s)}\big] \label{vira}
\end{align}
where the $q$-factorials are defined by
\bea
(q)_n=\prod_{k=1}^n (1-q^k),\qquad (q)_\infty=\prod_{k=1}^\infty (1-q^k)\label{qfactorials}
\eea
The minimal models are unitary if $m=m'-1$ and nonunitary if $m<m'-1$. In this section we consider the finitized characters of the $2\times 2$ fused minimal models with $m<\half m'$.

\subsection{Finitized bosonic characters}

From extensive numerics, we find empirically that the normalized finitized characters associated with the $\mbox{RSOS}(m,m')_{2\times 2}$ lattice models admit a bosonic form
\bea
\widehat{\ch}_{r,s}^{m,m'\!;(N)}(q)={X}_{abc}^{m,m'\!;(N)}(q)=\!\!\sum_{k=-\infty}^\infty \!\!\Big[q^{k(k m m'+m' r-ms)}T_{km'+\frac{b-a}{2}}^{(N)}(q)-q^{(km+r)(km'+s)}T_{km'+\frac{b+a}{2}}^{(N)}(q)\Big]
\label{finChars}
\eea
where
\bea
a=s,\qquad b=c=\rho_\mu(r),\qquad \mu=\mbox{$s$ mod 2}
\label{rsabc}
\eea
The $q$-trinomial coefficients~\cite{SeatonScott} are 
\bea
T_k^{(N)}(q)=T_{-k}^{(N)}(q)=\left[N\atop k\right]_2^{(0)}=\sum_{j=0}^N q^{j(j+k)} \qTrinomial N j{j\!+\!k}
\label{qtrin}
\eea
where the $q$-multinomial coefficients are defined in terms of $q$-factorials by
\be
\qTrinomial n\ell m=\begin{cases}
\frac{(q)_n}{(q)_\ell(q)_m(q)_{n\!-\!\ell\!-\!m}}& \ell,m,n\!-\!\ell\!-\!m\in \mathbb{Z}_{\ge 0}\\
0,&\mbox{otherwise}\end{cases}
\ee
In the limit $q\to1$, the $q$-multinomials reduce to multinomial coefficients
\be
\lim_{q\to1}\qTrinomial n\ell m=\trinomial n\ell m \label{eq:qto1}
\ee
which ensures the correct counting of states.

To arrive at the bosonic forms (\ref{finChars}), we used the fact that the $q$-trinomial coefficient $T_{k}^{(N)}(q)$ is a $q$-deformed counting of weighted $N$ step $2\times 2$ fused paths on the $A_{\infty}$ Dynkin diagram. This counting only respects the constraints $\sigma_{i+1}-\sigma_i = 0, \pm 2$, with $2k$ height difference between the initial and final state $\sigma_N - \sigma_0 = 2k$. This ensures correct counting, as $j+k$ is the number of up steps, $j$ is the number of down steps, and $N-2j-k$ is the number of flat steps. Thus, given that the local energy functions are periodically extended, $T_{k}^{(N)}(q)$ gives the correct one dimensional sums (without restricting to the fused $A_{m'}$). The required bosonic form of the one dimensional sum on $A_{m'}$ is then obtained by summing and subtracting the respective generalized paths. These bosonic forms (\ref{finChars}) were checked against the ground state one-dimensional sums with $b=c$ for all values of $r,s$ using Mathematica~\cite{Wolfram} out to size $N=12$ for $(m, m') =(2, 5), (2, 7), (3, 7), (3, 8)$, size $N = 11$ for $(m, m') = (2, 9), (4, 9)$, size $N = 10$ for
$(m, m') = (3, 10)$ and size $N = 9$ for $(m, m') = (2, 11), (3, 11), (4, 11), (5, 11)$. It would be of interest to obtain the bosonic expressions for $X_{abc}^{(N)}(q)$ more generally for $c=b,b\pm 2$ and to prove that these expressions satisfy the CTM recursions (\ref{recursion}).

Using (\ref{qtrin}), it is straightforward to show that the finitized characters also satisfy the Kac symmetry
\bea
\hat{X}_{abc}^{m,m'\!;(N)}(q)=\hat{X}_{m'-a,m'-b,m'-c}^{m,m'\!;(N)}(q)
\eea
Since the shaded band structure is symmetric, the Kac labels on the right are $(r,s)=(m-r,m'-s)$.

\subsection{$N\to\infty$ limit}

The finitized characters also agree with the full characters in the thermodynamic limit $N\to\infty$. 
For fixed $k\in{\Bbb Z}$, the modified $q$-trinomials satisfy
\begin{equation}
\lim_{N\to\infty}T_k^{(N)}(q) =\lim_{N\to\infty}
\sum_{j=0}^\infty q^{j(j+k)}\qTrinomial{N}{j}{j+k}
=\frac{1}{(q)_\infty}
\label{eq:A.limgpoly}
\end{equation}
To establish this we take the limit
inside the sum and use the elementary result
\begin{equation}
\lim_{N\to\infty} \qTrinomial{N}{j}{j+k}
=\lim_{N\to\infty}\;
\frac{(q)_N}
{(q)_j \, (q)_{j+k} \, (q)_{N-2j-k}}
=\frac{1}{(q)_j \, (q)_{j+k}}
\end{equation}
to obtain
\begin{equation}
\lim_{N\to\infty} \sum_{j=0}^\infty 
q^{j(j+k)}\qTrinomial{N}{j}{j+k}
=\sum_{j=0}^\infty 
\frac{q^{j(j+k)}}
{(q)_j\,  (q)_{j+k}}
=\frac{1}{(q)_\infty}
\end{equation}
The last equality follows by setting $z=q^{k+1}$ in the $q$-analogue of Kummer's theorem
(see (2.2.8) of \cite{Andrews})
\begin{equation}
\sum_{j=0}^\infty 
\frac{q^{j(j-1)}z^j}
{(1-q)\ldots (1-q^j)(1-z)(1-zq)\ldots (1-zq^{j-1})}
=\prod_{j=0}^\infty \frac{1}{1-zq^j}
\end{equation}
It follows that the limit of the finitized characters (\ref{finChars}) precisely reproduces the $n=1$ Virasoro characters
\bea
\lim_{N\to\infty} q^{-\frac{c}{24}+\Delta_{r,s}} \widehat{\ch}_{r,s}^{m,m'\!;(N)}(q)=\ch_{r,s}^{m,m'}(q)
\eea

\def\sfloor#1{\lfloor #1\rfloor}
\subsection{Logarithmic limit}
\label{SectLogLimit}

Following \cite{RasLogLimit} and \cite{PR2013}, the Kac characters of the logarithmic minimal models ${\cal LM}(p,p')$~\cite{PRZ} and their $n\times n$ fusion hierarchies~\cite{PRT2014} are given by taking the {\em logarithmic limit}. Symbolically, 
\bea
\lim_{m,m'\to\infty, \  {m'\over m}\to {p'\over p}+} {\cal M}(m,m')_{2\times 2}={\cal LM}(p,p'),\qquad 1\le p<\half p',\quad \mbox{$p,p'$ coprime}\label{logLimit}
\eea
The (one-sided) limit is taken through coprime pairs $(m,m')$ with ${m'\over m}>{p'\over p}$ and ${p'\over p}\ge 2$. The one-sided limit is needed to ensure the sequences of minimal model ground states converge to the correct logarithmic minimal model ground states. Formally, the logarithmic limit is taken in the continuum scaling limit after the thermodynamic limit. 
The equality indicates the identification of the spectra of the chiral CFTs. In principle, the Jordan cells appearing in the reducible yet indecomposable representations of the logarithmic minimal models should emerge in this limit but there are subtleties~\cite{RasLogLimit}.

Since finitized characters give the spectrum generating functions for finite truncated sets of conformal energies, the logarithmic limit can be applied directly to finitized characters. Assuming $m<\half m'$, $0<|q|<1$ and taking the logarithmic limit of the finitized characters ({\ref{finChars}) gives the finitized characters of ${\cal LM}(p,p')_{2\times 2}$ for $p<\half p'$
\begin{align}
\widehat{\chi}_{r,s}^{p,p'\!,(N)}(q)=\lim_{m,m'\to\infty, \  {m'\over m}\to {p'\over p}+}\widehat{\ch}_{r,s}^{m,m';(N)}(q) &= T_{\frac{b-a}{2}}^{(N)}(q)-q^{rs}\,T_{\frac{b+a}{2}}^{(N)}(q)
\label{blogchiT}
\end{align}
where $r,s,a,b,c$ are related by (\ref{rsabc}). Taking the thermodynamic limit, using (\ref{eq:A.limgpoly}), gives
\bea
\lim_{N\to\infty}q^{-\frac{c}{24}+\Delta_{r,s}} \widehat{\chi}_{r,s}^{p,p'\!,(N)}(q)=q^{-\frac{c}{24}+\Delta_{r,s}} \frac{1-q^{rs}}{(q)_\infty}=\chi_{r,s}^{p,p'}(q)
\eea
which agrees with the Kac characters of the logarithmic minimal models ${\cal LM}(p,p')$.

%
%

\section{Conclusion}

Using the one-dimensional sums arising from Baxter's off-critical Corner Transfer Matrix (CTM) formalism, we have argued that, for $m'>2m$, the RSOS$(m,m')_{1\times 1}$ and RSOS$(m,m')_{2\times 2}$ lattice models lie in the same universality class described by the nonunitary minimal CFT ${\cal M}(m,m')$. This result holds even though, in general, RSOS$(m,m')_{1\times 1}$ and RSOS$(m,m')_{2\times 2}$ are distinct lattice models. More specifically, we have conjectured the explicit bosonic form of the finitized characters and, for modest system sizes $N$, checked that these agree with the ground state one-dimensional sums. 
In the case $m'=2m+1$, we have further shown that the ground state one-dimensional sums of RSOS$(m,m')_{2\times 2}$ agree with those of Jacob and Mathieu~\cite{JacobM} based on half-integer RSOS paths. This connection with a Yang-Baxter integrable lattice model nicely explains the remarkable observed properties of these half-integer one-dimensional sums. The more general methods used here allow these observations to be extended to all RSOS$(m,m')_{2\times 2}$ lattice models.

%
\section*{Acknowledgments}
\vskip.1cm
\noindent
GZF thanks Jan de Gier and the School of Mathematics and Statistics for hospitality. Part of this work was carried out during visits of PAP to the APCTP as an ICTP Visiting Scholar. AVO is supported by a Melbourne University International Postgraduate Research Scholarship. We thank Elena Tartaglia for help in the early stages of this project.


\appendix

\section{Elliptic Functions}
\label{AppA}

In this Appendix we list definitions and relevant properties of the elliptic functions used in this paper. 

For a complex number $u$, the standard elliptic theta function $\vartheta_1(u,t)$~\cite{GR} is commonly expressed in terms the nome $t$, $|t|<1$, as the following infinite product:
\begin{align}
\vartheta_1(u,t)&=2t^{1/4}\sin u
\prod_{n=1}^\infty (1-2t^{2n}\cos 2u+t^{4n})(1-t^{2n})
\end{align}
Its conjugate modulus transformation, which relates theta functions of nome $t=e^{-\varepsilon}$ to those of nome $t'=e^{-\frac{\pi^2}{\varepsilon}}$, can be written as
\begin{align}
\vartheta_1(u,e^{-\varepsilon})&=\sqrt\frac{\pi}{\varepsilon}\, e^{-(u-\pi/2)^2/\varepsilon}E(e^{-2 \pi u/\varepsilon}, e^{-2\pi^2/\varepsilon} )
\end{align}
The elliptic $\vartheta_1(u)=\vartheta_1(u,t)$ function satisfies the fundamental identity
\begin{align}
&\vartheta_1(u+x)\vartheta_1(u-x)\vartheta_1(v+y)\vartheta_1(v-y) - \vartheta_1(u+y)\vartheta_1(u-y)\vartheta_1(v+x)\vartheta_1(v-x) \nn
&\qquad\qquad\qquad= \vartheta_1(x-y)\vartheta_1(x+y)\vartheta_1(u+v)\vartheta_1(u-v)
\end{align}

\section{Yang-Baxter Equation of Critical Fused RSOS$(m,m')_{2\times 2}$ Models}
\label{AppB}

In this Appendix we discuss the algebraic structure of the solution to the Yang-Baxter equations for the critical $2\times 2$ RSOS models. Following \cite{PRT2014}, the face transfer operators can be written as
\bea
\mathbb{X}_j(u)=\frac{s(\lambda-u)s(2\lambda-u)}{s(2\lambda)}I+s(u)s(\lambda-u)X_j+\frac{s(u)s(u+\lambda)}{s(2\lambda)} E_j,\qquad s(u)=\frac{\sin u}{\sin\lambda}
\label{eq:fusedFace}
\eea
where the identity $I$ and the generalized monoids $E_j$ and $X_j$   
\bea
\psset{unit=.6cm}
I\;=\ 
\begin{pspicture}[shift=-.9](0,0)(2,2)
\psellipse[linewidth=1pt,linecolor=black](0,1)(.1,.2)
\psellipse[linewidth=1pt,linecolor=black](1,0)(.2,.1)
\psellipse[linewidth=1pt,linecolor=black](2,1)(.1,.2)
\psellipse[linewidth=1pt,linecolor=black](1,2)(.2,.1)
\psarc[linewidth=1pt,linecolor=black](2,0){.8}{90}{180}
\psarc[linewidth=1pt,linecolor=black](2,0){1.2}{90}{180}
\psarc[linewidth=1pt,linecolor=black](0,2){.8}{270}{0}
\psarc[linewidth=1pt,linecolor=black](0,2){1.2}{270}{0}
\end{pspicture}\ =\,
\begin{pspicture}[shift=-1.27](-.4,-.4)(2.4,2.4)
\multiput(0,0)(0,1){2}{\psline[linewidth=1.5pt,linecolor=blue](-.2,.5)(2.2,.5)}
\psellipse[linewidth=1.5pt,linecolor=blue](-.2,1)(.1,.5)
\psellipse[linewidth=1.5pt,linecolor=blue](2.2,1)(.1,.5)
\multiput(0,0)(1,0){2}{\psline[linewidth=1.5pt,linecolor=blue](.5,-.2)(.5,2.2)}
\psellipse[linewidth=1.5pt,linecolor=blue](1,-.2)(.5,.1)
\psellipse[linewidth=1.5pt,linecolor=blue](1,2.2)(.5,.1)
\facegrid{(0,0)}{(2,2)}
\rput[bl](0,0){\loopa}
\rput[bl](1,0){\loopa}
\rput[bl](0,1){\loopa}
\rput[bl](1,1){\loopa}
\end{pspicture}\qquad
E_j\;=\ 
\begin{pspicture}[shift=-.9](0,0)(2,2)
\psellipse[linewidth=1pt,linecolor=black](0,1)(.1,.2)
\psellipse[linewidth=1pt,linecolor=black](1,0)(.2,.1)
\psellipse[linewidth=1pt,linecolor=black](2,1)(.1,.2)
\psellipse[linewidth=1pt,linecolor=black](1,2)(.2,.1)
\psarc[linewidth=1pt,linecolor=black](0,0){.8}{0}{90}
\psarc[linewidth=1pt,linecolor=black](0,0){1.2}{0}{90}
\psarc[linewidth=1pt,linecolor=black](2,2){.8}{180}{270}
\psarc[linewidth=1pt,linecolor=black](2,2){1.2}{180}{270}
\end{pspicture}\ 
=\,\begin{pspicture}[shift=-1.27](-.4,-.4)(2.4,2.4)
\multiput(0,0)(0,1){2}{\psline[linewidth=1.5pt,linecolor=blue](-.2,.5)(2.2,.5)}
\psellipse[linewidth=1.5pt,linecolor=blue](-.2,1)(.1,.5)
\psellipse[linewidth=1.5pt,linecolor=blue](2.2,1)(.1,.5)
\multiput(0,0)(1,0){2}{\psline[linewidth=1.5pt,linecolor=blue](.5,-.2)(.5,2.2)}
\psellipse[linewidth=1.5pt,linecolor=blue](1,-.2)(.5,.1)
\psellipse[linewidth=1.5pt,linecolor=blue](1,2.2)(.5,.1)
\facegrid{(0,0)}{(2,2)}
\rput[bl](0,0){\loopb}
\rput[bl](1,0){\loopb}
\rput[bl](0,1){\loopb}
\rput[bl](1,1){\loopb}
\end{pspicture}\qquad
X_j\;=\ 
\begin{pspicture}[shift=-.9](0,0)(2,2)
\psellipse[linewidth=1pt,linecolor=black](0,1)(.1,.2)
\psellipse[linewidth=1pt,linecolor=black](1,0)(.2,.1)
\psellipse[linewidth=1pt,linecolor=black](2,1)(.1,.2)
\psellipse[linewidth=1pt,linecolor=black](1,2)(.2,.1)
\psarc[linewidth=1pt,linecolor=black](0,0){.8}{0}{90}
\psarc[linewidth=1pt,linecolor=black](2,0){.8}{90}{180}
\psarc[linewidth=1pt,linecolor=black](0,2){.8}{270}{0}
\psarc[linewidth=1pt,linecolor=black](2,2){.8}{180}{270}
\end{pspicture}\ 
=\,\begin{pspicture}[shift=-1.27](-.4,-.4)(2.4,2.4)
\multiput(0,0)(0,1){2}{\psline[linewidth=1.5pt,linecolor=blue](-.2,.5)(2.2,.5)}
\psellipse[linewidth=1.5pt,linecolor=blue](-.2,1)(.1,.5)
\psellipse[linewidth=1.5pt,linecolor=blue](2.2,1)(.1,.5)
\multiput(0,0)(1,0){2}{\psline[linewidth=1.5pt,linecolor=blue](.5,-.2)(.5,2.2)}
\psellipse[linewidth=1.5pt,linecolor=blue](1,-.2)(.5,.1)
\psellipse[linewidth=1.5pt,linecolor=blue](1,2.2)(.5,.1)
\facegrid{(0,0)}{(2,2)}
\psarc[linewidth=1.5pt,linecolor=blue](0,0){.5}{0}{90}
\psarc[linewidth=1.5pt,linecolor=blue](2,0){.5}{90}{180}
\psarc[linewidth=1.5pt,linecolor=blue](2,2){.5}{180}{270}
\psarc[linewidth=1.5pt,linecolor=blue](0,2){.5}{270}{360}
\end{pspicture}
\eea
generate the $2\times 2$ fused Temperley-Lieb (TL) algebra. This algebra is a one-parameter specialization of the two-parameter BMW algebra. The properties of the generator $X_j$ were studied by a number of authors~\cite{FendleyRead2002,Fendley2006,FendleyJacobsen2008,FendleyKruskal2010}. Here it is useful to replace the generator $X_j$ with the generator
\bea
\psset{unit=.6cm}
\Xi_j\;=\ 
\begin{pspicture}[shift=-.9](0,0)(2,2)
\psellipse[linewidth=1pt,linecolor=black](0,1)(.1,.2)
\psellipse[linewidth=1pt,linecolor=black](1,0)(.2,.1)
\psellipse[linewidth=1pt,linecolor=black](2,1)(.1,.2)
\psellipse[linewidth=1pt,linecolor=black](1,2)(.2,.1)
\psellipse[linewidth=1pt,linecolor=black,rot=135](1,1)(.64,.13)
\psarc[linewidth=1pt,linecolor=black](0,0){.8}{0}{90}
\psarc[linewidth=1pt,linecolor=black](2,0){.8}{90}{180}
\psarc[linewidth=1pt,linecolor=black](0,2){.8}{270}{0}
\psarc[linewidth=1pt,linecolor=black](2,2){.8}{180}{270}
\end{pspicture}\ 
=\,\begin{pspicture}[shift=-1.27](-.4,-.4)(2.4,2.4)
\multiput(0,0)(0,1){2}{\psline[linewidth=1.5pt,linecolor=blue](-.2,.5)(2.2,.5)}
\psellipse[linewidth=1.5pt,linecolor=blue](-.2,1)(.1,.5)
\psellipse[linewidth=1.5pt,linecolor=blue](2.2,1)(.1,.5)
\multiput(0,0)(1,0){2}{\psline[linewidth=1.5pt,linecolor=blue](.5,-.2)(.5,2.2)}
\psellipse[linewidth=1.5pt,linecolor=blue](1,-.2)(.5,.1)
\psellipse[linewidth=1.5pt,linecolor=blue](1,2.2)(.5,.1)
\facegrid{(0,0)}{(2,2)}
\psellipse[linewidth=1.5pt,linecolor=blue,rot=135](1,1)(.92,.15)
\psarc[linewidth=1.5pt,linecolor=blue](0,0){.5}{0}{90}
\psarc[linewidth=1.5pt,linecolor=blue](2,0){.5}{90}{180}
\psarc[linewidth=1.5pt,linecolor=blue](2,2){.5}{180}{270}
\psarc[linewidth=1.5pt,linecolor=blue](0,2){.5}{270}{360}
\end{pspicture}\;=\;X_j-\beta^{-1} E_j,\qquad \Xi_j E_j=X_jE_j-\beta^{-1}E_j^2=0
\eea
where
\bea
\psset{unit=.75cm}
\beta=\beta_1=\begin{pspicture}[shift=-.4](0,0)(1,1)
\pscircle(.5,.5){.5}
\end{pspicture}
=x+x^{-1}=2\cos\lambda,\qquad x=e^{i\lambda},\qquad \beta_{n-1}=[x]_n=\frac{x^n-x^{-n}}{x-x^{-1}}=\frac{\sin n\lambda}{\sin\lambda}
\label{fugacities}
\eea
The generator $\Xi_j$ is obtained by cabling the two central strings of $X_j$ by decomposing the identity into orthogonal projectors
\bea
I\ = 
\psset{unit=.4cm}
\begin{pspicture}[shift=-.85](0,0)(2,2)
\psline[linewidth=1pt](.4,0)(.4,2)
\psline[linewidth=1pt](1.6,0)(1.6,2)
\end{pspicture}
=
\begin{pspicture}[shift=-.85](0,0)(2,2)
\psline[linewidth=1pt](.4,0)(.4,2)
\psline[linewidth=1pt](1.6,0)(1.6,2)
\psellipse[linewidth=1pt](1,1)(.6,.2)
\end{pspicture}
+{\beta}^{-1}
\begin{pspicture}[shift=-.85](0,0)(2,2)
\psbezier[linewidth=1pt,showpoints=false](.4,0)(.5,1)(1.5,1)(1.6,0)
\psbezier[linewidth=1pt,showpoints=false](.4,2)(.5,1)(1.5,1)(1.6,2)
\end{pspicture}\;=\;p_j+\beta^{-1} e_j
\eea
The face transfer operator can now be written as
\bea
\mathbb{X}_j(u)=\frac{s(\lambda-u)s(2\lambda-u)}{s(2\lambda)}I+s(u)s(\lambda-u)\,\Xi_j+\frac{s(2u)}{s(2\lambda)} E_j
\label{eq:fusedFace}
\eea

\def\TLia{
\raisebox{-2cm}{\begin{pspicture}(0,.2)(2,4)
\psline[linewidth=1.5pt,linecolor=blue](0.65,0.3)(0.65,1.5)
\psline[linewidth=1.5pt,linecolor=blue](1.35,0.3)(1.35,1.5)
\psline[linewidth=1.5pt,linecolor=blue](0.65,3.2)(0.65,4.1)
\psline[linewidth=1.5pt,linecolor=blue](1.35,3.2)(1.35,4.1)
\pscircle[linewidth=1.5pt,linecolor=blue](1,2.225){0.53}
\rput(0.8,1.4){\Esquare}
\rput(0.8,2.825){\Esquare}
\rput(0.65,0.1){\scriptsize$j$}
\rput(1.35,0.1){\scriptsize$j\!+\!1$}
\psline[linestyle=dashed](0.3,1.5)(0.3,2.95)
\psline[linestyle=dashed](1.7,1.5)(1.7,2.95)
\pscircle[fillstyle=solid,fillcolor=black](1,2.225){0.06}
\rput(1,0.7){\scriptsize$b$}
\rput(0.2,1.5){\scriptsize$a$}
\rput(1.8,1.5){\scriptsize$a$}
\rput(0.8,2.2){\scriptsize$d$}
\rput(0.2,2.9){\scriptsize$a$}
\rput(1.8,2.9){\scriptsize$a$}
\rput(1,3.75){\scriptsize$c$}
\end{pspicture}}
}

\def\TLia{
\raisebox{-2cm}{\begin{pspicture}(0,.2)(2,4)
\psline[linewidth=1.5pt,linecolor=blue](0.65,0.3)(0.65,1.5)
\psline[linewidth=1.5pt,linecolor=blue](1.35,0.3)(1.35,1.5)
\psline[linewidth=1.5pt,linecolor=blue](0.65,3.2)(0.65,4.1)
\psline[linewidth=1.5pt,linecolor=blue](1.35,3.2)(1.35,4.1)
\pscircle[linewidth=1.5pt,linecolor=blue](1,2.225){0.53}
\rput(0.8,1.4){\Esquare}
\rput(0.8,2.825){\Esquare}
\rput(0.65,0.1){\scriptsize$j$}
\rput(1.35,0.1){\scriptsize$j\!+\!1$}
\psline[linestyle=dashed](0.3,1.5)(0.3,2.95)
\psline[linestyle=dashed](1.7,1.5)(1.7,2.95)
\pscircle[fillstyle=solid,fillcolor=black](1,2.225){0.06}
\rput(1,0.7){\scriptsize$b$}
\rput(0.2,1.5){\scriptsize$a$}
\rput(1.8,1.5){\scriptsize$a$}
\rput(0.8,2.2){\scriptsize$d$}
\rput(0.2,2.9){\scriptsize$a$}
\rput(1.8,2.9){\scriptsize$a$}
\rput(1,3.75){\scriptsize$c$}
\end{pspicture}}
}

\def\XXia{
\raisebox{-2cm}{\begin{pspicture}(0,.2)(2,4)
\psline[linewidth=1.5pt,linecolor=blue](0.65,0.3)(0.65,4.1)
\psline[linewidth=1.5pt,linecolor=blue](1.35,0.3)(1.35,4.1)
\rput(0.8,1.4){\Xsquare}
\rput(0.8,2.825){\Xsquare}
\rput(0.65,0.1){\scriptsize$j$}
\rput(1.35,0.1){\scriptsize$j\!+\!1$}
\psline[linestyle=dashed](0.3,1.5)(0.3,2.95)
\psline[linestyle=dashed](1.7,1.5)(1.7,2.95)
\pscircle[fillstyle=solid,fillcolor=black](1,2.225){0.06}
\rput(1,0.7){\scriptsize$b$}
\rput(0.2,1.5){\scriptsize$a$}
\rput[L](1.85,1.55){\scriptsize$a'$}
\rput(0.8,2.2){\scriptsize$d$}
\rput(0.2,2.9){\scriptsize$a$}
\rput[L](1.85,2.95){\scriptsize$a'$}
\rput(1,3.75){\scriptsize$c$}
\end{pspicture}}
}

\def\TLiiA{
\raisebox{-1.9cm}{\begin{pspicture}(0,.2)(2.8,4.2)
\psline[linecolor=blue,linewidth=1.5pt](0.65,.4)(.65,1.2)
\psline[linecolor=blue,linewidth=1.5pt](1.35,.4)(1.35,1.2)
\psline[linecolor=blue,linewidth=1.5pt](0.65,3.2)(.65,4)
\psline[linecolor=blue,linewidth=1.5pt](1.35,3.2)(1.35,4)
\psline[linecolor=blue,linewidth=1.5pt](2.075,.4)(2.075,4)
\rput(0.8,1.4){\Esquare}
\rput(0.8,2.814){\Esquare}
\rput(1.51,2.11){\Esquare}
\rput(.65,.2){\scriptsize{$j$}}
\rput(1.35,.2){\scriptsize{$j\!+\!1$}}
\rput(2.05,.2){\scriptsize{$j\!+\!2$}}
\psarc[linewidth=1.5pt,linecolor=blue](1,2.225){0.505}{135}{225}
\rput(1,0.7){\scriptsize$b$}
\rput(1.8,1.4){\scriptsize$a$}
\rput(.2,1.5){\scriptsize$a$}
\rput(.2,2.95){\scriptsize$a$}
\rput(1.8,3){\scriptsize$a$}
\rput(.8,2.2){\scriptsize$d$}
\rput(2.55,2.2){\scriptsize$d$}
\rput(1,3.8){\scriptsize$c$}
\psline[linestyle=dashed](0.3,1.5)(0.3,2.95)
\end{pspicture}}
}

\def\XiiA{
\raisebox{-1.9cm}{\begin{pspicture}(0,.2)(2.8,4.2)
\psline[linecolor=blue,linewidth=1.5pt](0.65,.4)(.65,1.2)
\psline[linecolor=blue,linewidth=1.5pt](1.35,.4)(1.35,1.2)
\psline[linecolor=blue,linewidth=1.5pt](0.65,3.2)(.65,4)
\psline[linecolor=blue,linewidth=1.5pt](1.35,3.2)(1.35,4)
\psline[linecolor=blue,linewidth=1.5pt](2.075,.4)(2.075,4)
\rput(0.8,1.4){\Esquare}
\rput(0.8,2.814){\Esquare}
\rput(1.51,2.11){\Xsquare}
\rput(.65,.2){\scriptsize{$j$}}
\rput(1.35,.2){\scriptsize{$j\!+\!1$}}
\rput(2.05,.2){\scriptsize{$j\!+\!2$}}
\psarc[linewidth=1.5pt,linecolor=blue](1,2.225){0.505}{135}{225}
\rput(1,0.7){\scriptsize$b$}
\rput(1.8,1.4){\scriptsize$a$}
\rput(.2,1.5){\scriptsize$a$}
\rput(.2,2.95){\scriptsize$a$}
\rput(1.8,3){\scriptsize$a$}
\rput(.8,2.2){\scriptsize$e$}
\rput(2.55,2.2){\scriptsize$d$}
\rput(1,3.8){\scriptsize$c$}
\psline[linestyle=dashed](0.3,1.5)(0.3,2.95)
\pscircle[fillstyle=solid,fillcolor=black](1,2.225){0.06}
\end{pspicture}}
}

\def\TLiiB{
\begin{pspicture}(2.2,2)
\rput[l](0.1,0){\Ejr baca}
\psline[linecolor=blue,linewidth=1.5pt](2.05,-1)(2.05,1.2)
\rput(2.05,-1.15){\scriptsize$j\!+\!2$}
\end{pspicture}
}

\def\TLiia{
\begin{pspicture}(5.4,4.2)
\psline[linecolor=blue,linewidth=1.5pt](1.45,.4)(1.45,1.2)
\psline[linecolor=blue,linewidth=1.5pt](2.15,.4)(2.15,1.2)
\psline[linecolor=blue,linewidth=1.5pt](1.45,3.2)(1.45,4)
\psline[linecolor=blue,linewidth=1.5pt](2.15,3.2)(2.15,4)
\psline[linecolor=blue,linewidth=1.5pt](.75,.4)(.75,4)
\rput(.8,0){
\rput(0.8,1.4){\Esquare}
\rput(0.8,2.814){\Esquare}
}
\rput(.9,2.11){\Esquare}
\rput(.75,.2){\scriptsize{$j-1$}}
\rput(1.45,.2){\scriptsize{$j$}}
\rput(2.15,.2){\scriptsize{$j\!+\!1$}}
\psarc[linewidth=1.5pt,linecolor=blue](1.825,2.225){0.505}{315}{45}
\rput(1.8,0.65){\scriptsize$a$}
\rput(1.1,1.35){\scriptsize$b$}
\rput(2.6,1.5){\scriptsize$b$}
\rput(2.6,2.95){\scriptsize$b$}
\rput(1.1,3.1){\scriptsize$b$}
\rput(.3,2.2){\scriptsize$c$}
\rput(2,2.2){\scriptsize$c$}
\rput(1.8,3.8){\scriptsize$d$}
\rput(3.1,2.205){=}
\rput(4.2,2.1){\Ejr abdb}
\psline[linestyle=dashed](2.5,1.5)(2.5,2.95)
\psline[linecolor=blue,linewidth=1.5pt](5.35,1.1)(5.35,3.3)
\rput(5.45,.95){\scriptsize$j\!+\!2$}
\end{pspicture}
}

\def\Ejr#1#2#3#4{\begin{pspicture}[shift=-1.17](-.2,-.8)(1.2,1.6)
\rput(0.707,-.3){\scriptsize $#1$}
\rput(1.514,.507){\scriptsize $#2$}
\rput(0.707,1.364){\scriptsize $#3$}
\rput(-.1,.507){\scriptsize $#4$}
\psline[linecolor=blue,linewidth=1.5pt](.35,-.6)(.35,1.6)
\psline[linecolor=blue,linewidth=1.5pt](1.05,-.6)(1.05,1.6)
\rput{45}(0.707,-0.2){
\facegrid{(0,0)}{(1,1)}
\pspolygon[fillstyle=solid,fillcolor=lightblue,linewidth=0pt](1,1)(0,1)(1,0)(1,1)
\psline[linestyle=dashed](0,1)(1,0)
\psarc[linewidth=1.5pt,linecolor=blue](0,0){0.5}{0}{90}
\psarc[linewidth=1.5pt,linecolor=blue](1,1){0.5}{180}{270}
}
\rput(0.35,-0.75){\scriptsize$j$}
\rput(1.05,-0.75){\scriptsize$j\!+\!1$}
\end{pspicture}
}

\def\Esquare{
\begin{pspicture}[shift=-1.17](-.2,-.8)(1.2,1.6)
\rput{45}(0.707,-0.2){
\facegrid{(0,0)}{(1,1)}
\pspolygon[fillstyle=solid,fillcolor=lightblue,linewidth=0pt](1,1)(0,1)(1,0)(1,1)
\psline[linestyle=dashed](0,1)(1,0)
\psarc[linewidth=1.5pt,linecolor=blue](0,0){0.5}{0}{90}
\psarc[linewidth=1.5pt,linecolor=blue](1,1){0.5}{180}{270}
}
\end{pspicture}
}

\def\Xsquare{
\begin{pspicture}[shift=-1.17](-.2,-.8)(1.2,1.6)
\rput{45}(0.707,-0.2){
\facegrid{(0,0)}{(1,1)}
\pspolygon[fillstyle=solid,fillcolor=lightblue,linewidth=0pt](1,1)(0,1)(1,0)(1,1)
\psline[linewidth=1.5pt,linecolor=blue](0,.5)(1,.5)
\psline[linewidth=1.5pt,linecolor=blue](.5,0)(.5,1)
}
\end{pspicture}
}

\def\TLLoop{
\begin{pspicture}[shift=-.77](1.8,1.8)
\rput[br]{-45}(0.18,1.614){\pspolygon[fillstyle=solid,fillcolor=lightlightblue,linewidth=0pt](0,0)(1,0)(1,1)(0,0)\psline(0,0)(1,0)(1,1)\psline[linestyle=dashed](0,0)(1,1)\psarc[linewidth=1.5pt,linecolor=blue](1,0){0.5}{90}{180}}
\rput{135}(1.6,0.2){\pspolygon[fillstyle=solid,fillcolor=lightblue,linewidth=0pt](0,0)(1,0)(1,1)(0,0)\psline(0,0)(1,0)(1,1)\psline[linestyle=dashed](0,0)(1,1)\psarc[linewidth=1.5pt,linecolor=blue](1,0){0.5}{90}{180}}
\psline[linestyle=dashed](0.2,0.2)(0.2,1.6)
\psline[linestyle=dashed](1.6,0.2)(1.6,1.6)
\pscircle[linecolor=blue,linewidth=1.5pt](0.9,0.9){0.53}
\rput(0.1,0.1){\scriptsize$a$}
\rput(1.7,0.1){\scriptsize$a$}
\rput(0.1,1.7){\scriptsize$a$}
\rput(1.7,1.7){\scriptsize$a$}
\rput(0.7,0.9){\scriptsize$b$}
\pscircle[fillstyle=solid,fillcolor=black](0.9,0.9){0.06}
\end{pspicture}
}

\def\XLoop{
\begin{pspicture}[shift=-.77](1.8,1.8)
\rput[br]{-45}(0.18,1.614){
\pspolygon[fillstyle=solid,fillcolor=lightlightblue,linewidth=0pt](0,0)(1,0)(1,1)(0,0)
\psline(0,0)(1,0)(1,1)
\psline[linestyle=dashed](0,0)(1,1)
}
\rput{135}(1.6,0.2){
\pspolygon[fillstyle=solid,fillcolor=lightblue,linewidth=0pt](0,0)(1,0)(1,1)(0,0)
\psline(0,0)(1,0)(1,1)
\psline[linestyle=dashed](0,0)(1,1)
\psline[linewidth=1.5pt,linecolor=blue](.5,0)(.5,.5)(1,.5)(1.5,0)(1.5,-.5)(1,-.5)(.5,0)
}
\psline[linestyle=dashed](0.2,0.2)(0.2,1.6)
\psline[linestyle=dashed](1.6,0.2)(1.6,1.6)
\rput(0.1,0.1){\scriptsize$a$}
\rput(1.7,0.1){\scriptsize$a$}
\rput(0.1,1.7){\scriptsize$a$}
\rput(1.7,1.7){\scriptsize$a$}
\rput(0.7,0.9){\scriptsize$b$}
\pscircle[fillstyle=solid,fillcolor=black](0.9,0.9){0.06}
\end{pspicture}
}

\def\XaLoop{
\begin{pspicture}[shift=-.77](1.8,1.8)
\rput[br]{-45}(0.18,1.614){
\pspolygon[fillstyle=solid,fillcolor=lightlightblue,linewidth=0pt](0,0)(1,0)(1,1)(0,0)
\psline(0,0)(1,0)(1,1)
}
\rput{135}(1.6,0.2){
\pspolygon[fillstyle=solid,fillcolor=lightblue,linewidth=0pt](0,0)(1,0)(1,1)(0,0)
\psline(0,0)(1,0)(1,1)
\psline[linewidth=1.5pt,linecolor=blue](.5,0)(.5,.5)(1,.5)(1.5,0)(1.5,-.5)(1,-.5)(.5,0)
}
\psline[linestyle=dashed](0.2,0.2)(0.2,1.6)
\psline[linestyle=dashed](1.6,0.2)(1.6,1.6)
\rput(0.1,0.1){\scriptsize$a\!\!+\!\!1$}
\rput(1.7,0.1){\scriptsize$a\!\!-\!\!1$}
\rput(0.1,1.7){\scriptsize$a\!\!+\!\!1$}
\rput(1.7,1.7){\scriptsize$a\!\!-\!\!1$}
\rput(0.7,0.9){\scriptsize$b$}
\pscircle[fillstyle=solid,fillcolor=black](0.9,0.9){0.06}
\end{pspicture}
}

\def\XbLoop{
\begin{pspicture}[shift=-.77](1.8,1.8)
\rput[br]{-45}(0.18,1.614){
\pspolygon[fillstyle=solid,fillcolor=lightlightblue,linewidth=0pt](0,0)(1,0)(1,1)(0,0)
\psline(0,0)(1,0)(1,1)
}
\rput{135}(1.6,0.2){
\pspolygon[fillstyle=solid,fillcolor=lightblue,linewidth=0pt](0,0)(1,0)(1,1)(0,0)
\psline(0,0)(1,0)(1,1)
\psline[linewidth=1.5pt,linecolor=blue](.5,0)(.5,.5)(1,.5)(1.5,0)(1.5,-.5)(1,-.5)(.5,0)
}
\psline[linestyle=dashed](0.2,0.2)(0.2,1.6)
\psline[linestyle=dashed](1.6,0.2)(1.6,1.6)
\rput(0.1,0.1){\scriptsize$a\!\!-\!\!1$}
\rput(1.7,0.1){\scriptsize$a\!\!+\!\!1$}
\rput(0.1,1.7){\scriptsize$a\!\!-\!\!1$}
\rput(1.7,1.7){\scriptsize$a\!\!+\!\!1$}
\rput(0.7,0.9){\scriptsize$b$}
\pscircle[fillstyle=solid,fillcolor=black](0.9,0.9){0.06}
\end{pspicture}
}

\def\EXLoop{
\begin{pspicture}[shift=-.77](1.8,1.8)
\rput[br]{-45}(0.18,1.614){
\pspolygon[fillstyle=solid,fillcolor=lightlightblue,linewidth=0pt](0,0)(1,0)(1,1)(0,0)
\psline(0,0)(1,0)(1,1)
\psline[linestyle=dashed](0,0)(1,1)
}
\rput{135}(1.6,0.2){
\pspolygon[fillstyle=solid,fillcolor=lightblue,linewidth=0pt](0,0)(1,0)(1,1)(0,0)
\psline(0,0)(1,0)(1,1)
\psline[linestyle=dashed](0,0)(1,1)
\psarc[linewidth=1.5pt,linecolor=blue](1,0){0.5}{90}{180}
\psline[linewidth=1.5pt,linecolor=blue](1,.5)(1.5,0)(1.5,-.5)(1,-.5)(.5,0)
}
\psline[linestyle=dashed](0.2,0.2)(0.2,1.6)
\psline[linestyle=dashed](1.6,0.2)(1.6,1.6)
\rput(0.1,0.1){\scriptsize$a$}
\rput(1.7,0.1){\scriptsize$a$}
\rput(0.1,1.7){\scriptsize$a$}
\rput(1.7,1.7){\scriptsize$a$}
\rput(0.7,0.9){\scriptsize$b$}
\pscircle[fillstyle=solid,fillcolor=black](0.9,0.9){0.06}
\end{pspicture}
}

\def\XELoop{
\begin{pspicture}[shift=-.77](1.8,1.8)
\rput[br]{-45}(0.18,1.614){
\pspolygon[fillstyle=solid,fillcolor=lightlightblue,linewidth=0pt](0,0)(1,0)(1,1)(0,0)
\psline(0,0)(1,0)(1,1)
\psline[linestyle=dashed](0,0)(1,1)
}
\rput{135}(1.6,0.2){
\pspolygon[fillstyle=solid,fillcolor=lightblue,linewidth=0pt](0,0)(1,0)(1,1)(0,0)
\psline(0,0)(1,0)(1,1)
\psline[linestyle=dashed](0,0)(1,1)
\psarc[linewidth=1.5pt,linecolor=blue](1,0){0.5}{-90}{0}
\psline[linewidth=1.5pt,linecolor=blue](1,-.5)(.5,0)(.5,.5)(1,.5)(1.5,0)
}
\psline[linestyle=dashed](0.2,0.2)(0.2,1.6)
\psline[linestyle=dashed](1.6,0.2)(1.6,1.6)
\rput(0.1,0.1){\scriptsize$a$}
\rput(1.7,0.1){\scriptsize$a$}
\rput(0.1,1.7){\scriptsize$a$}
\rput(1.7,1.7){\scriptsize$a$}
\rput(0.7,0.9){\scriptsize$b$}
\pscircle[fillstyle=solid,fillcolor=black](0.9,0.9){0.06}
\end{pspicture}
}

\def\TLUII{
\begin{pspicture}(2.6,1)
\rput(0,-.8){
\pspolygon[fillstyle=solid,fillcolor=lightlightblue,linewidth=0pt](.2,1.614)(.9,.9)(1.614,1.614)(.2,1.614)
\rput(1.4,.8){\Esquare}
\pspolygon[fillstyle=solid,fillcolor=lightblue,linewidth=0pt](1.6,.2)(.2,.2)(.9,.9)(1.6,.2)
\psline[linestyle=dashed](1.614,0.2)(.2,.2)(.2,1.614)(1.614,1.614)
\psline(.2,.2)(.9,.9)(.2,1.614)
\psarc[linecolor=blue,linewidth=1.5pt](0.9,.9){.51}{45}{315}
\rput(.1,.1){\scriptsize$a$}
\rput(1.7,.1){\scriptsize$a$}
\rput(.7,.9){\scriptsize$d$}
\rput(2.4,.9){\scriptsize$d$}
\rput(.1,1.7){\scriptsize$a$}
\rput(1.7,1.7){\scriptsize$a$}
}
\end{pspicture}
}

\def\XUII{
\begin{pspicture}(2.6,1)
\rput(0,-.8){
\pspolygon[fillstyle=solid,fillcolor=lightlightblue,linewidth=0pt](.2,1.614)(.9,.9)(1.614,1.614)(.2,1.614)
\rput(1.4,.8){\Xsquare}
\pspolygon[fillstyle=solid,fillcolor=lightblue,linewidth=0pt](1.6,.2)(.2,.2)(.9,.9)(1.6,.2)
\psline[linestyle=dashed](1.614,0.2)(.2,.2)(.2,1.614)(1.614,1.614)
\psline(.2,.2)(.9,.9)(.2,1.614)
\psarc[linecolor=blue,linewidth=1.5pt](0.9,.9){.51}{45}{315}
\rput(.1,.1){\scriptsize$a$}
\rput(1.7,.1){\scriptsize$a$}
\rput(.7,.9){\scriptsize$e$}
\rput(2.4,.9){\scriptsize$d$}
\rput(.1,1.7){\scriptsize$a$}
\rput(1.7,1.7){\scriptsize$a$}
\pscircle[fillstyle=solid,fillcolor=black](.9,.9){0.06}
}
\end{pspicture}
}

\def\TLib{
\raisebox{-1.22cm}{\begin{pspicture}(0,-1.2)(2,1.3)
\rput(.8,0){\Ejr {b}a{c}a}
\end{pspicture}}
}

\def\Xia{
\raisebox{-1.22cm}{\begin{pspicture}(0,-1.2)(1.8,1.3)
\rput(.8,0){\Xjr {b}{a}{c}{a}}
\end{pspicture}}
}
\def\Xiaa{
\raisebox{-1.22cm}{\begin{pspicture}(0,-1.2)(1.8,1.3)
\rput(.8,0){\Xjr {b}{\raisebox{3pt}{$\,\,a'$}}{c}{a}}
\end{pspicture}}
}

\def\Xib{
\raisebox{-1.22cm}{\begin{pspicture}(-.1,-1.2)(2.1,1.3)
\rput(.8,0){\Xjr {b}{\ \ \ a\!\!-\!\!1}{c}{\;a\!\!+\!\!1\ \ \ }}
\end{pspicture}}
}
\def\Xibb{
\raisebox{-1.22cm}{\begin{pspicture}(-.1,-1.2)(2.1,1.3)
\rput(.8,0){\Xjr {b}{\ \ \ a\!\!+\!\!1}{c}{\;a\!\!-\!\!1\ \ \ }}
\end{pspicture}}
}

\def\Xjr#1#2#3#4{\begin{pspicture}[shift=-1.17](-.2,-.8)(1.2,1.6)
\rput(0.707,-.3){\scriptsize $#1$}
\rput(1.514,.507){\scriptsize $#2$}
\rput(0.707,1.364){\scriptsize $#3$}
\rput(-.1,.507){\scriptsize $#4$}
\psline[linecolor=blue,linewidth=1.5pt](.35,-.6)(.35,1.6)
\psline[linecolor=blue,linewidth=1.5pt](1.05,-.6)(1.05,1.6)
\rput{45}(0.707,-0.2){
\facegrid{(0,0)}{(1,1)}
\pspolygon[fillstyle=solid,fillcolor=lightblue,linewidth=0pt](1,1)(0,1)(1,0)(1,1)
\psline[linestyle=dashed](0,1)(1,0)
}
\rput(0.35,-0.75){\scriptsize$j$}
\rput(1.05,-0.75){\scriptsize$j\!+\!1$}
\psline[linecolor=blue,linewidth=1.5pt](.34,.14)(1.05,.9)
\psline[linecolor=blue,linewidth=1.5pt](.34,.9)(1.05,.14)

\end{pspicture}
}

In the RSOS representation, the matrices representing $E_j$ and $\beta\,\Xi_j$ admit the following non-zero rank-1 factorized blocks
\bea
E_j:\qquad \TLib=\;\vec e_{b,a,a}^T \tilde{\vec e}_{a,a,c}
\eea 
\bea
\beta\,\Xi_j:\qquad \Xia=\vec x_{b,a,a}^T \tilde{\vec x}_{a,a,c},\qquad  \Xib=\;\Xibb=\vec y_{b,a,a}^T \tilde{\vec y}_{a,a,c}
\eea 
Here $T$ denotes the transpose and the triangle weights are given by the row vectors
\begin{subequations}
\begin{align}
\vec e_{b,a,a}&=\frac{1}{S_{a-1}S_{a+1}}(S_{a-1},S_a,S_{a+1}),\qquad& \tilde{\vec e}_{a,a,c}&=\frac{1}{S_a}(S_{a+1}S_{a+2},S_{a-1}S_{a+1},S_{a-2}S_{a-1})\\
\vec x_{b,a,a}&=\frac{1}{S_{a-1}S_{a+1}}(-S_{a-1}^2,S_{2a},S_{a+1}^2),\qquad& \tilde{\vec x}_{a,a,c}&=\frac{1}{S_a^2}(-S_aS_{a+2},S_{2a},S_{a-2}S_a)\\
\vec y_{b,a,a}&=(1,1),\qquad& \tilde{\vec y}_{a,a,c}&=\frac{1}{S_a}(S_{a+2},S_{a-2})
\end{align}
\end{subequations}
where, for fixed $a$, the vector entries are labelled by $b,c=a+2,a,a-2$ and $b,c=a+1,a-1$ respectively. 
Using the relations
\begin{subequations}
\begin{align}
\disp\tilde{\vec e}_{a,a,b}\cdot\vec e_{b,a,a}&=\TLLoop=\frac{S_{a+2}+S_a+S_{a-2}}{S_a}=x^2+1+x^{-2}=\beta_2\\
\disp\tilde{\vec x}_{a,a,b}\cdot\vec x_{b,a,a}&=\XLoop=\frac{S_{a-1}^2S_aS_{a+2}+S_{2a}^2+S_{a-2}S_aS_{a+1}^2}{S_{a-1}S_a^2 S_{a+1}}=x^2+x^{-2}=\frac{\beta_3}{\beta}\\
\disp\tilde{\vec y}_{a,a,b}\cdot\vec y_{b,a,a}&=\XaLoop=\XbLoop=\frac{S_{a+2}+S_{a-2}}{S_a}=x^2+x^{-2}=\frac{\beta_3}{\beta}\\
\disp\tilde{\vec e}_{a,a,b}\cdot\vec x_{b,a,a}&=\EXLoop=\frac{-S_{a-1}S_{a+2}+S_{2a}+S_{a-2}S_{a+1}}{S_a}=0\\
\disp\tilde{\vec x}_{a,a,b}\cdot\vec e_{b,a,a}&=\XELoop=\frac{-S_{a-1}S_{a+2}+S_{2a}+S_{a-2}S_{a+1}}{S_{a-1}S_aS_{a+1}}=0
\end{align}
\end{subequations}
it follows that, after suitable normalization, $E_j$ and $\Xi_j$ are commuting orthogonal idempotents. This is seen graphically as
\bea
E_j^2=\TLia =\beta_2\, \TLib\!\!=\beta_2 E_j,\qquad
\beta^2\Xi_j^2=\XXia =\frac{\beta_3}{\beta}\, \Xiaa=\frac{\beta_3}{\beta}\,\Xi_j
\eea
Similarly, the relations $E_j E_{j\pm 1} E_{j} = E_j$ and $E_j \,\Xi_{j\pm 1} E_{j}=E_j$ follows graphically as
\bea
E_j E_{j+1} E_{j} =\TLiiA = \TLiiB=E_j,\qquad\qquad
\TLUII =1
\eea
\bea
\beta E_j \,\Xi_{j\pm 1} E_{j} =\XiiA = \frac{\beta_3}{\beta} \TLiiB=\frac{\beta_3}{\beta}\,E_j,\qquad\qquad
\XUII = \frac{\beta_3}{\beta}
\eea

In addition, setting $Y_j=\beta \Xi_j$, the generators satisfy the following cubic relations in accord with (3.34) of \cite{TP}
\begin{subequations}
\begin{align}
E_j Y_{j\pm 1}E_j&=\tfrac{\beta_3}{\beta}\,E_j\\
Y_jE_{j\pm1}E_j&=(Y_{j\pm1}+E_{j\pm1}-1)E_j\\
Y_jY_{j\pm1}E_j&=\big(\tfrac{\beta_3}{\beta}-1\big)(Y_{j\pm1}+E_{j\pm1}-1)E_j\\
(Y_j+E_j)E_{j\pm1}(Y_j+E_j)&=(Y_{j\pm1}+E_{j\pm1})E_j(Y_{j\pm1}+E_{j\pm1})\\
Y_jY_{j\pm1}Y_j-Y_{j\pm1}Y_jY_{j\pm1}&=\beta^2(E_{j\pm1}Y_j-E_jY_{j\pm1}+Y_jE_{j\pm1}-Y_{j\pm1}E_j+E_j-E_{j\pm1})+Y_j-Y_{j\pm1}
\end{align}
\end{subequations}

Let $Z_j=Y_j+E_j$, then
\begin{subequations}
\begin{align}
E_j Z_{j\pm 1}E_j&=\beta_2\,E_j\\
Z_jE_{j\pm1}E_j&=Z_{j\pm1}\,E_j\\
Z_jZ_{j\pm1}E_j&=((\beta_2-1)Z_{j\pm1}+1)E_j\\
Z_j\,E_{j\pm1}Z_j&=Z_{j\pm1}\,E_j\,Z_{j\pm1}\\
Y_jY_{j\pm1}Y_j-Y_{j\pm1}Y_jY_{j\pm1}&=\beta^2(E_{j\pm1}Y_j-E_jY_{j\pm1}+Y_jE_{j\pm1}-Y_{j\pm1}E_j+E_j-E_{j\pm1})+Y_j-Y_{j\pm1}
\end{align}
\end{subequations}

Expanding the Yang-Baxter equation 
\bea
\mathbb{X}_j(u)\mathbb{X}_{j\pm1}(u+v)\mathbb{X}_j(v)=\mathbb{X}_{\j\pm1}(v)\mathbb{X}_j(u+v)\mathbb{X}_{j\pm1}(u)
\eea
in terms of the face operators (\ref{eq:fusedFace}) as a multivariable Laurent polynomial in $z=e^{i u}$ and $w=e^{i v}$, equating coefficients and using these cubic relations, it follows that the Yang-Baxter equations is satisfied.


\end{document}